\documentclass[
  aps,
  prd,
  nofootinbib,
  onecolumn,
  twocolumn,
  preprintnumbers,
  superscriptaddress,
  floatfix,
  10pt
]{revtex4-2}

\usepackage{placeins}
\usepackage{silence} 
\usepackage[
  pdftex,
  colorlinks=true,
  pdfpagemode=UseNone,
  bookmarksopen=true,
]{hyperref}

\usepackage[utf8]{inputenc}
\usepackage{amsmath} 
\usepackage{amssymb} 
\usepackage{calc}
\usepackage{color}
\usepackage{dcolumn} 
\usepackage{graphicx}
\usepackage[normalem]{ulem} 

\newif\ifdocomment
\docommentfalse
\docommenttrue
\ifdocomment
\newcommand{\comment}[1]{\textcolor{red}{[#1]}}
\newcommand{\highlight}[1]{\textcolor{red}{#1}}
\newcommand{\str}[1]{}
\newcommand{\postsubmissionStrike}[1]{\sout{#1}}

\else
\newcommand{\comment}[1]{\PackageError{comment}{unresolved comment}{}}
\newcommand{\highlight}[1]{\PackageError{comment}{unresolved highlight}{}}
\newcommand{\str}[1]{}
\newcommand{\postsubmissionStrike}[1]{}

\fi

\newcommand{\kmax}{\ensuremath{k_{\rm max}}}
\newcommand{\tc}{\ensuremath{t_{c}}}
\newcommand{\tz}{\ensuremath{t_{0}}}

\newcommand{\Nuc}{\ensuremath{N}}
\newcommand{\Deut}{\text{D}}
\newcommand{\z}{\ensuremath{z}}
\newcommand{\zexp}{\z\text{ expansion}}
\newcommand{\pvalue}{\ensuremath{p}\text{ value}}
\newcommand{\pv}{\ensuremath{p}} 
\newcommand{\Qtmin}{\ensuremath{Q^{2}_{\rm min}}}

\newcommand{\rasq}{\ensuremath{r_{A}^{2}}}

\newcommand{\GeV}{\ensuremath{{\rm GeV}}}
\newcommand{\GeVsq}{\ensuremath{{\rm GeV}^{2}}}
\newcommand{\GeVc}{\ensuremath{{\rm GeV}/c}}
\newcommand{\fmsq}{\ensuremath{{\rm fm}^{2}}}

\newcommand{\rfr}[1]{Ref.~\cite{#1}}
\newcommand{\eqn}[1]{Eqn.~(\ref{#1})}
\newcommand{\fgr}[1]{Fig.~\ref{#1}}
\newcommand{\tbl}[1]{Tab.~\ref{#1}}
\newcommand{\sct}[1]{Sect.~\ref{#1}}
\newcommand{\CAPrfr}[1]{Reference~\cite{#1}}

\newcommand{\rfrs}{Refs.}
\newcommand{\eqns}{Eqns.}
\newcommand{\fgrs}{Figs.}
\newcommand{\tbls}{Tabs.}
\newcommand{\scts}{Sects.}

\newcommand{\anl}{\text{ANL}}
\newcommand{\bnl}{\text{BNL}}
\newcommand{\fnal}{\text{FNAL}}
\newcommand{\bebc}{\text{BEBC}}
\newcommand{\minerva}{\text{MINERvA}}

\begin{document}
\title{The Nucleon Axial Form Factor from Elementary Target Data}

\preprint{LLNL-JRNL-2014317,FERMILAB-PUB-25-0912-LBNF-T}

\newcommand{\Rutgers}{Rutgers, The State University of New Jersey, Piscataway, New Jersey 08854, USA}
\newcommand{\Hampton}{Hampton University, Dept. of Physics, Hampton, VA 23668, USA}
\newcommand{\Dortmund}{Institute of Physics, Dortmund University, 44221, Germany }
\newcommand{\Otterbein}{Department of Physics, Otterbein University, 1 South Grove Street, Westerville, OH, 43081 USA}
\newcommand{\JMU}{James Madison University, Harrisonburg, Virginia 22807, USA}
\newcommand{\Florida}{University of Florida, Department of Physics, Gainesville, FL 32611}
\newcommand{\UCIrvine}{Department of Physics and Astronomy, University of California, Irvine, Irvine, California 92697-4575, USA}
\newcommand{\CBPF}{Centro Brasileiro de Pesquisas F\'{i}sicas, Rua Dr. Xavier Sigaud 150, Urca, Rio de Janeiro, Rio de Janeiro, 22290-180, Brazil}
\newcommand{\PUCP}{Secci\'{o}n F\'{i}sica, Departamento de Ciencias, Pontificia Universidad Cat\'{o}lica del Per\'{u}, Apartado 1761, Lima, Per\'{u}}
\newcommand{\INRM}{Institute for Nuclear Research of the Russian Academy of Sciences, 117312 Moscow, Russia}
\newcommand{\Jlab}{Jefferson Lab, 12000 Jefferson Avenue, Newport News, VA 23606, USA}
\newcommand{\Pittsburgh}{Department of Physics and Astronomy, University of Pittsburgh, Pittsburgh, Pennsylvania 15260, USA}
\newcommand{\Guanajuato}{Campus Le\'{o}n y Campus Guanajuato, Universidad de Guanajuato, Lascurain de Retana No. 5, Colonia Centro, Guanajuato 36000, Guanajuato M\'{e}xico.}
\newcommand{\Athens}{Department of Physics, University of Athens, GR-15771 Athens, Greece}
\newcommand{\Tufts}{Physics Department, Tufts University, Medford, Massachusetts 02155, USA}
\newcommand{\WM}{Department of Physics, William \& Mary, Williamsburg, Virginia 23187, USA}
\newcommand{\FNAL}{Fermi National Accelerator Laboratory, Batavia, Illinois 60510, USA}
\newcommand{\Purdue}{Department of Chemistry and Physics, Purdue University Calumet, Hammond, Indiana 46323, USA}
\newcommand{\MCLA}{Massachusetts College of Liberal Arts, 375 Church Street, North Adams, MA 01247}
\newcommand{\UMD}{Department of Physics, University of Minnesota -- Duluth, Duluth, Minnesota 55812, USA}
\newcommand{\Northwestern}{Northwestern University, Evanston, Illinois 60208}
\newcommand{\UNI}{Facultad de Ciencias F\'{i}sicas, Universidad Nacional Mayor de San Marcos, CP 15081, Lima, Per\'{u}}
\newcommand{\Rochester}{Department of Physics and Astronomy, University of Rochester, Rochester, New York 14627 USA}
\newcommand{\Austin}{Department of Physics, University of Texas, 1 University Station, Austin, Texas 78712, USA}
\newcommand{\USM}{Departamento de F\'{i}sica, Universidad T\'{e}cnica Federico Santa Mar\'{i}a, Avenida Espa\~{n}a 1680 Casilla 110-V, Valpara\'{i}so, Chile}
\newcommand{\Geneva}{University of Geneva, 1211 Geneva 4, Switzerland}
\newcommand{\Chicago}{Enrico Fermi Institute, University of Chicago, Chicago, IL 60637 USA}
\newcommand{\OregonState}{Department of Physics, Oregon State University, Corvallis, Oregon 97331, USA}
\newcommand{\oxford}{Oxford University, Department of Physics, Oxford, OX1 3PJ United Kingdom}
\newcommand{\umiss}{University of Mississippi, Oxford, Mississippi 38677, USA}
\newcommand{\upenn}{Department of Physics and Astronomy, University of Pennsylvania, Philadelphia, PA 19104}
\newcommand{\AMU}{Department of Physics, Aligarh Muslim University, Aligarh, Uttar Pradesh 202002, India}
\newcommand{\wroclaw}{University of Wroclaw, plac Uniwersytecki 1, 50-137 Wroa\l{}aw, Poland}
\newcommand{\Mohali}{Department of Physical Sciences, IISER Mohali, Knowledge City, SAS Nagar, Mohali - 140306, Punjab, India}
\newcommand{\CINVESTAV}{Departamento de Fisica Col. San Pedro Zacatenco, 07360 Mexico, DF, Av. Instituto PolitÃ©cnico Nacional, Mexico}
\newcommand{\york}{York University, Department of Physics and Astronomy, Toronto, Ontario, M3J 1P3 Canada}
\newcommand{\ND}{Department of Physics and Astronomy, University of Notre Dame, Notre Dame, Indiana 46556, USA}
\newcommand{\ICL}{The Blackett Laboratory,  Imperial College London,  London SW7 2BW, United Kingdom}
\newcommand{\warwick}{Department of Physics, University of Warwick, Coventry, CV4 7AL, UK}
\newcommand{\qmul}{G O Jones Building, Queen Mary University of London, 327 Mile End Road, London E1 4NS, UK}
\newcommand{\LLNL}{Nuclear and Chemical Sciences Division, Lawrence Livermore National Laboratory, Livermore, California 94550, USA}
\newcommand{\kentucky}{Department of Physics and Astronomy, University of Kentucky, Lexington, Kentucky 40506, USA}

\newcommand{\mmooreThanks}{now at SLAC National Accelerator Laboratory and Stanford University Department of Physics, Menlo Park, CA 94025, USA}
\newcommand{\ricfregianThanks}{now at Department of Physics and Astronomy, University of California at Davis, Davis, CA 95616, USA}
\newcommand{\kleykampThanks}{now at Department of Physics and Astronomy, University of Mississippi, Oxford, MS 38677}
\newcommand{\adrianThanks}{Now at Department of Physics, Drexel University, Philadelphia, Pennsylvania 19104, USA}
\newcommand{\aolivierThanks}{now at Argonne National Laboratory, Lemont, IL 60439 USA}
\newcommand{\lazazuetareyesThanks}{now at Syracuse University, Syracuse, NY 13244, USA}


\author{A.S.~Meyer}                       \affiliation{\LLNL}
\author{T.~Cai}                           \affiliation{\york}  \affiliation{\Rochester}
\author{M.~Moore}\thanks{\mmooreThanks}   \affiliation{\Rochester}
\author{S.~Akhter}                        \affiliation{\AMU}
\author{Z.~Ahmad~Dar}                     \affiliation{\WM}  \affiliation{\AMU}
\author{M.~Sajjad~Athar}                  \affiliation{\AMU}
\author{M.~Betancourt}                    \affiliation{\FNAL}
\author{H.~Budd}                          \affiliation{\Rochester}
\author{G.~Caceres}\thanks{\ricfregianThanks}  \affiliation{\CBPF}
\author{D.S.~Correia}                     \affiliation{\CBPF}
\author{G.A.~D\'{i}az~}                   \affiliation{\FNAL}  \affiliation{\Rochester}
\author{J.~Felix}                         \affiliation{\Guanajuato}
\author{A.M.~Gago}                        \affiliation{\PUCP}
\author{H.~Gallagher}                     \affiliation{\Tufts}
\author{P.K.~Gaur}                         \affiliation{\AMU}
\author{S.M.~Gilligan}                    \affiliation{\OregonState}
\author{R.~Gran}                          \affiliation{\UMD}
\author{E.~Granados}                       \affiliation{\Guanajuato}  \affiliation{\Guanajuato}
\author{D.A.~Harris}                      \affiliation{\york}  \affiliation{\FNAL}
\author{A.L.~Hart}                        \affiliation{\qmul}
\author{R.J.~Hill}                        \affiliation{\kentucky} \affiliation{\FNAL}
\author{J.~Kleykamp}\thanks{\kleykampThanks}  \affiliation{\Rochester}
\author{A.~Klustov\'{a}}                  \affiliation{\ICL}
\author{M.~Kordosky}                      \affiliation{\WM}
\author{D.~Last}                          \affiliation{\Rochester}  \affiliation{\upenn}
\author{A.~Lozano}\thanks{\adrianThanks}  \affiliation{\CBPF}
\author{S.~Manly}                         \affiliation{\Rochester}
\author{W.A.~Mann}                        \affiliation{\Tufts}
\author{K.S.~McFarland}                   \affiliation{\Rochester}
\author{O.~Moreno}                        \affiliation{\WM}  \affiliation{\Guanajuato}
\author{J.K.~Nelson}                      \affiliation{\WM}
\author{A.~Olivier}\thanks{\aolivierThanks}  \affiliation{\ND}\affiliation{\Rochester} 
\author{V.~Paolone}                       \affiliation{\Pittsburgh}
\author{G.N.~Perdue}                      \affiliation{\FNAL}  \affiliation{\Rochester}
\author{C.~Pernas}                        \affiliation{\WM}
\author{M.A.~Ram\'{i}rez}                 \affiliation{\upenn}  \affiliation{\Guanajuato}
\author{R.D.~Ransome}                     \affiliation{\Rutgers}
\author{D.~Ruterbories}                   \affiliation{\Rochester}
\author{H.~Schellman}                     \affiliation{\OregonState}
\author{C.J.~Solano~Salinas}              \affiliation{\UNI}
\author{N.H.~Vaughan}                     \affiliation{\OregonState}
\author{M.O.~Wascko}                      \affiliation{\oxford}  \affiliation{\ICL}
\author{L.~Zazueta}\thanks{\lazazuetareyesThanks}  \affiliation{\WM}


\date{\today}

\collaboration{The MINERvA Collaboration}\ \noaffiliation

\begin{abstract}
Precise neutrino-nucleon amplitudes are essential ingredients for predicting neutrino event rates in current and upcoming long-baseline neutrino oscillation experiments. A common neutrino interaction with a low reaction threshold and with most of the energy carried by two final state particles is quasielastic scattering, for which the nucleon axial form factor, $F_{A}(Q^{2})$, is a dominant source of uncertainty. Improvements to the nucleon axial form factor rely on neutrino scattering data with elementary targets to reduce or eliminate the need for nuclear modeling systematics. This work examines constraints on the nucleon axial form factor that can be achieved from datasets of neutrino scattering on deuterium targets, Lattice QCD predictions, and from the recent hydrogen target data from the MINERvA Collaboration. Significant tension is found between hydrogen and deuterium target data, suggesting that extractions from deuterium underestimate both the central value and uncertainty of the form factor. Parameterizations for and uncertainties of the nucleon axial form factor using the $z$ expansion are provided.
\end{abstract}
\maketitle

\section{Introduction}

Explorations of the properties of neutrinos and their flavor oscillation
 are entering a new era of precision.
Current and upcoming flagship long baseline neutrino oscillation experimental programs
 will measure neutrino interactions with exceptional precision,
 thereby enabling new insights about the neutrino mass hierarchy
 and CP violation in the leptonic sector~\cite{
NOvA:2007rmc,
JUNO:2015sjr,
DUNE:2015lol,
Hyper-Kamiokande:2018ofw,
DUNE:2020ypp}.
In support of the substantial ongoing experimental efforts,
 new theoretical guidance is needed to help meet the ambitious
 precision requirements of next-generation neutrino scattering experiments.
With help from improved theory constraints, these experiments can maximize
 their potential for investigating the physics of neutrinos.

Improving the theoretical description of neutrino-nucleus cross sections is not a simple task.
Practical limitations prevent the creation of a neutrino beam
 that is simultaneously narrowly-peaked around a single energy
 and sufficiently intense for a high-statistics measurement.
Modern neutrino oscillation experiments therefore cover a range of energies
 with several different physical mechanisms at play.
To extract a spectrum of neutrino interactions at both a near and far detector,
 which in turn gives access to the oscillation probability as a function of energy,
 distributions of neutrinos must be statistically reconstructed
 under the assumption of a model that encompasses all of the known interaction physics.
It is therefore of interest to understand and quantify several neutrino interaction
 processes over a range of incident energies.

Isolated exclusive interaction channels that have both large cross sections
 and large uncertainties are the primary targets for theoretical improvements.
The most prominent of these is neutrino quasielastic scattering,
 in which a neutrino scatters off of a quasi-free nucleon within a bound nuclear system.
Neutrino quasielastic scattering is a primary signal measurement process of
 long baseline neutrino oscillation experiments
 and the dominant process at play for small neutrino energies.
The weak interactions characteristic of neutrino scattering
 are sensitive to a nucleon axial current that is not probed
 by electromagnetic interactions.
The axial current interaction is therefore not nearly as well-constrained
 by experimental scattering measurements as the vector current interactions
 that are a part of the electromagnetic interaction.

Despite being less constrained than the vector form factors,
 the axial form factor has historically been quoted with very small uncertainties.
Those small uncertainties result from the use of the dipole parameterization,
 \begin{equation}
 F_A(Q^2)=\frac{g_A}{(1+Q^2/M_A^2)^2},
 \end{equation}
 which has a single free parameter, $M_A$, the ``axial mass".
This parameterization has insufficient freedom to describe the set of shapes
 consistent with theoretical constraints and existing
 experimental data~\cite{Bernard:2001rs,Bodek:2007ym}.
Reanalysis with neutrino-deuterium scattering data and a form factor
 with more parameterized freedom revealed that the form factor uncertainty
 should be at least an order of magnitude larger to accommodate
 the full range of possible form factor shapes~\cite{Meyer:2016oeg}.
There is also recent evidence from Monte Carlo tunes suggesting
 that the tuned neutrino quasielastic cross sections are larger than
 previous fits for the axial from factor from the deuterium
 data~\cite{MicroBooNE:2021ccs,GENIE:2022qrc}.
\newcommand{\lqcd}{\text{LQCD}}%
This evidence appeared simultaneously
 with first principles calculations of Lattice Quantum Chromodynamics (\lqcd),
 which suggest that the axial form factor is underestimated
 at large momentum transfer by almost 30\%~\cite{
 Meyer:2022mix,Tomalak:2023pdi}.
A primary interest is to quantify modern constraints of the nucleon axial form factor
 accurately and with theoretically robust uncertainties.
For this purpose, sources that are free or nearly-free nucleon targets are most useful.
This includes neutrino-deuterium scattering experimental data 
 analyzed in \rfr{Meyer:2016oeg},
 which in this work are studied in conjunction with
 the addition of data from the \bebc{} neutrino-deuterium
 experiment~\cite{Wachsmuth:1979th,Barlag:1984uga,Allasia:1990uy}.
This analysis also uses the antineutrino-hydrogen scattering experimental data
 of \rfr{MINERvA:2023avz}, which is measured on a hydrocarbon target
 with kinematic separation of the events on hydrogen from the carbon background.
Complementary constraints originating from \lqcd{}~are
 also explored in a sister paper~\cite{Meyer:inprep},
 which will be compared to and combined with experimental sources in the present work.

Additional constraints
 on the slope of the form factor near zero momentum transfer
 could be obtained from additional pion electroproduction data~\cite{Bernard_2001}
 and muonic hydrogen~\cite{Hill:2017wgb}.
The slope can be extracted from pion electroproduction data
 by comparing to a prediction obtained from expanding the amplitude
 for small $Q^{2}$ and $M_{\pi}$ in Heavy Baryon Chiral Perturbation Theory
 at one-loop order, such as in \rfr{Bernard:1992ys},
 where the slope shows up as a low energy constant in the effective theory.
The pion electroproduction amplitudes are typically computed close
 to the pion production threshold, around $Q^{2}\sim0.1~\GeVsq$~\cite{%
 Blomqvist:1996tx,A1:1999kwj,Hilt:2013fda},
 and at this scale corrections can enter at the level of 10\%~\cite{Bernard:1992ys}.
For more details on the relationship between
 invariant amplitudes and the differential cross sections,
 see for example \rfr{Hilt:2013fda}.
The muonic hydrogen capture rate is also sensitive to the axial form factor
 at a momentum transfer fixed by kinematics,
 around a value of $Q^{2}\sim0.01~\GeVsq$~\cite{%
 Czarnecki:2007th,MuCap:2012lei,MuCap:2015boo}.
Constraints on the axial radius from muon capture enter at about
 the 50\% level~\cite{Hill:2017wgb}, with an uncertainty budget dominated by statistics.
Studies of constraints on the slope from these other sources
 are outside the scope of this work.

The sections in this paper are organized as follows.
In \sct{sec:fitdetails}, details of the experimental datasets and fits
 employed in this work are discussed.
\sct{sec:fitting} discusses the particulars of fitting data and evaluation
 of systematic uncertainties based on these fits.
\sct{sec:results} contains the final results from fitting
 including quantified uncertainties.
\sct{sec:discussion} discusses the results of this work and their implications
 on experiments that involve neutrino scattering.

\section{Fit Details}
\label{sec:fitdetails}

This section focuses on the details associated with fitting data
 from neutrino quasielastic scattering off of both hydrogen and deuterium targets.
The primary observable of interest is the flux-integrated neutrino-nucleon
 quasielastic scattering cross section.
The differential cross section is used to fit the nucleon axial form factor, $F_{A}$,
 as a function of the spacelike 4-momentum transfer squared, $Q^{2}$.

After introducing the formalism in \sct{sec:differentialcrosssection},
 the parameterization for $F_{A}$ used in this work
 is outlined in
 \scts~\ref{sec:zexpansion}~and~\ref{sec:zexpansion_regularization}.
\sct{sec:datasets} gives an overview of the datasets and systematic
 corrections applied to those datasets that are relevant for this work.
The remaining \scts~\ref{sec:handling_event}~and~\ref{sec:handling_differential}
 deal with the specifics for making theory predictions of the flux-integrated
 differential cross section from the available information about the datasets.

\subsection{Differential Quasielastic Scattering Cross Section}
\label{sec:differentialcrosssection}

The neutrino-nucleon quasielastic cross section for an isosymmetric,
 unpolarized target is given as a function of $Q^{2}$ and neutrino energy $E_{\nu}$ by
\cite{LlewellynSmith:1971uhs,Formaggio:2012cpf}
\begin{align}
 \frac{d\sigma_{N}}{dQ^{2}}(&Q^{2},E_{\nu}) =
 \frac{G_{F}^{2} M_{N}^{2} |V_{ud}|^{2}}{8\pi E_{\nu}^{2}}
 \Biggr[
 \nonumber\\&
 A(Q^{2}) \pm \frac{(s-u)}{M_{N}^{2}} B(Q^{2})
 +\frac{(s-u)^{2}}{M_{N}^{4}} C(Q^{2})
 \Biggr],
 \label{eq:differential_crosssection}
\end{align}
 where the sign on the $B$ term is taken to be positive (negative)
 for neutrino (antineutrino) scattering
 and the neutrino is assumed to be massless.
In this expression,
 $G_{F}$ is the Fermi constant,
 $M_{N}$ is the average nucleon mass,
 $V_{ud}$ is the CKM matrix element,
 and the kinematic expansion parameter is
\begin{align}
 s-u = 4M_{N}E_{\nu} -Q^{2} -m_{\ell}^{2},
\end{align}
 which depends on the lepton mass $m_{\ell}$.
Neglecting second-class currents,
 the expressions $A$, $B$, and $C$ are given in terms of form factors,
\begin{align}
 A &=
 \frac{m_{\ell}^{2} + Q^{2}}{M_{N}^{2}}
 \Biggr[
 \nonumber\\&
 (1+\tau) F_{A}^{2}
 + \tau (F_{1} +F_{2})^{2}
 - (F_{1} -\tau F_{2})^{2}
 \nonumber\\&
 -\frac{m_{\ell}^{2}}{4M_{N}^{2}} \biggr(
 (F_{1}+F_{2})^{2} + (F_{A} +2F_{P})^{2} -4(1+\tau)F_{P}^{2}
 \biggr) \Biggr],
 \\
 B &= 4\tau F_{A} (F_{1} + F_{2}),
 \\
 C &= \frac{1}{4} (F_{A}^{2} +F_{1}^{2} + \tau F_{2}^{2}).
 \label{eq:differential_xsec_c}
\end{align}
These expressions use the dimensionless kinematic parameter $\tau = Q^{2}/4M_{N}^{2}$.
In the limit of high $E_{\nu}$,
 the kinematic prefactors on the $A$ and $B$ terms
 become small compared to that of $C$,
 and \eqn{eq:differential_crosssection}
 becomes essentially independent of $E_{\nu}$.

The four relevant form factors are
 the vector form factors $F_{1}$ and $F_{2}$,
 the induced pseudoscalar form factor $F_{P}$,
 and the axial form factor $F_{A}$,
 which all have functional dependence on $Q^{2}$.
The vector form factors are taken from \rfrs~\cite{Bradford:2006yz}~and~\cite{Borah:2020gte}\footnote{%
 This builds on a previous work by Ye {\it et al.} in \rfr{Ye:2017gyb}
 by optimizing the form factors for use at low $Q^{2}$.},
 with the latter being used as the default choice.
The vector form factors are assumed to be precisely known,
 and the impact of their uncertainty will be assessed in \sct{sec:systematics_vector}.
\newcommand{\pcac}{PCAC}%
\newcommand{\ppd}{PPD}%
By imposing the Partially Conserved Axial Current (\pcac)~\cite{Adler:1964um}
 and Pion Pole Dominance (\ppd)~\cite{Adler:1965ga} constraints,
 the induced pseudoscalar current is related to the axial form factor
 with the expression
\begin{align}
 F_{P} = \frac{2M_{N}^{2}}{M_{\pi}^{2} + Q^{2}} F_{A}.
 \label{eq:pionpoledominance}
\end{align}
The \pcac{} relation imposes a Ward Identity-like
 relation between the divergence of the axial current
 and the pseudoscalar current, proportional to the sum
 of the quark masses rather than the difference.
\ppd{} then assumes that the new pseudoscalar form factor
 that appears is dominated by the pion pole through
 the Goldberger-Treiman relation, which
 imposes a constraint connecting the pseudoscalar
 form factor to the induced pseudoscalar
 and allows the pseudoscalar form factor to be removed.

The deviation of the induced pseudoscalar form factor from the combined
 \pcac{} and \ppd{} relation is not expected to be significant.
This assertion is supported by various studies with
 \lqcd~\cite{
 Jang:2019vkm,
 RQCD:2019jai,
 Park:2021ypf,
 Djukanovic:2022wru,
 Jang:2023zts,
 Alexandrou:2025tfv},
 which observe no statistically significant deviation
 from the \ppd{} assumption in the continuum limit.
Considering that all contributions of $F_{P}$ to the
 differential cross section are also suppressed by a
 factor of $(m_{\ell}/M_{N})^{2}$,
 the potential bias introduced assuming the
 \pcac{} and \ppd{} constraints is expected to be small.
It follows then that
 the leading contribution to the uncertainty
 of the quasielastic differential cross section
 is attributed to the nucleon axial form factor $F_{A}$.

\subsection{\texorpdfstring{\z{}}{z} Expansion Parameterization}
\label{sec:zexpansion}

The \zexp{} is formulated as a conformal mapping to a small expansion parameter~%
 \cite{Hill:2010yb,Bhattacharya:2011ah}.
The transformation from 4-momentum transfer $Q^{2}$ to \z{} is given as
\begin{equation}
 z(Q^{2};\tc,\tz)
 = \frac{\sqrt{\tc+Q^{2}} - \sqrt{\tc-\tz}}{\sqrt{\tc+Q^{2}} + \sqrt{\tc-\tz}}.
 \label{eq:zparameterization}
\end{equation}
The parameter $\tc \leq 9 M_{\pi}^{2}$ is bounded by the
 particle production threshold, which is limited by $3\pi$ production
 for the axial channel.
The value $Q^{2}_{z=0}$ that satisfies $z(Q^{2}_{z=0};\tc,\tz) = 0$
 is determined by the selection of $Q^{2}_{z=0} = -\tz$.
The parameter $\tz$ is chosen for convenience and can be adjusted to
 decrease the maximum value of $|z|$ over the entire range of $Q^{2}$
 probed by neutrino scattering experiments.
The value of \z{} satisfies the inequality $|z|<1$ within
 the kinematic range $Q^{2} \in (-\tc,\infty)$,
 guaranteeing a power series with a small expansion parameter
 within a range of momentum transfers relevant for quasielastic scattering.
For brevity, the dependence of \z{} on $Q^{2}$, \tc,~and~\tz~will be omitted in subsequent equations.

Having formulated the transformation from $Q^{2}$ to \z{},
 the form factors are now expressed as a power series in the parameter \z{} as
\begin{equation}
 F_{A}(z) = \sum_{k=0}^{\infty} a_{k} z^{k}.
\end{equation}
In theory the sum contains an infinite number of terms,
 but in practice the sum is truncated at a finite order \kmax.
To regulate the large momentum transfer behavior of the form factor,
 a sequence of sum rules is enforced with the constraints
\begin{align}
 &\Big( \frac{\partial}{\partial z} \Big)^{n} F_{A}(z) \Big|_{z=1}
 = 0, \, n \in\{0,\dots,3\}
 \nonumber\\
 \implies &
 \sum_{k=n}^{\kmax} k(k-1)\dots(k-n+1) \, a_{k} = 0 .
 \label{eq:sumrules_derivatives}
\end{align}
These sum rules have minimal impact on the shape of the form factor
 within the range of data, but prevent the form factor from exhibiting
 unbounded behavior in the $Q^{2}\to\infty$ limit.
This is useful for extrapolating the form factor outside of the range of data,
 as is needed for many Monte Carlo generators, while still retaining reasonable
 functional behavior.
One additional sum rule is added to fix the intercept of the form factor
 to the axial coupling value in the PDG\footnote{%
 This work uses the convention $g_{A} > 0$,
 whereas PDG reports the axial coupling with a convention $g_{A} < 0$.
 Note that \rfr{Meyer:2016oeg} also uses a convention with $g_{A} < 0$,
 which amounts to a sign flip of all of the \zexp{} coefficients.}~\cite{PDG2024},
\begin{align}
 &F_{A}(Q^{2}=0) = g_{A}
 \nonumber\\
 \implies &
 -g_{A} +\sum_{k=0}^{\kmax} \, a_{k} z_{0}^{k} = 0
 \label{eq:sumrules_intercept}
\end{align}
 where $z_0 = z(Q^{2}=0)$.
This constraint is taken to be exact since the uncertainty on $g_{A}$
 is below the precision of the axial form factor probed by experiments.
With the full set of four derivative sum rules and axial coupling constraint,
 a fit to a \zexp{} with \kmax~coefficients will have $\kmax-4$ free parameters.
The formulas needed to implement the sum rules with full precision are explicitly given in Appendix~\ref{sec:soYouWantToImplementTheSumRuleConstraint}.

\subsection{\texorpdfstring{\z{}}{z} Expansion Regularization}
\label{sec:zexpansion_regularization}

The power series coefficients of the \zexp{} are constrained by unitarity
 to be bounded and decreasing with increasing order~\cite{Hill:2010yb}.
The relative size of the coefficients is therefore not expected to be too large.
This was the motivation for introducing a $\chi^{2}$ penalty term in
 \rfr{Meyer:2016oeg} of the form
\begin{equation}
 \chi^{2}_{\rm penalty}(\lambda) =
 \lambda \sum_{k=1}^{\kmax} \Biggr| \frac{a_{k}}{a_{0}\sigma_{k}} \Biggr|^{2},
 \label{eq:zexp_penalty}
\end{equation}
 where the nominal $\lambda$ was previously taken to be 1.
The size of the nominal prior widths on the coefficient ratios was taken to be
\begin{equation}
 \sigma_{k} = {\rm min}\big[ 5, 25/k \big]
\end{equation}
 reflecting a modest prior width for low-order coefficients
 and more restriction on higher order coefficients.

In this work, the same penalty term as in \eqn{eq:zexp_penalty}
 is employed to regulate the relative size of coefficients
 and the choice of width $\sigma_{k}$ is retained.
However, the value of $\lambda$ is selected by imposing an L-curve heuristic,
 taking the optimal value of $\lambda$ for a particular fit to be the point
 of minimum radius of curvature (or $\lambda=0$ in cases where no minimum is observed).
The final choice of $\lambda$ is selected by finding a compromise that accommodates
 several different fit choices
 (including \kmax, \tz, and included datasets)
 as described in \sct{sec:differentialcrosssection}.

\subsection{Datasets and Corrections}
\label{sec:datasets}

\subsubsection{Experimental Datasets}

\begin{table*}[t]
 \begin{tabular}{c|c|c|c|c|c}
 References & Dataset & Interaction & Dist. Type & Data Cuts & Flux Unc.
 \\
 \hline &&&&\\[-1em]
 \cite{Mann:1973pr,Barish:1977qk,Barish:1978pj,Miller:1982qi}
 & \anl  & $\nu_{\mu}+{\rm D}$ & $dN/dQ^{2}$ & $Q^{2}\geq\{0.06,0.20\}~\GeVsq$ & $dN/dE$
 \\
 \cite{Baker:1981su}
 & \bnl  & $\nu_{\mu}+{\rm D}$ & $dN/dQ^{2}$ & $Q^{2}\geq\{0.06,0.20\}~\GeVsq$ & $dN/dE$
 \\
 \cite{Kitagaki:1983px}
 & \fnal & $\nu_{\mu}+{\rm D}$ & $dN/dQ^{2}$ & $Q^{2}\geq\{0.06,0.20\}~\GeVsq$ & $dN/dE$
 \\
 \cite{Wachsmuth:1979th,Barlag:1984uga,Allasia:1990uy}
 & \bebc & $\nu_{\mu}+{\rm D}$ & $d\sigma/dQ^{2}$ & --- & Scaled
 \\
 \cite{MINERvA:2022vmb,MINERvA:2023avz}
 & \minerva & $\bar{\nu}_{\mu}+p$ & $d\sigma/dQ^{2}$
 & $1.5\leq p_{\mu} \leq 20~{\rm GeV}$, $\theta_{\mu}\leq 20^{\circ}$ & Covariance
 \end{tabular}
 \caption{
 The list of datasets considered in this work.
 The columns of the table indicate
  the references for the dataset,
  the label used for the dataset,
  the scattering interaction considered,
  the type of distribution reported in the reference,
  the implementation of the flux uncertainty,
  and
  the cuts applied to the dataset.
 \label{tab:datasets}
 }
\end{table*}

There are five experimental datasets that were studied in this work.
Three of these datasets (\anl, \bnl, and \fnal) were previously considered in \rfr{Meyer:2016oeg}.
The relevant information about each dataset is listed in \tbl{tab:datasets}.

The first two columns of \tbl{tab:datasets} indicate the references for that dataset
 and the label applied to those data.
The third column contains the scattering interaction that was considered,
 either corresponding to scattering with
 hydrogen ($\bar{\nu}_{\mu}+p$, \minerva~only)
 or
 deuterium ($\nu_{\mu}+{\rm D}$).
The fourth column contains the type of distribution reported for the dataset,
 either as an event distribution ($dN/dQ^{2}$)
 or as a differential cross section ($d\sigma/dQ^{2}$).
The way that each of these event distributions is handled will
 be discussed throughout the rest of this section.

The fifth column is the cuts applied to the data.
There are again three options, related to the kinematics:
 ``$Q^{2}$'' indicates that the low-$Q^{2}$ bins are removed
 to avoid systematics associated with low momentum transfer.
Further attempts to handle the low-$Q^{2}$ behavior
 are explained in Sections~\ref{sec:correction_acceptance}~and~\ref{sec:correction_deuterium}.
Two cuts to the 4-momentum transfer were applied,
 taking $0.06$~and~$0.20~\GeVsq$ to match the historical choices in \rfr{Meyer:2016oeg},
 and the consequences of these cuts discussed in more detail in \sct{sec:minimumQ2}.
The \minerva~dataset applies two other cuts:
 ``$p_{\mu}$,'' a cut on the outgoing muon momentum,
 and ``$\theta_{\mu}$,'' a cut on the angle of the outgoing muon momentum.
The outgoing momentum for the \minerva~dataset
 was restricted to the range $1.5\leq p_{\mu}\leq20~{\rm GeV}/c$
 and the outgoing angle to be $\theta_{\mu}\leq20^{\circ}$.
The remaining sixth column is described in the next subsection.

\subsubsection{Flux Uncertainty}

The last column of \tbl{tab:datasets} is the method that was applied
 to capture the uncertainty due to the neutrino flux.
There are three options listed.
For the \anl, \bnl, and \fnal~datasets, labeled with ``$dN/dE$,''
 the flux uncertainty is applied to the bins of the neutrino energy
 event distribution.
Each bin in the energy event distribution is allowed to float independently
 with a series of nuisance parameters $\eta_{i}$ as
\begin{equation}
 \Biggr[\frac{dN}{dE_{\nu}}(\vec{\eta})\Biggr]_{i}
 = \Biggr[\frac{dN}{dE_{\nu}}\Biggr]_{i}
 + \eta_{i}\, \Biggr[\delta\frac{dN}{dE_{\nu}}\Biggr]_{i}
 \label{eq:correction_flux}
\end{equation}
 where $\delta[{dN}/{dE_{\nu}}]_{i}$ is the statistical uncertainty
 on the event distribution in bin $i$.
Each nuisance parameter $\eta_{i}$ is priored with a Gaussian penalty of $0\pm1$.

In the case of the ``scaled'' flux uncertainty for \bebc,
 all bins in the $d\sigma/dQ^{2}$ distribution
 were allowed to float with a single normalization across all bins,
\begin{equation}
 {\cal N}(\tilde{\eta}) = 1 + \tilde{\eta}\,d{\cal N},
 \label{eq:normalization_priored}
\end{equation}
 where $d{\cal N}$ is the relative flux uncertainty over all bins
 and $\tilde{\eta}$ is a nuisance parameter that is fit with the data.
The nominal value of $d{\cal N}$ is chosen to be 0.10,
 and is compared to a value of 0.20 in \sct{sec:compatibility_minerva_bebc}.
The nuisance parameter was also given a Gaussian penalty of $0\pm1$.

The ``covariance'' label for \minerva~indicates that
 the flux uncertainty was included in the covariance matrix
 for the $d\sigma/dQ^{2}$ distribution and so is not considered separately.

\subsubsection{Normalizations}
\label{sec:normalizations}

The \anl, \bnl, and \fnal~datasets, lack sufficient information
 to absolutely normalize the event distributions.
The available distributions lack correlations between
 the event $E_{\nu}$ and $Q^{2}$.
Without this information, there is a complicated interplay
 between the applied $Q^{2}$ cut, the differential cross section
 with its deuterium correction, and the flux uncertainty.
When bins are removed from the differential cross section,
 the neutrino energy distribution must also be reduced to accurately implement the removal of low $Q^2$ events.
The exact amount that each of the energy distribution bins
 should be reduced depends on the fraction of events
 that would fall into the $Q^{2}$ bins in question,
 which in turn depends on the assumed nucleon form factors
 and deuterium corrections for those $Q^{2}$ bins.
It is therefore difficult to propagate the available flux uncertainties
 to the binned predictions.

To address this concern,
 the \anl, \bnl, and \fnal{} datasets are fit with
 a normalization ${\cal N}$ that is allowed to float
 without an imposed prior.
The flux uncertainty on the \bebc~dataset is captured through
 a floating normalization that is priored via \eqn{eq:normalization_priored}.
For the \minerva~dataset, the normalization is fixed to 1
 as information about the flux uncertainty is embedded in the
 provided covariance.

\subsubsection{Efficiency Corrections}
\label{sec:correction_acceptance}

The deuterium bubble chamber experiments suffered from reduced efficiency -- referred to as ``acceptance" in \rfr{Miller:1982qi} --
 when tracks were not prominent enough to measure accurately.
This reduced efficiency primarily affected the low-$Q^{2}$ region
 and was accounted for with the efficiency correction
\begin{align}
 f(Q^{2};\xi)
 &= \big[ \epsilon(Q^{2}) + \xi \delta \epsilon(Q^{2}) \big]^{-1}
\nonumber\\
 &= \frac{1}{\epsilon(Q^{2})} \Biggr[ 1 + \xi \frac{\delta \epsilon(Q^{2})}{\epsilon(Q^{2})} \Biggr]^{-1}.
 \label{eq:correction_acceptance}
\end{align}
The efficiency correction $(\epsilon)$ and its uncertainty $(\delta\epsilon)$ used in this work was taken from
 Figure 1 of \rfr{Miller:1982qi}.
The same form of the correction was applied to the \anl, \bnl, and \fnal~datasets
 with an independent $\xi$ nuisance parameter for each dataset.

\subsubsection{Deuterium Corrections}
\label{sec:correction_deuterium}

For experiments with a deuterium target,
 a correction is needed to relate the
 free nucleon quasielastic cross section ($\sigma_{N}$)
 to the deuterium cross section ($\sigma_{\rm D}$).
This correction is typically assumed (as is the case in the present work)
 to be characterized by a ratio that depends only on $Q^{2}$,
\begin{equation}
 \frac{d\sigma_{\rm D}}{dQ^{2}}(E_{\nu},Q^{2})
 \approx R(Q^{2}) \frac{d\sigma_{N}}{dQ^{2}}(E_{\nu},Q^{2}).
 \label{eq:correction_deuterium_simple}
\end{equation}
The approach of marginalizing the deuterium correction into
 multiplicative corrections was discussed in \rfrs~\cite{Singh:1971md,Shen:2012xz}.

The ratio $R(Q^{2})$ should in general depend on the energy transfer $\omega$,
 and consequently $E_{\nu}$, as well.
This can be seen by assuming an oversimplified approximation
 where the presence of a spectator proton inside of a deuterium nucleus
 forces the neutron to move with some Fermi momentum
 $\vec{p}$ relative to the two-nucleon center of mass system.
Then the relationship between 4-momentum transfer squared
 and the energy transfer in quasielastic interactions takes the form
\begin{equation}
 \omega = \frac{Q^{2} + 2\vec{p}\cdot\vec{q}}{2 E_{N}}.
\end{equation}
In the special case of $\vec{p}=0$, that yields
 an exact correspondence $\omega = Q^{2} / 2 M_{N}$.
If a Lorentz boost were applied to cancel out the Fermi motion,
 then the effective $E_{\nu}$ as seen by the neutron at rest
 would be modified even though the $Q^{2}$ remains the same.
Unless the distribution of neutron momenta within the deuterium
 nucleus conspired to cancel all $\omega$ dependence,
 both $Q^{2}$ and $\omega$ must therefore play separate roles
 in the deuterium correction.
\newcommand{\tpth}{\text{2p2h}}
This kind of effect would be realized in nature by,
 for example, two-particle two-hole (\tpth) interactions
 in the deuterium nucleus.
The short-range interactions characterizing the \tpth{}
 are not included in the deuterium corrections of
 Singh~\cite{Singh:1971md} given in $R$
 but would modify the $\omega$ dependence at fixed $Q^{2}$,
 especially at low $Q^{2}$.
Ideally the ratio could be promoted to depend on $E_{\nu}$ as well,
 choosing $E_{\nu}$ over $\omega$ because of the
 existing dependence on $E_{\nu}$ via the flux,
\begin{equation}
 \frac{d\sigma_{\rm D}}{dQ^{2}}(E_{\nu},Q^{2})
 = R(E_{\nu},Q^{2}) \frac{d\sigma_{N}}{dQ^{2}}(E_{\nu},Q^{2}).
 \label{eq:correction_deuterium_dream}
\end{equation}

There are a number of concerns regarding the deuterium data,
 many of which relate to the low-$Q^{2}$ data.
One immediate worry is that the deuterium effects are poorly understood,
 namely that the energies involved in neutrino scattering are sensitive
 to short-range interactions between nucleons inside the deuterium nucleus.
Since no correction simultaneously exploring the $\omega$ and $Q^{2}$
 dependence of the deuterium nucleus is readily available in the literature,
 this study is relegated to future works.
The final quoted results from this work will not combine deuterium data
 with nucleon and hence will not be subject to the choice
 that is made for the form of $R$.

\subsection{Handling Event Distributions}
\label{sec:handling_event}

The \anl, \bnl, and \fnal~datasets report event distributions,
 which require different handling than flux-integrated differential
 cross sections.
The relation used to compute the theory prediction
 for the event distribution in bin $i$ is
\begin{align}
 \Delta Q^{2}_{i} \, \Biggr[ \frac{dN_{\Deut}}{d Q^{2}} \Biggr]_{i}
 =& {\cal N}
 \int_{{\rm bin~}i} dQ^{2}
 \int dE_{\nu}
 \Biggr[
 \nonumber\\
 &f(Q^{2}) R(Q^{2}) \,
 \frac{d\Phi}{d E_{\nu}} (E_{\nu}) \,
 \frac{d\sigma_{N}}{d Q^{2}}(E_{\nu},Q^{2}) \Biggr] .
 \label{eq:theory_eventdistribution}
\end{align}
This relation depends on
 the normalization ${\cal N}$ (discussed in \sct{sec:normalizations}),
 the efficiency correction $f$ (\eqn{eq:correction_acceptance}),
 the deuterium correction $R$ (\eqn{eq:correction_deuterium_simple}),
 and the flux $\Phi$, which is determined from the event distribution
 over neutrino energy
\begin{equation}
 \frac{d\Phi}{d E_{\nu}}
 =
 \frac{1}{\sigma(E_{\nu})} \Biggr[ \frac{dN}{d E_{\nu}} \Biggr].
 \label{eq:flux}
\end{equation}

The cross section that appears in the denominator of \eqn{eq:flux}
 is obtained by integrating the differential cross section over $Q^{2}$,
\begin{equation}
 \sigma(E_{\nu}) = \int dQ^{2} \frac{d\sigma_{N}}{d Q^{2}}(E_{\nu},Q^{2}).
\end{equation}
This integration was carried out numerically with fixed
 free nucleon form factor parameterizations.
The fits were allowed to run to completion
 with the fixed parameterizations, and then the parameterizations
 were updated to allow for variations in the form factors.
After 4 iterations, the fit parameters had saturated
 such that further updates resulted in negligible shifts.

The neutrino flux for the \fnal{} dataset
 is at a high enough energy that the differential
 cross section becomes independent of $E_{\nu}$,
 as discussed in the context of
 \eqns~(\ref{eq:differential_crosssection})%
 --(\ref{eq:differential_xsec_c}).
This means that the energy event distribution,
 $d\Phi/dE_{\nu}$, in \eqn{eq:theory_eventdistribution}
 factorizes into a rescaling of the
 normalization factor ${\cal N}$.
This simplification does not extend to the
 \anl{} and \bnl{} datasets, which have event distributions
 that span much lower energy ranges.
For these experiments, altering the neutrino flux
 amounts to uncontrolled changes to the form factor shape.

The integration over $Q^{2}$ and $E_{\nu}$
 depends on both continuous functions as well as on binned datasets.
The bin widths are sufficiently coarse that the
 continuous functions can change substantially over a single bin,
 which can lead to systematic effects due to binning.
To suppress this effect,
 the data bins are subdivided to a finer resolution,
 and the binned data are kept constant over the entire coarse bin width.
The functions are then computed at the bin center of each fine bin
 and summed over all bins.

To be explicit, the integration over the event distribution
 to obtain the flux is approximated as
\begin{align}
 \Biggr[ \int& d E_{\nu} \, \frac{d\Phi}{dE_{\nu}}(E_{\nu})
 \, \frac{d\sigma_{N}}{d Q^{2}}(E_{\nu},Q^{2}) \Biggr]_{\text{bin}~j}
 \nonumber\\&
 \to
 \Biggr[ \frac{dN}{d E_{\nu}} \Biggr]_{j}
 \Biggr[ \sum_{\text{bin}~k \in \text{bin}~j}
 \frac{\Delta E_{\nu,k}}{\sigma(E_{\nu,k})}
 \frac{d\sigma_{N}}{d Q^{2}}(E_{\nu,k},Q^{2}) \Biggr] ,
\end{align}
 where $\Delta E_{\nu,k}$ is
 the bin width of the fine bin $k$ within coarse bin $j$
 and $E_{\nu,k}$ is the energy at the center of fine bin $k$.
A similar procedure is applied for the integral over $Q^{2}$ bins,
\begin{align}
 \Biggr[ \int& dQ^{2}
 f(Q^{2}) R(Q^{2}) \,
 \frac{d\sigma_{N}}{d Q^{2}}(E_{\nu},Q^{2}) \Biggr]_{{\rm bin~}i}
 \nonumber\\
 &\to
 \sum_{\text{bin}~k \in \text{bin}~i}
 \Delta Q^{2}_{k} \, f(Q^{2}_{k}) R(Q^{2}_{k}) \,
 \frac{d\sigma_{N}}{d Q^{2}}(E_{\nu},Q^{2}_{k}).
 \label{eq:fineresolution_Q2}
\end{align}
In both cases,
 a resolution 10 times finer than the coarse binning
 was sufficiently fine that further subbinning
 had a negligible effect on fit results.

\subsection{Handling Differential Cross Sections}
\label{sec:handling_differential}

The \bebc~and~\minerva~datasets both report
 flux-integrated differential cross sections that can be absolutely normalized.
Both \bebc{} and \minerva{} also span ranges of $E_{\nu}$
 where the differential cross section is approximately
 energy independent.
As a consequence, modifications to the flux are largely absorbed
 by any uncertainty associated with the absolute normalization
 of the data.
For both of these experiments,
 the flux-integrated differential cross section
 $d\tilde{\sigma}_{(\Nuc,\Deut)}/dQ^{2}$ for bin $i$
 with either a nucleon (\Nuc) or deuterium (\Deut) target is
\begin{align}
 \Delta Q^{2}_{i} \, \Biggr[ \frac{d\tilde{\sigma}_{(\Nuc,\Deut)}}{d Q^{2}} \Biggr]_{i}
 =
 &
 \frac{\cal N}{\tilde{\Phi}}
 \int_{{\rm bin~}i} dQ^{2} R(Q^{2})
 \int dE_{\nu}
 \Biggr[
 \nonumber\\&
 \frac{d\Phi}{dE_{\nu}}(E_{\nu})
 \frac{d\sigma_{N}}{d Q^{2}}(E_{\nu},Q^{2}) \Biggr] .
 \label{eq:theory_fluxintegrateddifferential}
\end{align}
The denominator includes the flux integrated over energy,
\begin{align}
 \tilde{\Phi} = \int dE_{\nu} \, \Biggr[ \frac{d\Phi}{dE_{\nu}}(E_{\nu}) \Biggr]
 = \sum_{j} \Delta E_{\nu,j} \Biggr[ \frac{d\Phi}{dE_{\nu}} \Biggr]_{j}.
\end{align}
The efficiency correction discussed in Sect.~\ref{sec:correction_acceptance} are not included in either the \minerva{} or \bebc{} datasets.

Like with the event distributions,
 the integrals in \eqn{eq:theory_fluxintegrateddifferential}
 are discretized on a finer resolution to reduce binning systematics.
The relevant integrals are
\begin{align}
 \Biggr[ \int& d E_{\nu} \, \frac{d\Phi}{dE_{\nu}}(E_{\nu})
 \, \frac{d\sigma_{N}}{d Q^{2}}(E_{\nu},Q^{2}) \Biggr]_{\text{bin}~j}
 \nonumber\\&
 \to
 \Biggr[ \frac{d\Phi}{d E_{\nu}} \Biggr]_{j}
 \Biggr[ \sum_{\text{bin}~k \in \text{bin}~j}
 \Delta E_{\nu,k} \,
 \frac{d\sigma_{N}}{d Q^{2}}(E_{\nu,k},Q^{2}) \Biggr]
\end{align}
 and again \eqn{eq:fineresolution_Q2} but with $f(Q^{2})=1$ fixed.

There are a few differences between \bebc{} and \minerva{} worth noting.
For the \bebc~dataset, the normalization factor ${\cal N}$
 is given in \eqn{eq:normalization_priored},
 which includes the flux uncertainty.
The deuterium correction $R(Q^{2})$ is the same as used in the
 other deuterium datasets, as discussed in \sct{sec:correction_deuterium}.
For the \minerva~dataset, the normalization is instead fixed to ${\cal N}=1$
 and the deuterium correction taken to be $R(Q^{2})=1$.
Any uncertainty in the flux is included in the covariance matrix reported
 by the \minerva~collaboration.
In addition, the antineutrino-nucleon differential cross section
 is used for $d\sigma_{N}/dQ^{2}$ rather than the
 neutrino-nucleon differential cross section used in \bebc.

\FloatBarrier

\section{Fitting}
\label{sec:fitting}

\begin{table}[btp]
\begin{tabular}{c|r|r}
Parameter & \rfr{Meyer:2016oeg} & This work \\
\hline
$g_A$ & $1.2723$
 & $1.2754$~\cite{PDG2024} \\
$t_{c}$ & $9\cdot(0.14~\GeV)^{2}$ & $9\cdot(0.134~\GeV)^{2}$ \\
$t_0$ & $-0.28~\GeVsq$ & $-0.50~\GeVsq$ \\
\# sum rules & 4 & 4 \\
\hline
$M_{\pi}$ & 0.1395702~\GeV & 0.1395702~\GeV \\
$m_{\mu}$ & $0.1057~\GeV$ & $0.1056583755~\GeV$ \\
$\mu_{p}$ & 2.7928 & 2.7928 \\
$\mu_{n}$ & $-1.9130$ & $-1.9130$ \\
$M_{N}$ & $0.9389~\GeV$ & $0.93891875434~\GeV$ \\
$\cos{\theta_{C}}$ & 0.9743 & 0.9743 \\
$G_{F}$ & $1.166\times10^{-5}~{\rm GeV}^{-2}$ & $1.166\times10^{-5}~{\rm GeV}^{-2}$ \\
Vector FFs & BBBA05~\cite{Bradford:2006yz} & Borah {\it et al.}~\cite{Borah:2020gte} \\
\end{tabular}
\caption{
A list of parameters needed to evaluate the differential cross sections.
The first column gives the fit parameter,
 the second column lists the parameters used in \rfr{Meyer:2016oeg},
 and the third column gives the parameters used in this work.
\label{tab:fitparameters}
}
\end{table}

In this section, the results of fits are discussed.
In particular, the compatibility of the input datasets
 is scrutinized to determine the most accurate depiction
 of the neutrino-nucleon quasielastic
 axial form factor and its uncertainty.

The default set of parameters used in this reference are listed in
\tbl{tab:fitparameters}.

\subsection{L--Curve Studies and Parameterization Selection}
\label{sec:lcurve}

One of the first tasks is to select a nominal parameterization
 to use for the form factor.
This includes the selection of \tz{} and \kmax{}
 for use in the following fits comparing datasets.
The sensitivity to these inputs will depend on how strongly
 prior constraints on parameters of the \zexp{} are enforced,
 and this strength is controlled by the parameter $\lambda$ in \eqn{eq:zexp_penalty}.

In previous work~\cite{Meyer:2016oeg},
 the penalty strength $\lambda$ was assumed based on arguments
 from unitarity constraints.
In this work, a data-driven L-curve approach~\cite{Lcurve,Lcurvetext} will be employed
 to select $\lambda$ in order to avoid bias in the fit parameters.
It is worth noting that either choice is an acceptable heuristic
 and ideally any results should not depend strongly on such a choice.

The L-curve approach compares the data contribution $\chi^{2}_{\rm data}$
 versus the penalty contribution $\chi^{2}_{\rm penalty}$
 under variations of the penalty $\lambda$ to find the point where the
 two competing $\chi^{2}$ constraints are balanced.
A penalty term that is too strong may introduce bias into the fit results,
 while a weak penalty with too many fit parameters could result in overfitting.
When some parameters are constrained primarily by the penalty term,
 a healthy L-curve will exhibit an L-shaped bend and the ``optimal'' choice
 is taken at the point with a minimum radius of curvature.
If few enough parameters are used, the $\chi^{2}_{\rm penalty}$ term may go to 0
 as $\lambda\to0$ without exhibiting a bend,
 indicating that no penalty term is required.
\newcommand{\dof}{\text{DoF}}%
Additional parameters are warranted as long as the corresponding
 change to the augmented $\chi^{2}$,
\begin{align}
  \chi^{2}_{\rm aug} &= \chi^{2}_{\rm data} + \chi^{2}_{\rm penalty} \nonumber\\
  &= \chi^{2}_{\rm data} + \lambda \sum_{k=1}^{\kmax} \Biggr| \frac{a_{k}}{a_{0}\sigma_{k}} \Biggr|^{2},
\end{align}
 is commensurate with the decrease in the
 fit degrees of freedom to result in an improved fit quality:
 for each additional fit parameter,
 $\chi^{2}_{\rm aug}/\dof$ should decrease,
 where \dof{} is the number of degrees of freedom.
For this work, $\chi^{2}_{\rm data}$ includes the squared residuals
 from the usual theory-data differences in the differential cross sections
 as well as statistical uncertainties for the flux (\eqn{eq:correction_flux})
 and priors for the efficiency corrections (\eqn{eq:correction_acceptance}).
$\chi^{2}_{\rm penalty}$ includes only the regularization for
 the \zexp{} parameters and is defined in \eqn{eq:zexp_penalty}.

\begin{figure*}[hbt!]
\begin{tabular}{ll}
\includegraphics[clip,trim={40 0 90 70},width=0.45\textwidth]{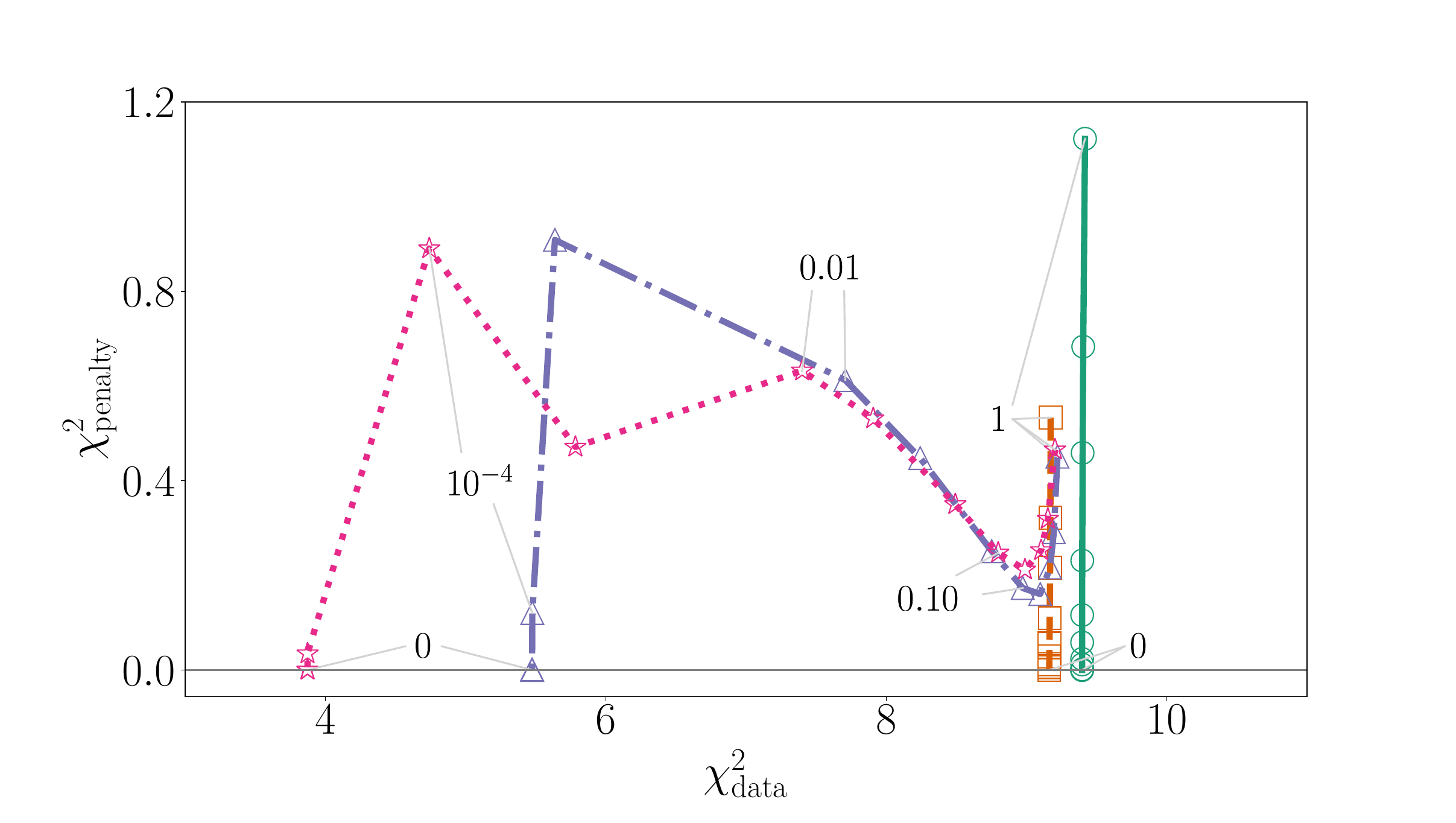}
&
\includegraphics[clip,trim={40 0 90 70},width=0.45\textwidth]{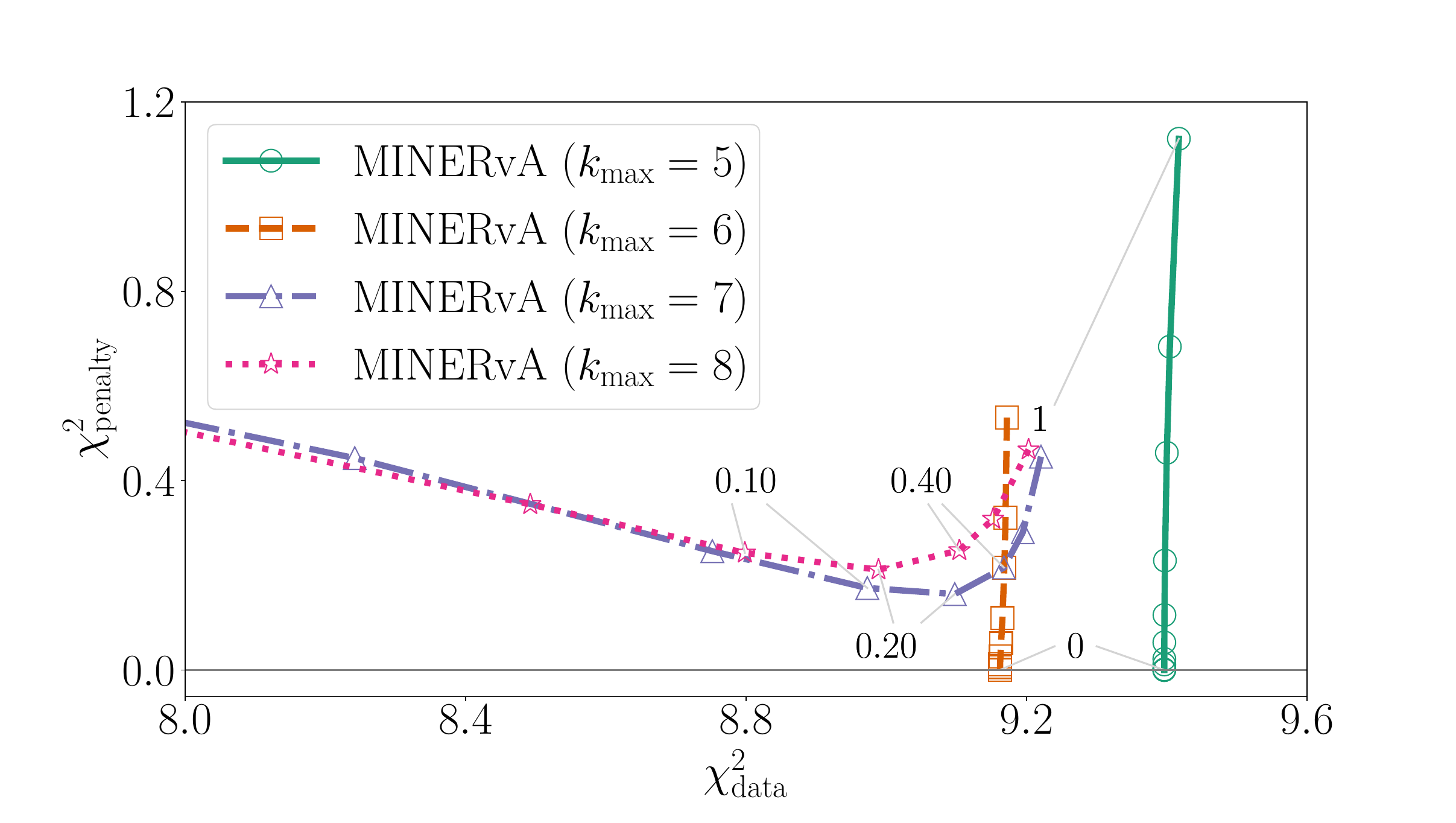}
\end{tabular}
\caption{
L-curve obtained from fitting only the \minerva{} dataset.
The left plot shows the full range $\lambda \in [0,1]$,
 and the right plot is zoomed in on the region with minimum radius of curvature.
The numbers superimposed on the plot indicate the value of
 $\lambda$.
\label{fig:lcurve_minerva}}
\end{figure*}

The fit L-curve for the \minerva{} dataset in isolation is plotted in \fgr{fig:lcurve_minerva}.
This plot uses $\tz=-0.5~\GeVsq$ with various choices of $\lambda$ and \kmax.
The $\kmax=5$ and $\kmax=6$ curves exhibit no L-shape,
 indicating that the parameterizations are sufficiently well constrained by data
 that no regularization term is needed.
The value of $\chi^{2}_{\rm data}$ for $\lambda=0$
 only decreases by 0.2 when transitioning from $\kmax=5$ to $\kmax=6$.
The $\kmax=6$ parameterization (with 2 free parameters)
 produces only a marginally better $\chi^{2}_{\rm data}$
 than the $\kmax=5$ parameterization (with 1 free parameter).
The corresponding goodness-of-fit, $\chi^{2}_{\rm aug}/\dof$,
 increases from $\kmax=5$ to $\kmax=6$, demonstrating a preference for $\kmax=5$.
The expected L-shape curve appears for $\kmax\geq7$ with a point of minimum curvature
 around $\lambda\approx0.2$.
Additional fluctuation is seen in the left panel of \fgr{fig:lcurve_minerva},
 in particular for $\kmax=8$ and $\lambda\approx10^{-3}$,
 indicating that the prior terms are not constraining enough
 to prevent the fit from falling into an unrealistic minimum.

\begin{figure*}[hbt!]
\begin{tabular}{ll}
\includegraphics[clip,trim={60 0 90 70},width=0.45\textwidth]{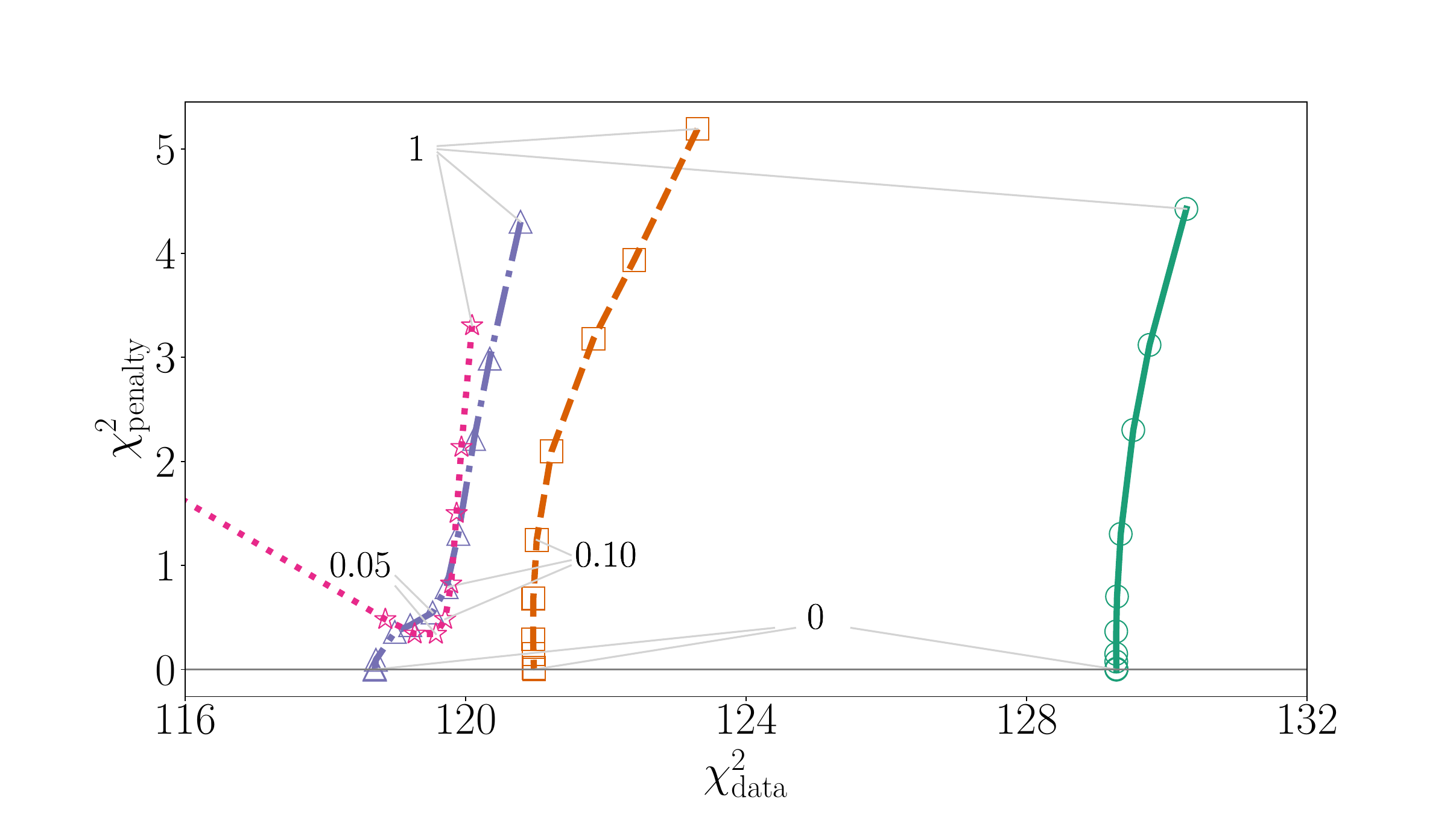}
&
\includegraphics[clip,trim={60 0 90 70},width=0.45\textwidth]{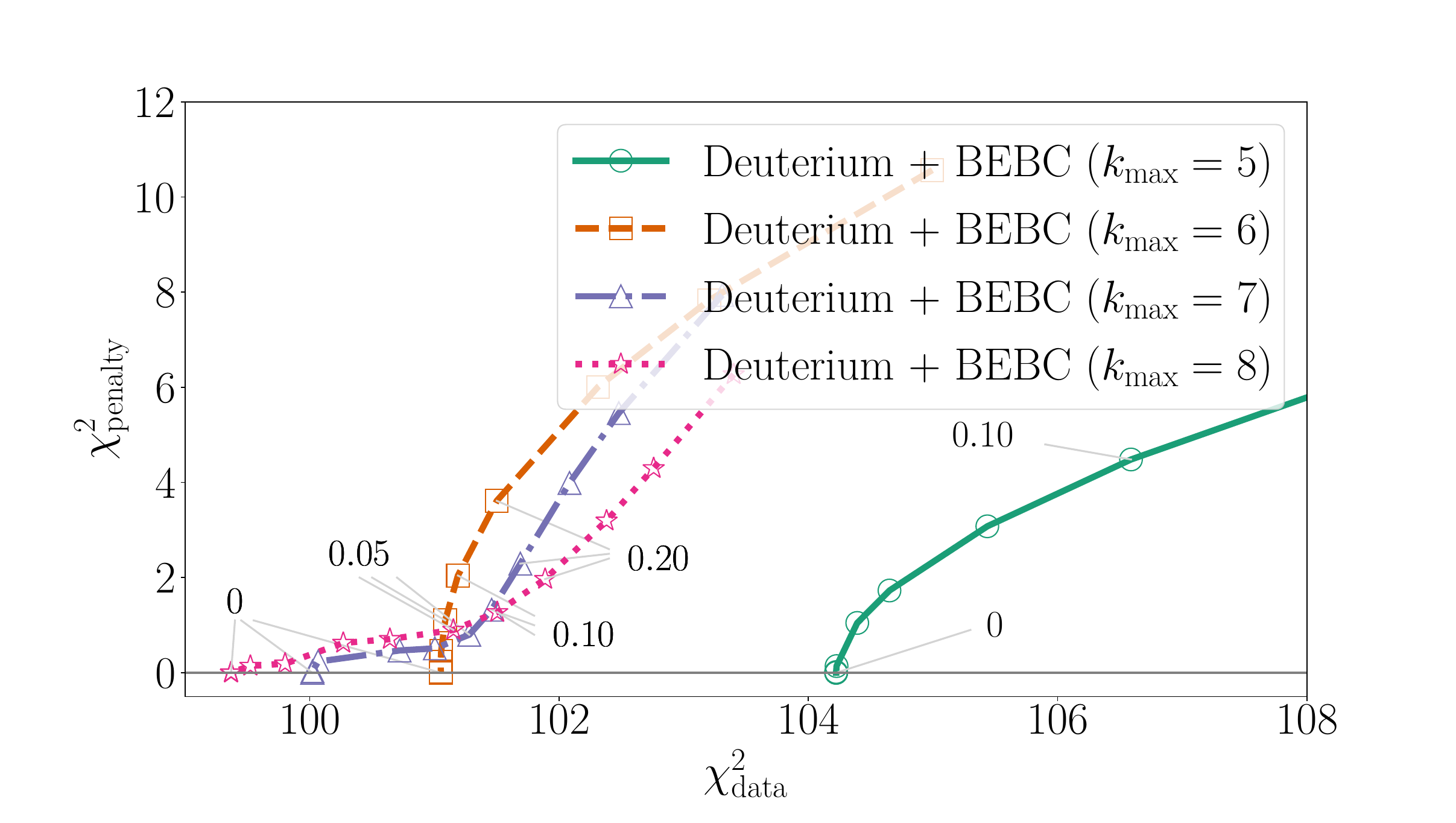}
\end{tabular}
\caption{
L-curve obtained from fitting all of the deuterium datasets together.
The left plot shows the $\Qtmin=0.06~\GeVsq$ cut,
 and the right plot the $\Qtmin=0.20~\GeVsq$ cut.
\label{fig:lcurve_deutx}}
\end{figure*}

\fgr{fig:lcurve_deutx} similarly shows L-curve plots for combined fits
 to all of the deuterium datasets.
The two panels show the effect of different $\Qtmin$ cuts
 on the L-curves for these datasets.
Like the \minerva{} dataset, the fits with $\kmax=5$
 and $\kmax=6$ do not exhibit a distinctive L-shape,
 indicating that no penalty $(\lambda=0)$ is preferred.
Unlike with the \minerva{} dataset alone,
 there is a more substantial decrease in $\chi^{2}_{\rm data}$
 for $\kmax=6$ versus $\kmax=5$.
These trends are exhibited for both $Q^{2}$ cuts.

In the left panel ($\Qtmin=0.06~\GeVsq$ cut),
 the $\kmax=7$ fit does exhibit a small L-shape around
 $\lambda\approx0.02$ before $\chi^{2}_{\rm penalty}$
 again decreases with $\lambda$ all the way to 0,
 indicating preference for a small regularization.
$\kmax=8$ has a distinct L-shape bend around $\lambda\approx0.1$
 that denotes a clear preference for regularization.
In the right panel ($\Qtmin=0.20~\GeVsq$ cut),
 both $\kmax=7$ and $\kmax=8$ behave as they do in
 the left panel, showing only slight preference
 for a nonzero $\lambda$.
For the $\Qtmin=0.06~\GeVsq$ cut,
 the fits prefer regularized
 $\kmax=7$ with $\lambda=0.05$,
 which has $\chi^{2}_{\rm aug}/\dof\approx 1.09$,
 over unregularized $\kmax=6$
 with $\chi^{2}_{\rm aug}/\dof\approx 1.12$.
This slight preference disappears for the $\Qtmin=0.20~\GeVsq$ cut,
 and again $\kmax=6$ with no regularization is the preferred fit strategy.

Although other parameterizations were tested,
 including variations of the choices for $\kmax$, $\tz$, $\lambda$,
 and the included datasets,
 no appreciable departures from the general pictures described in
 \fgrs~\ref{fig:lcurve_minerva}~and~\ref{fig:lcurve_deutx} were seen.
In isolation, the \bebc{} dataset L-curve behaves much like the \minerva{} dataset
 does in \fgr{fig:lcurve_minerva}.
The L-curves for the deuterium event distributions without
 the \bebc{} dataset are very similar to those seen with
 all of the deuterium datasets together, shown in \fgr{fig:lcurve_deutx}.
Including all five datasets together again reproduces the same picture.
The values of $\tz\in\{0, -0.28, -0.50\}~\GeVsq$
 were tested to look for different preferred choices of values
 for $\kmax$ and $\lambda$, but no additional concerns arise for these choices.

Based on the considerations in this section,
 there are a few compromises that perform
 acceptably across all fit data.
The two choices that will be considered henceforth,
 both with $\tz=-0.50~\GeVsq$, are
\begin{enumerate}
 \item $\kmax=6$ with no regularization ($\lambda=0$), and
 \item $\kmax=7$ with $\lambda=0.1$.
\end{enumerate}
The former has the advantage of not being subject
 to dependence on the choice of regularization,
 while the latter gives a form factor parameterization with
 more freedom and presumably more conservative uncertainties.
Both choices produce fits with similar fit quality in general,
 although the only instances where the $\kmax=7$ fits are preferred
 occur for the $\Qtmin=0.06~\GeVsq$ cut.
No occurrences for $\Qtmin=0.20~\GeVsq$ were found
 where the change in $\chi^{2}_{\rm data}$ was sufficient to offset
 the decrease in the degrees of freedom.
The value $\tz=-0.50~\GeVsq$ is chosen to better captures
 some of the sensitivity to higher $Q^{2}$ from the \minerva{} result.
This choice of \tz{} restricts the range $0\leq Q^{2} \lesssim 2.55~\GeVsq$
 containing the majority of the events to satisfy $|z|\lesssim 0.34$,
 and has a maximal $z(Q^{2}=10~\GeVsq) \approx 0.60$.

\subsection{\texorpdfstring{$\chi^{2}$}{Chi-squared} Compatibility Tests}
\label{sec:deltachi2_compatibility}

The difference between data $\chi^{2}$ values
 are used to compute \pvalue{}s in order to assess the
 compatibility between different fit assumptions.
A 1-degree of freedom $\chi^{2}$ test is constructed from the difference
\begin{align}
 \Delta\chi^{2} = \chi^{2}_{A+B} -\chi^{2}_{A} -\chi^{2}_{B},
 \label{eq:deltachi2}
\end{align}
 where $\chi^{2}_{A}$, $\chi^{2}_{B}$, and $\chi^{2}_{A+B}$
 are respectively the $\chi^{2}_{\rm data}$ from fits to
 dataset $A$, dataset $B$, and both datasets $A$ and $B$ together.
Two datasets that produce a \pvalue{} $\geq0.05$ are considered to be compatible
 with each other and may be combined in a single fit without concern.

\subsection{Minimum \texorpdfstring{$Q^2$}{Q2} Cuts and Regularization Dependence}
\label{sec:minimumQ2}

\subsubsection{Event Distribution Datasets}
\label{sec:deuterium_event_distributions}

In the \anl, \bnl, and \fnal{} deuterium datasets,
 the low-$Q^{2}$ region requires the most care due to its
 sensitivity to systematic effects.
There are a few considerations to be conscious of:
\begin{enumerate}
 \item The tracks from the struck proton can be too short to
 reconstruct reliably at low $Q^{2}$, leading to poor efficiency
 and inaccurate characterization of the kinematics.
 \CAPrfr{Miller:1982qi} estimates that the track reconstruction efficiency
 is about $89\pm7\%$ for events in the range $0.05\leq Q^{2} \leq 0.10~\GeVsq$.
 The efficiency correction has been added to alleviate some of this effect.  Although this problem will be worst at low $Q^2$, resolution is in principle a concern at all $Q^2$, and the papers describing the deuterium bubble chamber measurements give little insight into the treatment, save a comment in the \fnal{} paper stating that they only consider events where $\Delta p/p < 0.5$~\cite{Kitagaki:1983px}.
 \item
 The spectator proton also cannot be reliably measured when it has small outgoing momentum.
 For momenta below $0.1~\GeVc$, nearly all of the spectator protons are missed.
 Fits are employed to extract spectator momenta for these
 ``two-prong'' events assuming values centered at 0 and
 prior widths of around $\pm50~{\rm MeV}/c$.
 The distributions are typically compared to a Hulth\'en wavefunction~\cite{Hulthen1942a,Hulthen1942b,Hulthen1957}
 to demonstrate the accuracy of the fit, but the majority of the events
 fall into the region where fits must be used for at least one of the proton tracks.
The Hulth\'en wavefunction underpredicts the number of spectator protons
 in the high-momentum region where both protons can be directly observed.
 This high-momentum discrepancy is attributed to final state interactions in the literature.
 \item The corrections due to deuterium effects are assumed
 to be largest at low $Q^{2}$, due to Pauli exclusion principle
 for the low-momentum outgoing protons.
 The corresponding deuterium effect essentially turns off above $Q^{2} \gtrsim 0.15~\GeVsq$ in corrections applied to the data.
 \item Even with the corrections discussed here,
 the differential cross section data still exhibit a turnover
 at low-$Q^{2}$ that is too sharp to be well-described by the fits.
\end{enumerate}

\begin{figure*}[hbt!]
\begin{tabular}{ll}
\includegraphics[width=0.45\textwidth]{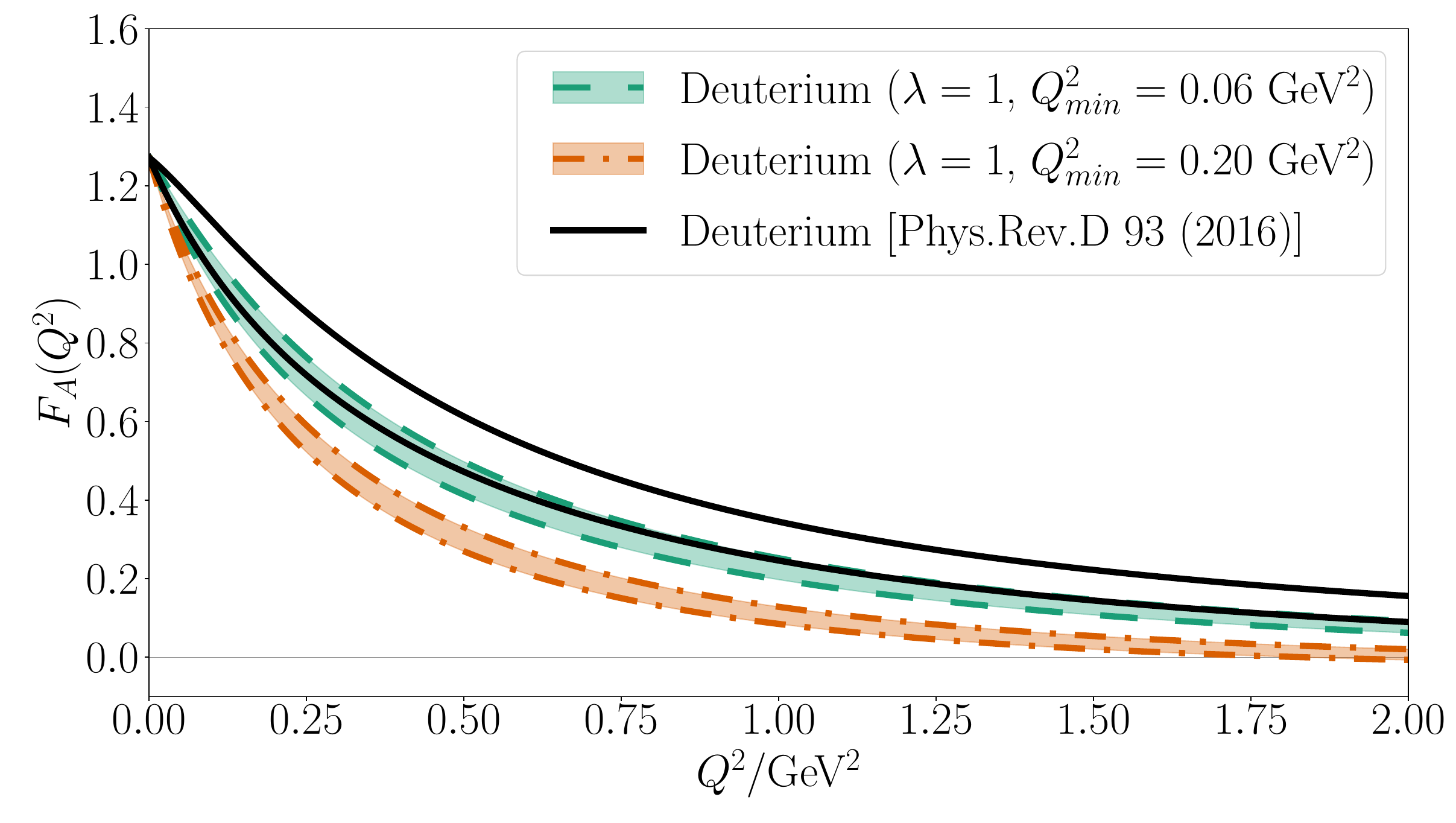}
&
\includegraphics[width=0.45\textwidth]{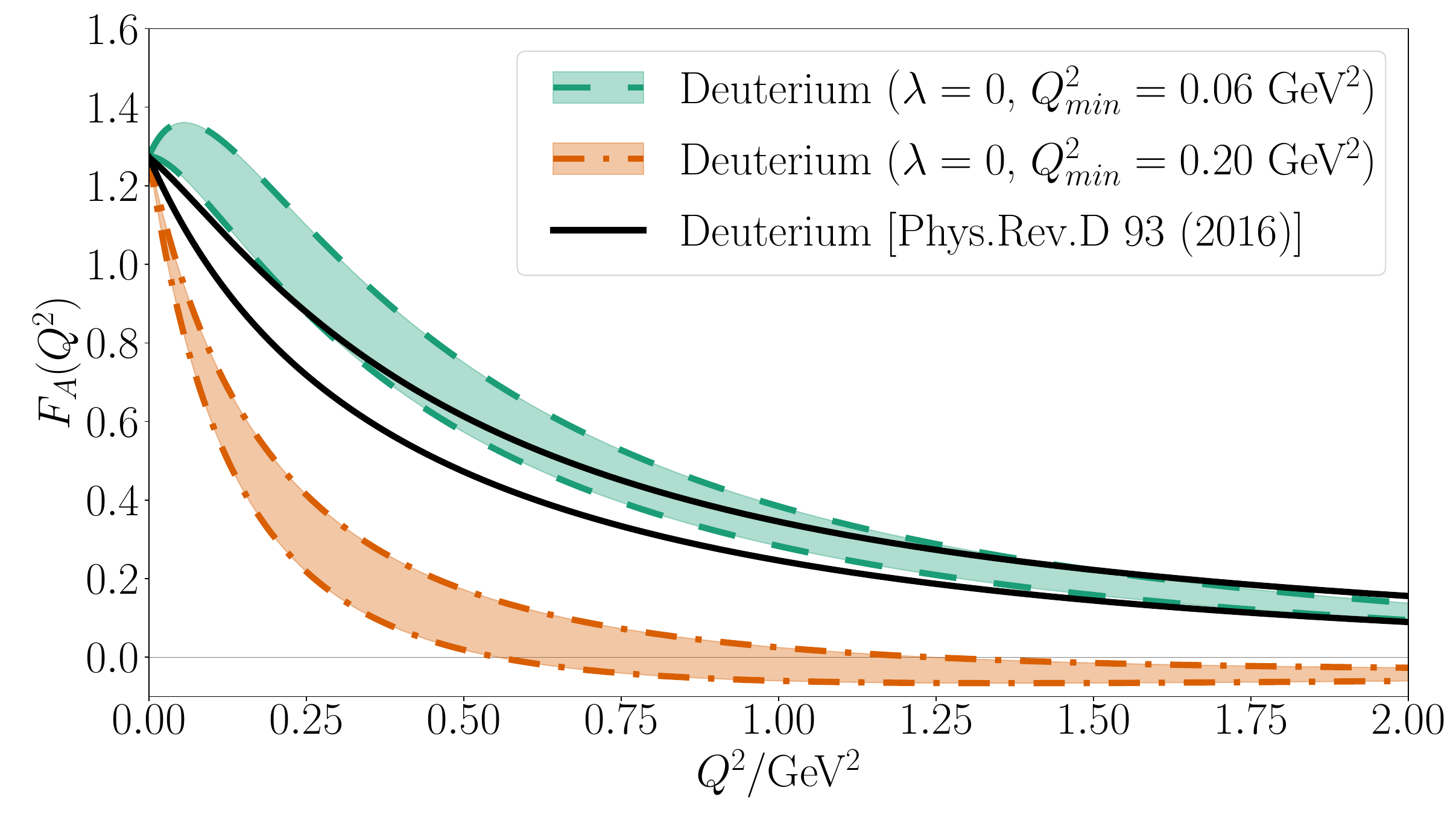}
\end{tabular}
\caption{
Plots of the axial form factor as a function of the 4-momentum transfer $Q^{2}$.
The left panel shows the regularized fits with $\lambda=1$ and $\kmax=7$.
The right panel shows the unregularized fits with $\lambda=0$ and $\kmax=6$.
The cut at $\Qtmin=0.06~\GeVsq$
 is given by the teal shaded region bounded by a dashed line,
 and the cut at $\Qtmin=0.20~\GeVsq$
 by the orange shaded region bounded by the dot-dashed line.
The result from \rfr{Meyer:2016oeg}, which has $\kmax=8$ and $\lambda=1$,
 is given by the unfilled region bounded by solid black lines.
\label{fig:lambda_dependence}
}
\end{figure*}

To test the dependence on systematics due to the inclusion of low-$Q^{2}$ region,
 two different ranges of $Q^{2}$ are considered for the
 \anl, \bnl, and \fnal{} deuterium datasets, as in \rfr{Meyer:2016oeg}.
The two $\Qtmin$ cuts applied
 both with full regularization ($\lambda=1$, left)
 and unregularized ($\lambda=0$, right) are plotted in \fgr{fig:lambda_dependence}.
The choice of $\Qtmin=0.06~\GeVsq$ was taken
 as the default in \rfr{Meyer:2016oeg},
 which appears in \fgr{fig:lambda_dependence}
 as the unfilled region bounded by two solid black lines.
This is most similar to the teal shaded region,
 which only differs from the previously published results
 by choice of vector form factors and some fixed parameter inputs.
This selection discards the first $Q^{2}$ bin,
 which is consistent with the practice employed
 in the original experimental publications.
The other cut, $\Qtmin=0.20~\GeVsq$, was selected
 in \rfr{Meyer:2016oeg} after considering several cuts
 and finding the minimum cut that would suppress
 sensitivity to \kmax.

The left panel of \fgr{fig:lambda_dependence}
 shows some moderate dependence on the minimum $Q^{2}$ cut.
This was previously reported in \rfr{Meyer:2016oeg},
 where the inability of the fit function to describe
 the low-$Q^{2}$ differential cross section data
 resulted in a significant degradation of the goodness-of-fit.
When the constraint from the regularization is removed
 (right panel of \fgr{fig:lambda_dependence}),
 this inconsistency gets considerably worse.
This is a consequence of an approximate degeneracy between
 the floating normalizations for the event distribution datasets
 (discussed in \sct{sec:normalizations})
 and the axial form factor shape.
Without the constraint of the \zexp{} parameter regularization,
 the $a_{k}$ parameters are allowed to float arbitrarily far from 0
 to deform the shape and better fit the curvature of the form factor.
The change in the form factor scale from modifying the curvature
 is then absorbed into the floating normalizations.

\subsubsection{Addition of \bebc{} Dataset}
\label{sec:addition_bebc}

A solution is needed for the degeneracy
 between the floating normalization and the axial form factor shape.
A solution that does not appeal to introduction
 of a regularization would be preferable to one that does.
Fortunately, the datasets with a flux-integrated differential cross section
 are absolutely normalized to within their flux uncertainty
 and are therefore not as sensitive to this degeneracy
 as the \anl, \bnl, and \fnal{} datasets fit with a floating normalization.
The \bebc{} dataset can therefore act as a source of constraint on the overall normalization
 of the event distribution datasets.

\begin{figure*}[hbt!]
\begin{tabular}{ll}
\includegraphics[width=0.45\textwidth]{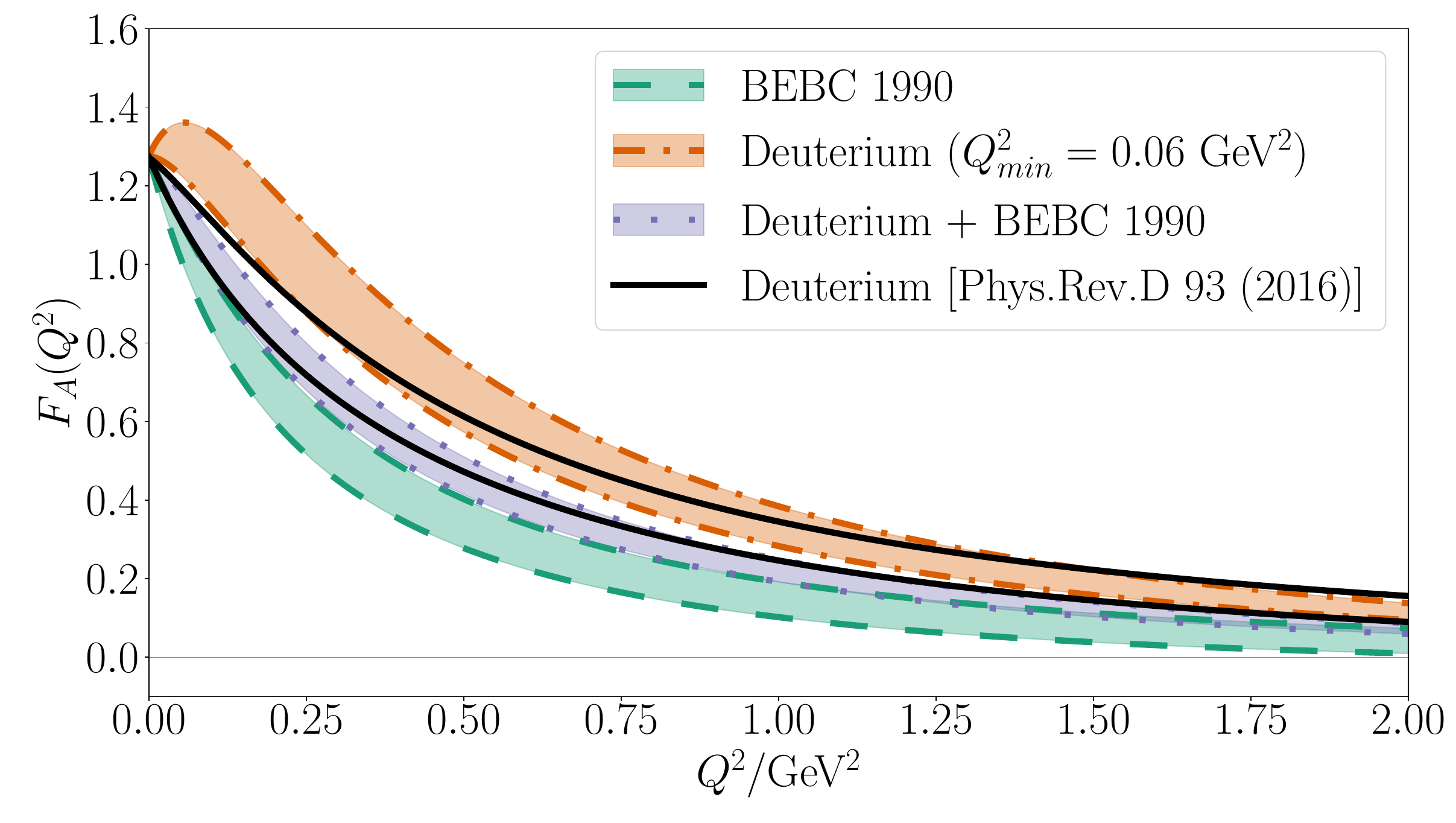}
&
\includegraphics[width=0.45\textwidth]{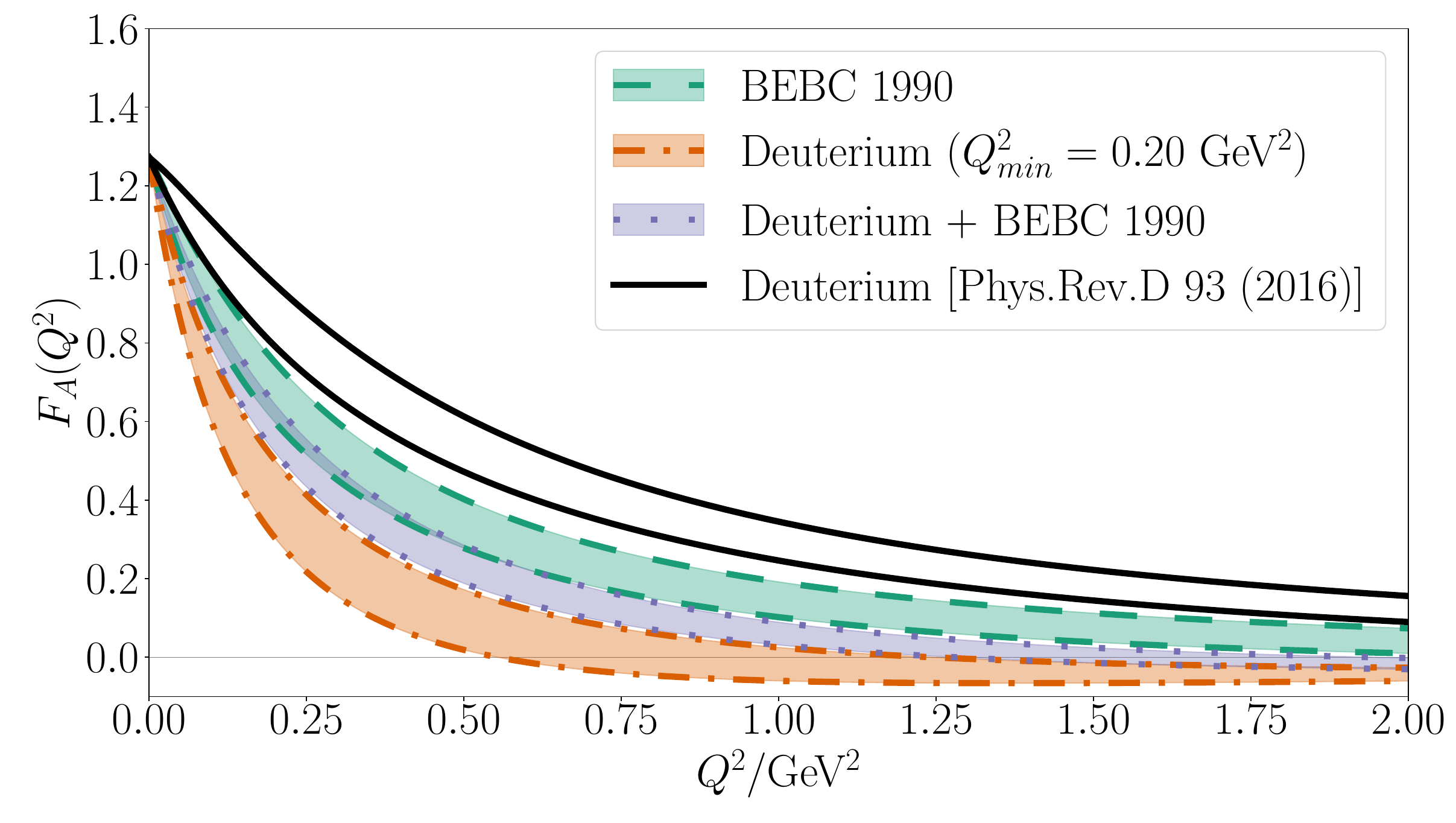}
\end{tabular}
\caption{
The fit axial form factors both
 before and after the addition of the \bebc{} dataset
 to the other event distributions.
In these fits, $\kmax=6$ is chosen and the \zexp{} parameters are unregularized ($\lambda=0$).
The left (right) panel shows the form factors obtained
 when the $0.06~\GeVsq$ ($0.20~\GeVsq$)
 cuts are applied to the event distribution datasets.
The teal shaded region bounded by dashed lines indicates the
 fit only to the \bebc{} dataset (the same in both panels).
The orange shaded region bounded by the dot-dashed line indicates the
 event-distribution datasets that were shown in the right panel
 of \fgr{fig:lambda_dependence}.
The blue-violet shaded region bounded by the dotted line is the combined
 fit including both the event-distribution datasets and the \bebc{} dataset.
The result from \rfr{Meyer:2016oeg} is given by the unfilled region bounded by solid black lines.
\label{fig:addition_bebc}}
\end{figure*}

The effect of the addition of the \bebc{} dataset
 to a combined fit with the other event distribution datasets
 is shown in \fgr{fig:addition_bebc}.
The addition of the \bebc{} dataset has a nontrivial impact
 on the constraint of the form factor normalization,
 partially resolving the degeneracy.
Without the \bebc{} dataset, the form factor shape
 at moderate $Q^{2}$ disagrees with the \bebc{} dataset by
 upwards of $2-3\sigma$.
After the addition of the \bebc{} dataset, the form factor
 is stable around the \bebc{} normalization within about $2\sigma$.
This situation is summarized in \fgr{fig:addition_bebc_summary},
 where the \bebc{} dataset and combined fits are shown together.

\begin{figure}[tbp]
\includegraphics[width=0.45\textwidth]{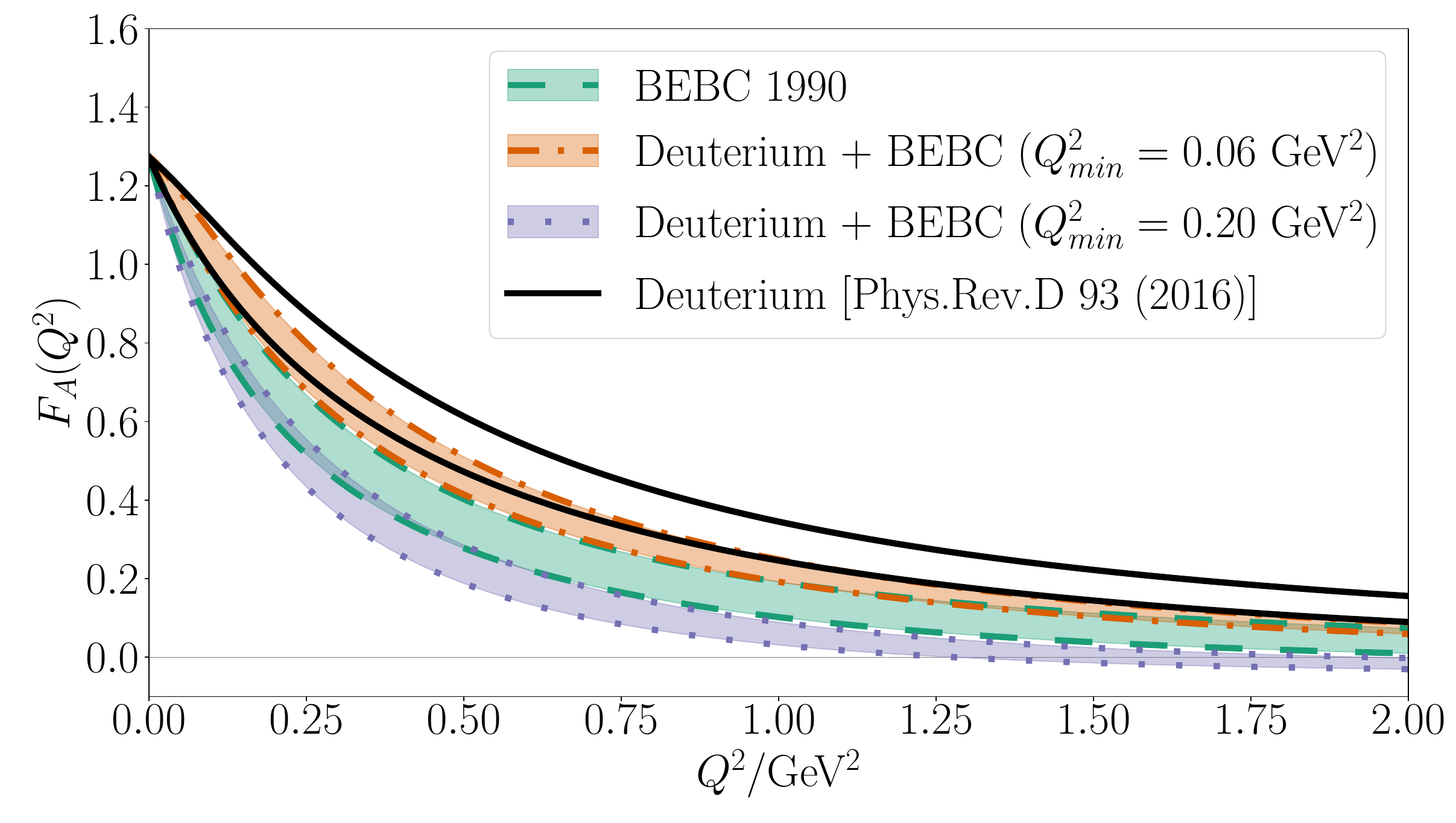}
\caption{
The fit axial form factors for the \bebc{} dataset in isolation
 (teal shading, bounded by dashed line)
 and the combined fit with the event distribution datasets.
The orange shaded region bounded by the dot-dashed line is
 the same joint-fit result shown in the left panel of \fgr{fig:addition_bebc}.
The blue-violet shaded region bounded by the dotted line is
 the same joint-fit result shown in the right panel of \fgr{fig:addition_bebc}.
\label{fig:addition_bebc_summary}}
\end{figure}

The results of the $\Delta\chi^{2}$ compatibility between
 the \bebc{} dataset and the other deuterium datasets are shown in
 \tbls~\ref{tab:deltachi2_bebc_deut_k6}~and~\ref{tab:deltachi2_bebc_deut_k7}.
In all cases, there is reasonable compatibility between the datasets.
Better compatibility seen for $\Qtmin=0.20~\GeVsq$,
 which has better goodness-of-fit than the $\Qtmin=0.06~\GeVsq$ fit,
 and for the regularized $\kmax=7$ fit,
 which has more fit parameters than the unregularized $\kmax=6$ fit.
The \bebc{} dataset is not subject to the \Qtmin{} cut
 and so the improvement in compatibility from changing the \Qtmin{} cut
 is entirely due to the removal of the poorly-described low-$Q^{2}$ region
 in the event distribution datasets.

\begin{table}[btp]
\begin{tabular}{l|rr|rr}
    & \multicolumn{2}{c|}{$\Qtmin=0.06~\GeVsq$}
    & \multicolumn{2}{c}{$\Qtmin=0.20~\GeVsq$}
 \\[.25em]
Fit
 & $\chi^{2}/\dof$ & $p_{\Delta \chi^{2}}$
 & $\chi^{2}/\dof$ & $p_{\Delta \chi^{2}}$
 \\[.25em]
\hline&&\\[-1em]
BEBC                           &      6.1/\phantom{10}6    &&      6.1/\phantom{10}6    \\[.25em]
Deuterium                      &    108.3/100              &&     91.8/\phantom{1}94    \\[.25em]
Deuterium+BEBC                 &    116.8/108              &&     99.5/102              \\[.25em]
$\Delta \chi^{2}$              &      2.5/\phantom{10}1    &                 0.11
&      1.6/\phantom{10}1    &                 0.20 \\[.25em]
\end{tabular}
\caption{
$\Delta \chi^{2}$ tests of compatibility for the
 deuterium event distributions and \bebc{} dataset.
These values are for fits with $\kmax=6$ and $\lambda=0$.
The two \Qtmin{} values are shown in the pairs of columns.
Within the left column of each pair of columns,
 the $\chi^{2}_{\rm data}$ for the two datasets in isolation are listed,
 then the $\chi^{2}_{\rm data}$ for the combined fit,
 and finally the $\Delta\chi^{2}$ from \eqn{eq:deltachi2}.
The right column of each pair shows the computed \pvalue{}
 for the $\Delta\chi^{2}$ with 1 degree of freedom.
\label{tab:deltachi2_bebc_deut_k6}
}
\end{table}

\begin{table}[btp]
\begin{tabular}{l|rr|rr}
    & \multicolumn{2}{c|}{$\Qtmin=0.06~\GeVsq$}
    & \multicolumn{2}{c}{$\Qtmin=0.20~\GeVsq$}
 \\[.25em]
Fit
 & $\chi^{2}/\dof$ & $p_{\Delta \chi^{2}}$
 & $\chi^{2}/\dof$ & $p_{\Delta \chi^{2}}$
 \\[.25em]
\hline&&\\[-1em]
BEBC                           &      6.1/\phantom{1}12    &&      6.1/\phantom{1}12    \\[.25em]
Deuterium                      &    108.3/106              &&     93.2/100              \\[.25em]
Deuterium+BEBC                 &    116.2/114              &&    100.2/108              \\[.25em]
$\Delta \chi^{2}$              &      1.8/\phantom{10}1    &                 0.18
 &      0.9/\phantom{10}1    &                 0.34 \\[.25em]
\end{tabular}
\caption{
The same as \tbl{tab:deltachi2_bebc_deut_k6},
 but for $\kmax=7$ and $\lambda=0.1$.
\label{tab:deltachi2_bebc_deut_k7}
}
\end{table}

\subsubsection{All Deuterium versus MINERvA}
\label{sec:deutx_minerva}

The $\Delta\chi^{2}$ tests from \sct{sec:addition_bebc}
 demonstrate that the complete set of four deuterium results
 are sufficiently compatible that they can be averaged together.
However, the two different \Qtmin{} cuts still exhibit
 a $\sim3\sigma$ shift due to the apparent degeneracy
 between normalization and form factor shape,
 seen in \fgr{fig:addition_bebc_summary}.
Ideally, if the \minerva{} hydrogen results are compatible
 with the complete set of deuterium results, these additional data
 would pin down the remnant degeneracy and remove this uncertainty,
 providing a precise axial form factor determination from all
 elementary target data sources.
The compatibility between these datasets will be explored in more detail
 in this subsection.

\begin{figure*}[hbt!]
\begin{tabular}{ll}
\includegraphics[width=0.45\textwidth]{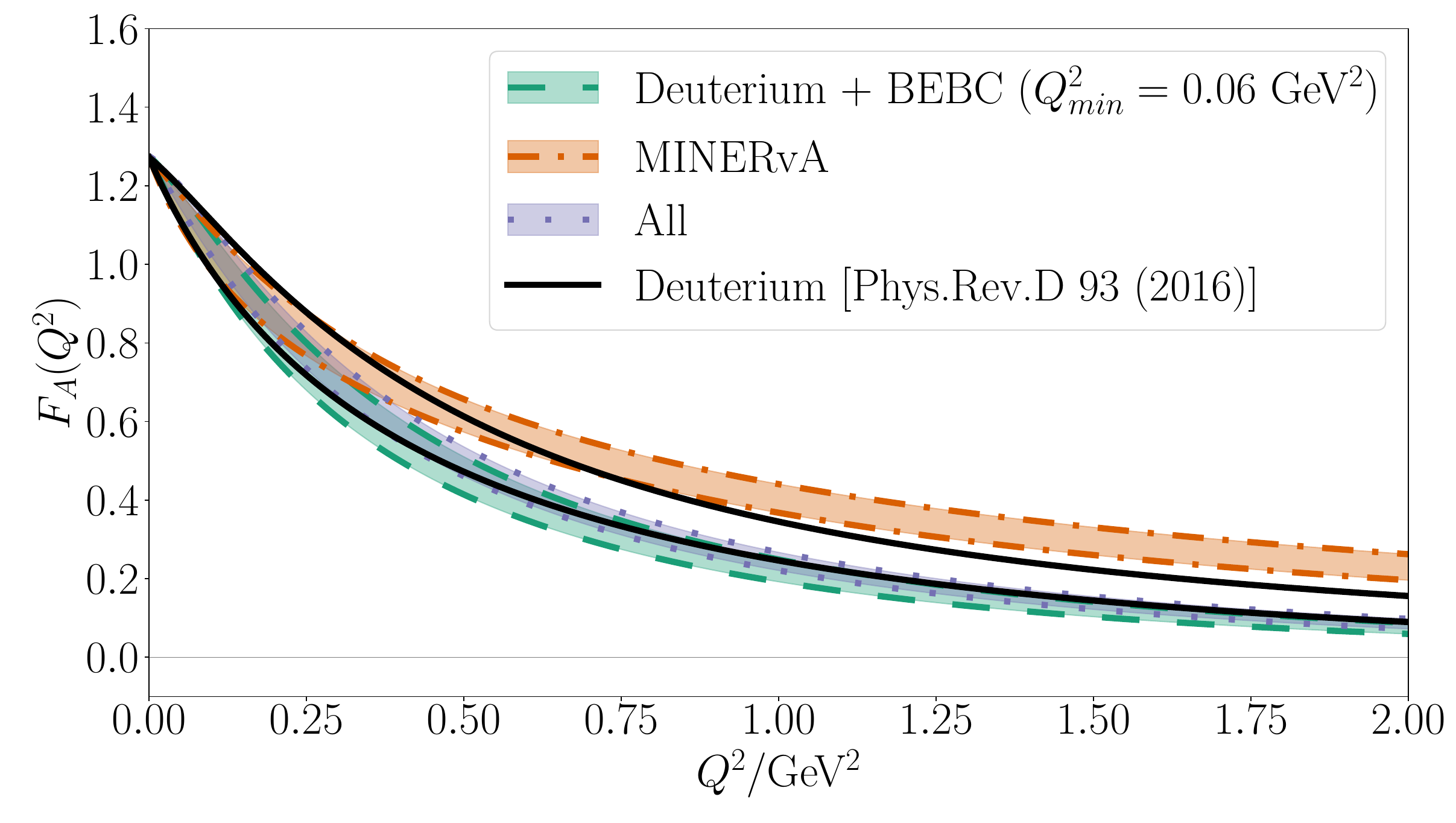}
&
\includegraphics[width=0.45\textwidth]{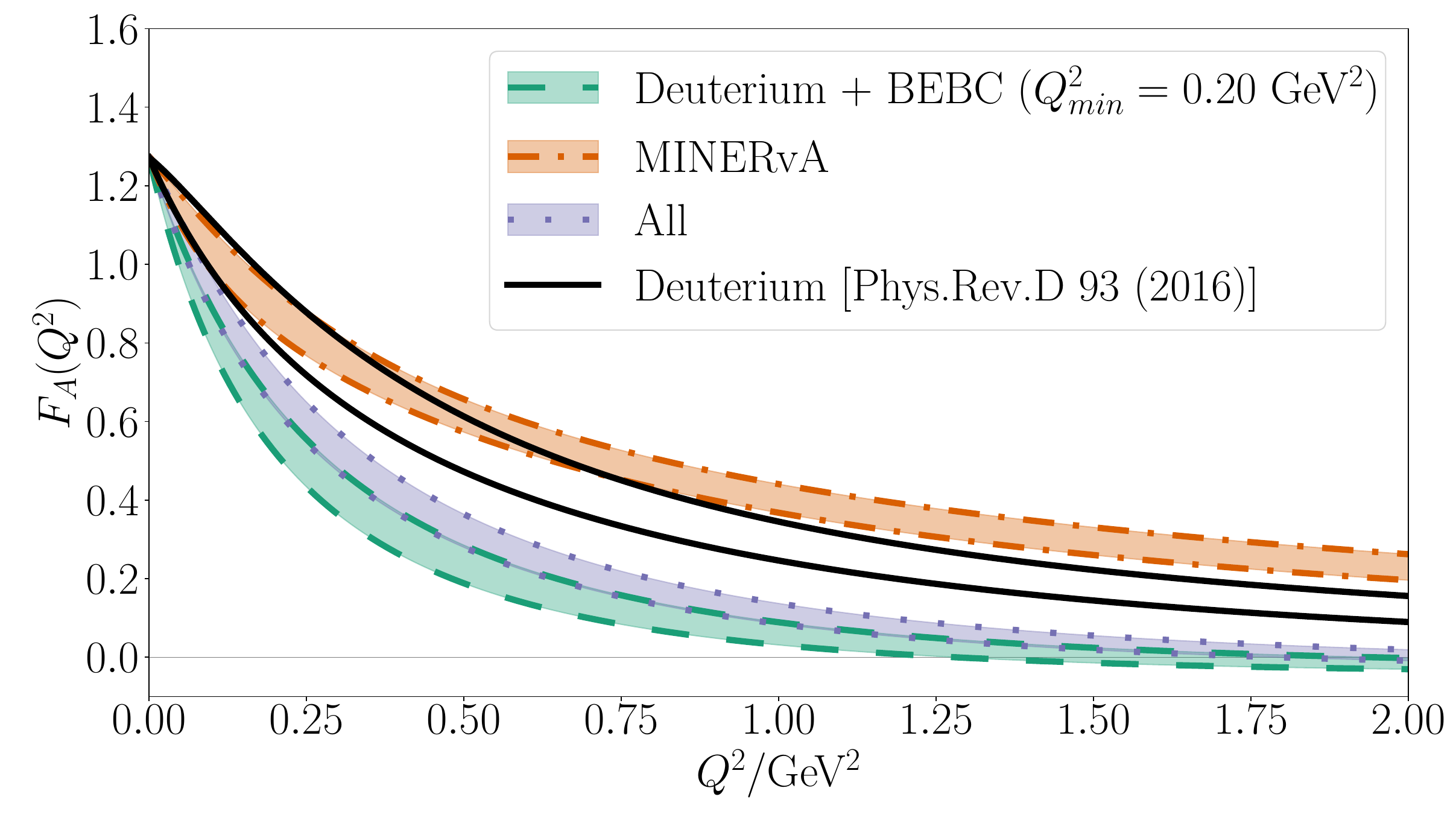}
\end{tabular}
\caption{
Axial form factor results for fits
 comparing the complete set of deuterium data versus the \minerva{} dataset.
In these fits, $\kmax=6$ is chosen and the \zexp{} parameters are unregularized ($\lambda=0$).
The left (right) panel shows the form factors obtained
 when the $0.06~\GeVsq$ ($0.20~\GeVsq$)
 cuts are applied to the event distribution datasets.
The teal shaded region bounded by dashed lines indicates the
 fit to the complete set of deuterium results, subject to the \Qtmin{} cut.
The orange shaded region bounded by the dot-dashed line shows the
 fit to the \minerva{} dataset in isolation (same in both panels).
The blue-violet shaded region bounded by the dotted line is the combined fit to all datasets.
The result from \rfr{Meyer:2016oeg} is given by the unfilled region bounded by solid black lines.
\label{fig:faq2_deutx_minerva}}
\end{figure*}

A comparison of the \minerva{} dataset to the rest of the deuterium datasets
 is shown in \fgr{fig:faq2_deutx_minerva}.
Both sets of data have comparable constraints on the axial form factor uncertainty.
However, a clear discrepancy between \minerva{} and the deuterium
 can be seen at moderate and high $Q^{2}$ values for both \Qtmin{} cuts.
\minerva{} prefers a slower falloff than the deuterium datasets,
 larger by $>2\sigma$ even for the lower \Qtmin{} cutoff.
The combined fit including all datasets is dominated by the deuterium fits,
 exhibiting a slight pull toward the \minerva{} but remaining mostly pinned
 to the deuterium result.

The $\Delta\chi^{2}$ compatibility tests comparing
 the \minerva{} dataset to the deuterium datasets
 are shown in
 \tbls{} \ref{tab:deltachi2_deutx_minerva_k6}
 and \ref{tab:deltachi2_deutx_minerva_k7}.
Although the \pvalue{}s are nearly acceptable for the
 $\Qtmin=0.06~\GeVsq$ fits,
 and in fact the $\kmax=7$ regularized fit in
 \tbl{tab:deltachi2_deutx_minerva_k7} rounds up to $p_{\Delta\chi^{2}}=0.05$,
 the agreement is artificial due to the additional
 poorly-described fit data included in the lower $Q^{2}$ cut range.
There are 6 data bins that are cut when increasing the cut from
 $\Qtmin=0.06~\GeVsq$ to $0.20~\GeVsq$.
The $\chi^{2}_{\rm data}$ of the ``All Deuterium''
 fit in \tbl{tab:deltachi2_deutx_minerva_k6}
 decreases by 17.3 when those low-$Q^{2}$ data are cut,
 which is not commensurate with the decrease in the number of bins.

\begin{table}[btp]
\begin{tabular}{l|rr|rr}
    & \multicolumn{2}{c|}{$\Qtmin=0.06~\GeVsq$}
    & \multicolumn{2}{c}{$\Qtmin=0.20~\GeVsq$}
 \\[.25em]
Fit
 & $\chi^{2}/\dof$ & $p_{\Delta \chi^{2}}$
 & $\chi^{2}/\dof$ & $p_{\Delta \chi^{2}}$
 \\[.25em]
\hline&&\\[-1em]
All Deuterium                  &    116.8/108              &&     99.5/102              \\[.25em]
MINERvA                        &      9.2/\phantom{1}11    &&      9.2/\phantom{1}11    \\[.25em]
All                            &    130.4/120              &&    121.1/114              \\[.25em]
$\Delta \chi^{2}$              &      4.4/\phantom{10}1    &  0.04
 &     12.4/\phantom{10}1    &  $4\times10^{-4}$ \\[.25em]
\end{tabular}
\caption{
$\Delta \chi^{2}$ tests of compatibility comparing
 all deuterium datasets versus the \minerva{} dataset.
The format of this table is the same as for \tbl{tab:deltachi2_bebc_deut_k6}.
These values are for fits with $\kmax=6$ and $\lambda=0$.
\label{tab:deltachi2_deutx_minerva_k6}
}
\end{table}

\begin{table}[btp]
\begin{tabular}{l|rr|rr}
    & \multicolumn{2}{c|}{$\Qtmin=0.06~\GeVsq$}
    & \multicolumn{2}{c}{$\Qtmin=0.20~\GeVsq$}
 \\[.25em]
Fit
 & $\chi^{2}/\dof$ & $p_{\Delta \chi^{2}}$
 & $\chi^{2}/\dof$ & $p_{\Delta \chi^{2}}$
 \\[.25em]
\hline&&\\[-1em]
All Deuterium                  &    116.2/114              &&    100.2/108              \\[.25em]
MINERvA                        &      9.0/\phantom{1}17    &&      9.0/\phantom{1}17    \\[.25em]
All                            &    129.2/126              &&    121.3/120              \\[.25em]
$\Delta \chi^{2}$              &      4.0/\phantom{10}1    &  0.05
&     12.1/\phantom{10}1    &  $5\times10^{-4}$ \\[.25em]
\end{tabular}
\caption{
The same as \tbl{tab:deltachi2_deutx_minerva_k6},
 except for fits with $\kmax=7$ and $\lambda=0.1$.
\label{tab:deltachi2_deutx_minerva_k7}
}
\end{table}

To further illustrate this incompatibility,
 consider the changes to $\chi^{2}$ from inclusion
 of the \minerva{} dataset.
Comparing the ``All Deuterium'' fits to the ``All'' fits
 in \tbl{tab:deltachi2_deutx_minerva_k6},
 introduction of the \minerva{} dataset (with 14 bins)
 increases $\chi^{2}_{\rm data}$ by
 only 13.6 for $\Qtmin=0.06~\GeVsq$,
 versus 21.6 for $\Qtmin=0.20~\GeVsq$.
The $\Delta\chi^{2}$ from each of these fit comparisons
 is almost entirely attributed to shifting the \minerva{}
 data to accommodate the deuterium datasets:
 the contribution of the \minerva{} dataset to
 $\chi^{2}_{\rm data}$ of the ``All'' fit
 is 13.8 for $\Qtmin=0.06~\GeVsq$
 and 19.2 for $\Qtmin=0.20~\GeVsq$
 (a difference from the \minerva{} fit $\chi^{2}_{\rm data}$
 in isolation by 4.6 and 10.0 respectively).

The $\Delta\chi^{2}$ difference for $\Qtmin=0.06~\GeVsq$
 is therefore less than for $\Qtmin=0.20~\GeVsq$ only because
 poorly-described low-$Q^{2}$ data tend to bias the
 axial form factor toward a shape that is not as far
 from the preferred shape of the \minerva{} dataset.
When these data are cut, the preferred ``All Deuterium''
 fit result shifts farther from the preferred \minerva{} fit result.
The \pvalue{} degrades more significantly when the \minerva{} data are added
 to the higher \Qtmin{} cut fit
 and the deuterium datasets are no longer considered to be compatible.
Given other doubts cast by corrections due to the spectator
 nucleon in the deuterium nucleus,
 the unknown absolute normalization of the data,
 and potentially poorly-characterized corrections to the low-$Q^{2}$ data,
 the deuterium datasets should be viewed with some skepticism.

This conclusion also comes in the historical context of predictions from \lqcd{} results~\cite{
 RQCD:2019jai,
 Park:2021ypf,
 Djukanovic:2022wru,
 Jang:2023zts,
 Alexandrou:2025tfv,
 Meyer:2022mix}.
\lqcd{} has consistently predicted a slower falloff of the axial form factor
 with respect to the other deuterium results,
 a conclusion that was presented
 concurrent with Monte Carlo tuning efforts with similar conclusions and
 before the public release of the \minerva{} dataset.
These statements from \lqcd{} come after significant investment from multiple groups,
 including internal self-consistency checks as well as
 nontrivial cross-checks between different collaborations.
Systematic effects from these calculations are understood and well-controlled.
More detailed discussion about these comparisons is deferred
 to the sister paper,~\rfr{Meyer:inprep}.

\subsubsection{\texorpdfstring{\minerva}{MINERvA}--\bebc{} Compatibility}
\label{sec:compatibility_minerva_bebc}

Since agreement between \minerva{} and the other deuterium datasets
 is contingent on inclusion of the poorly-fit low $Q^{2}$ bins,
 the \minerva{} dataset is considered to be inconsistent with the deuterium datasets.
The lost correlations between $Q^{2}$ and $E_{\nu}$
 of these historical deuterium event data and the
 absence of nuclear corrections in the \minerva{} dataset
 lend credence to the \minerva{} as the more accurate estimate
 of the form factor and its uncertainty.
However, the \bebc{} dataset,
 which is presented as a differential cross section
 rather than an unnormalized event distribution
 and has a larger overall uncertainty when including the flux uncertainty,
 might still be sufficiently compatible with the \minerva{} dataset to consider separately.
This possibility is considered in this subsection.

The $\Delta\chi^{2}$ tests for \bebc{} and \minerva{}
 are given in \tbl{tab:deltachi2_bebc_minerva}.
In these fits, a 10\% flux uncertainty is assumed from the \bebc{} dataset.
The same trend that was observed in \sct{sec:deutx_minerva}
 is seen again when the other deuterium datasets are removed.
The combined fit to both \bebc{} and \minerva{}
 also produces a modest increase in $\chi^{2}_{\rm data}$
 over the two datasets individually, with nearly all of the
 increase coming from the \minerva{} contribution to the $\chi^{2}$.
For completeness, the $\kmax=6$ unregularized combined fit
 contributes $\chi^{2}_{\rm data}\sim14.3$ from the \minerva{} dataset,
 in contrast to only $\chi^{2}_{\rm data}\sim6.8$ from \bebc{}
 (compared to 9.2 and 6.1, respectively,
 from the second column of \tbl{tab:deltachi2_bebc_minerva}).
\tbl{tab:deltachi2_bebc_minerva_n20} provides the same tests,
 but relaxes the \bebc{} normalization uncertainty to 20\%.
No substantial difference is seen for these fits,
 and the \pvalue{}s for $\Delta\chi^{2}$ still fall short
 of the acceptable range for compatibility.

\begin{table}[btp]
\begin{tabular}{l|rr|rr}
    & \multicolumn{2}{c|}{$\kmax=6$ $\lambda=0.0$}
    & \multicolumn{2}{c}{$\kmax=7$, $\lambda=0.1$}
 \\[.25em]
Fit
 & $\chi^{2}/\dof$ & $p_{\Delta \chi^{2}}$
 & $\chi^{2}/\dof$ & $p_{\Delta \chi^{2}}$
 \\[.25em]
\hline&&\\[-1em]
BEBC                           &      6.1/\phantom{10}6    &&      6.1/\phantom{1}12    \\[.25em]
MINERvA                        &      9.2/\phantom{1}11    &&      9.0/\phantom{1}17    \\[.25em]
MINERvA+BEBC                   &     21.2/\phantom{1}18    &&     20.8/\phantom{1}24    \\[.25em]
$\Delta \chi^{2}$              &      5.9/\phantom{10}1    &  0.01
&      5.6/\phantom{10}1    &  0.02 \\[.25em]
\end{tabular}
\caption{
Table for $\Delta\chi^{2}$ tests of compatibility for \bebc{} and \minerva{} datasets,
 in the same format as \tbl{tab:deltachi2_bebc_deut_k6}.
The two pairs of columns are for unregularized $\kmax=6$ and regularized $\kmax=7$ fits.
These fits assume a 10\% uncertainty on the flux normalization of the \bebc{} dataset.
\label{tab:deltachi2_bebc_minerva}
}
\end{table}

\begin{table}[btp]
\begin{tabular}{l|rr|rr}
    & \multicolumn{2}{c|}{$\kmax=6$ $\lambda=0.0$}
    & \multicolumn{2}{c}{$\kmax=7$, $\lambda=0.1$}
 \\[.25em]
Fit
 & $\chi^{2}/\dof$ & $p_{\Delta \chi^{2}}$
 & $\chi^{2}/\dof$ & $p_{\Delta \chi^{2}}$
 \\[.25em]
\hline&&\\[-1em]
BEBC                           &      6.0/\phantom{10}6    &&      6.1/\phantom{1}12    \\[.25em]
MINERvA                        &      9.2/\phantom{1}11    &&      9.0/\phantom{1}17    \\[.25em]
MINERvA+BEBC                   &     19.6/\phantom{1}18    &&     19.2/\phantom{1}24    \\[.25em]
$\Delta \chi^{2}$              &      4.5/\phantom{10}1    &  0.03
&      4.1/\phantom{10}1    &  0.04 \\[.25em]
\end{tabular}
\caption{
The same as \tbl{tab:deltachi2_bebc_minerva},
 but assuming a 20\% uncertainty on the flux normalization of the \bebc{} dataset.
\label{tab:deltachi2_bebc_minerva_n20}
}
\end{table}

This concludes the comparison of different datasets.
Although the deuterium datasets are in agreement with each other,
 there is also a disagreement between the hydrogen dataset from \minerva{}
 and the other deuterium datasets.
This inconsistency implies that the form for $R(Q^{2})$,
 or ratio of the deuterium to hydrogen cross section,
 is different from what is assumed.
This shape dependence will be nontrivial to sort out
 because the differential cross section bins with $Q^{2}$
 are smeared out under the integration over the neutrino flux
 at low $E_{\nu}$.
More neutrino scattering measurements using hydrogen and deuterium targets
 will be necessary to sort out deuterium effects from flux effects.
Until the time of these measurements,
 \lqcd{} would be useful as an additional source of constraint
 on the nucleon axial form factor.

\subsection{Systematics}

In this section,
 systematics of the form factors are studied.
\sct{sec:systematics_parameterization} examines the effects
 of changing \tz{}, \kmax{}, and $\lambda$ over the fits.
\newcommand{\bbba}{BBBA05}%
\sct{sec:systematics_vector} reports the changes and uncertainties
 given when replacing the vector form factor parameterization
 from \rfr{Bradford:2006yz} (referred to as the \bbba{} parameterization)
 with the \zexp{} parameterization from \rfr{Borah:2020gte}.

\subsubsection{Parameterization Choices}
\label{sec:systematics_parameterization}

In principle, the parameterization choice should be insensitive to the choice
 of fit parameters \tz, \kmax, and $\lambda$.
No statistically significant shifts are observed under the different fit choices
 made in this work.

\fgr{fig:faq2_minerva_kmax} shows the effects of choosing between the different
 truncations for the axial form factor power series.
The value of $\lambda$ is adjusted based on the preferred choices
 from the L-curve studies of \sct{sec:lcurve}.
With the adjustment of $\lambda$, overfitting is avoided and the two results
 can be compared to each other directly.
The two fits agree very well over the entire interesting $Q^{2}$
 range with minimal deviation.

\begin{figure}[tbp]
\includegraphics[width=0.45\textwidth]{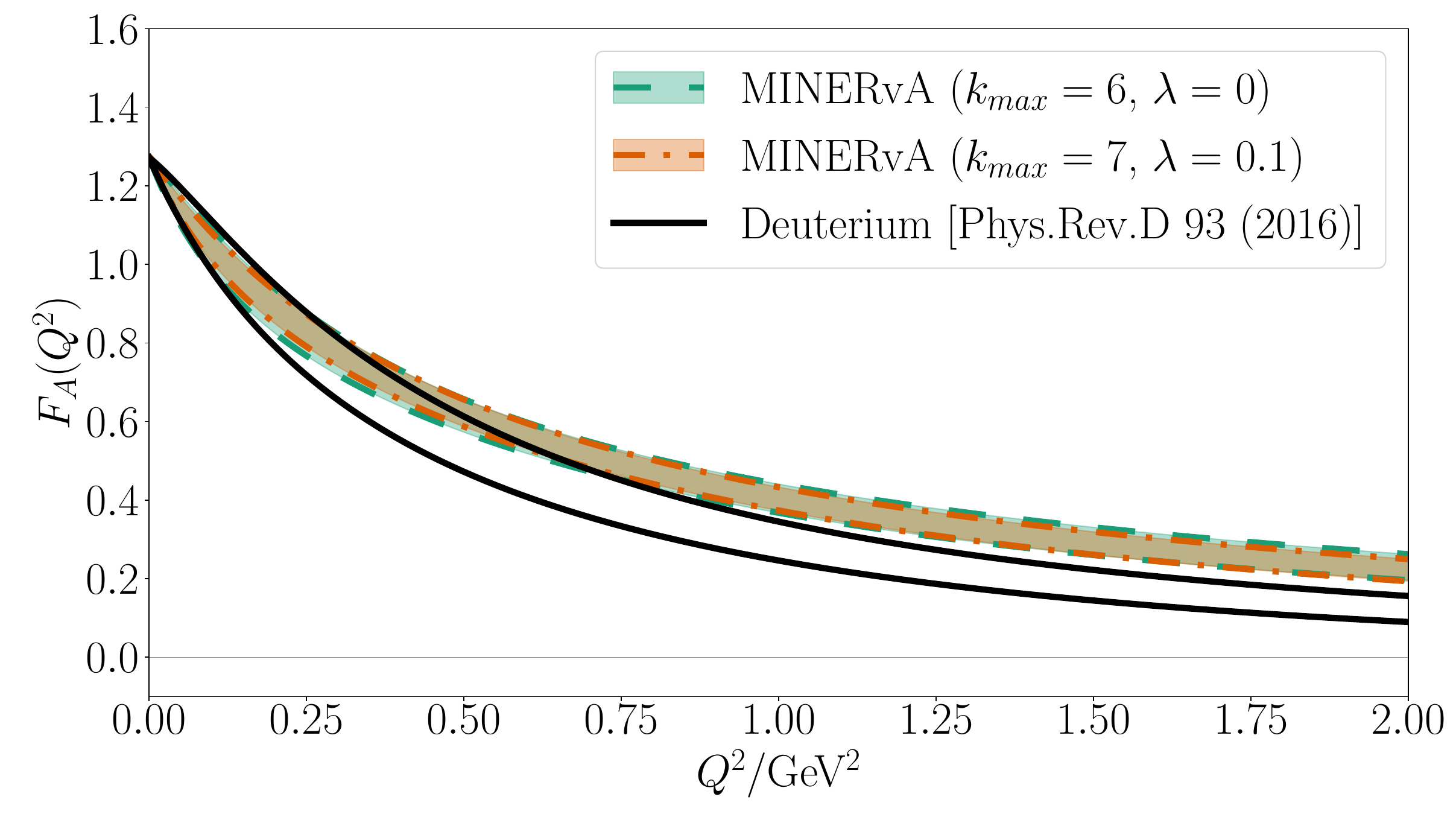}
\caption{
The \minerva{} dataset fit to different choices of \kmax{} and $\lambda$.
The $\kmax=6$ fit is left unregularized ($\lambda=0$),
 while $\kmax=7$ requires a modest regularization parameter
 of $\lambda=0.1$, as described in the text of \sct{sec:lcurve}.
For reference, the form factor result from \rfr{Meyer:2016oeg}
 is plotted as an unfilled region bounded by black lines.
\label{fig:faq2_minerva_kmax}
}
\end{figure}

The selection of \tz{} should be selected as a compromise
 to minimize the maximum value of $|z|$ in a range of $Q^{2}$
 that is interesting for experiments.
The choice $\tz=-0.50~\GeVsq$ used in this work
 more appropriately cover the range $0 < Q^{2} < 6.0~\GeVsq$
 measured by the \minerva{} experiment than the choice
 $\tz=-0.28~\GeVsq$ used in \rfr{Meyer:2016oeg}.
The maximum value of $|z|$ (at $Q^{2}=6.0~\GeVsq$)
 for this choice is around 0.51,
 versus $|z|\approx0.58$ for $\tz=-0.28~\GeVsq$
 and $|z|\approx0.78$ for $\tz=0$.
This choice is also optimized for evaluations within the
 range $0 < Q^{2} < 2.0~\GeVsq$, where $|z|\lesssim0.33$,
 which is relevant for applications to long baseline neutrino oscillation experiments.

\fgr{fig:faq2_minerva_t0} shows the shifts of using a different \tz{} expansion point
 for the \zexp{} parameterizations.
For $\tz=-0.50~\GeVsq$ and $\tz=-0.28~\GeVsq$,
 unregularized parameterizations with $\kmax=6$ are appropriate
 to describe the form factor shape.
When $\tz=0$, the range of $|z|$ is large enough such that a
 $\kmax=6$ parameterization is no longer sufficient to describe the shape.
The data are also not constraining enough to avoid overfitting without
 introduction of a light regularization term as well.
However, when both of these considerations are taken into account,
 the final form factor shape is consistent with the other two choices of \tz.

\begin{figure}[tbp]
\includegraphics[width=0.45\textwidth]{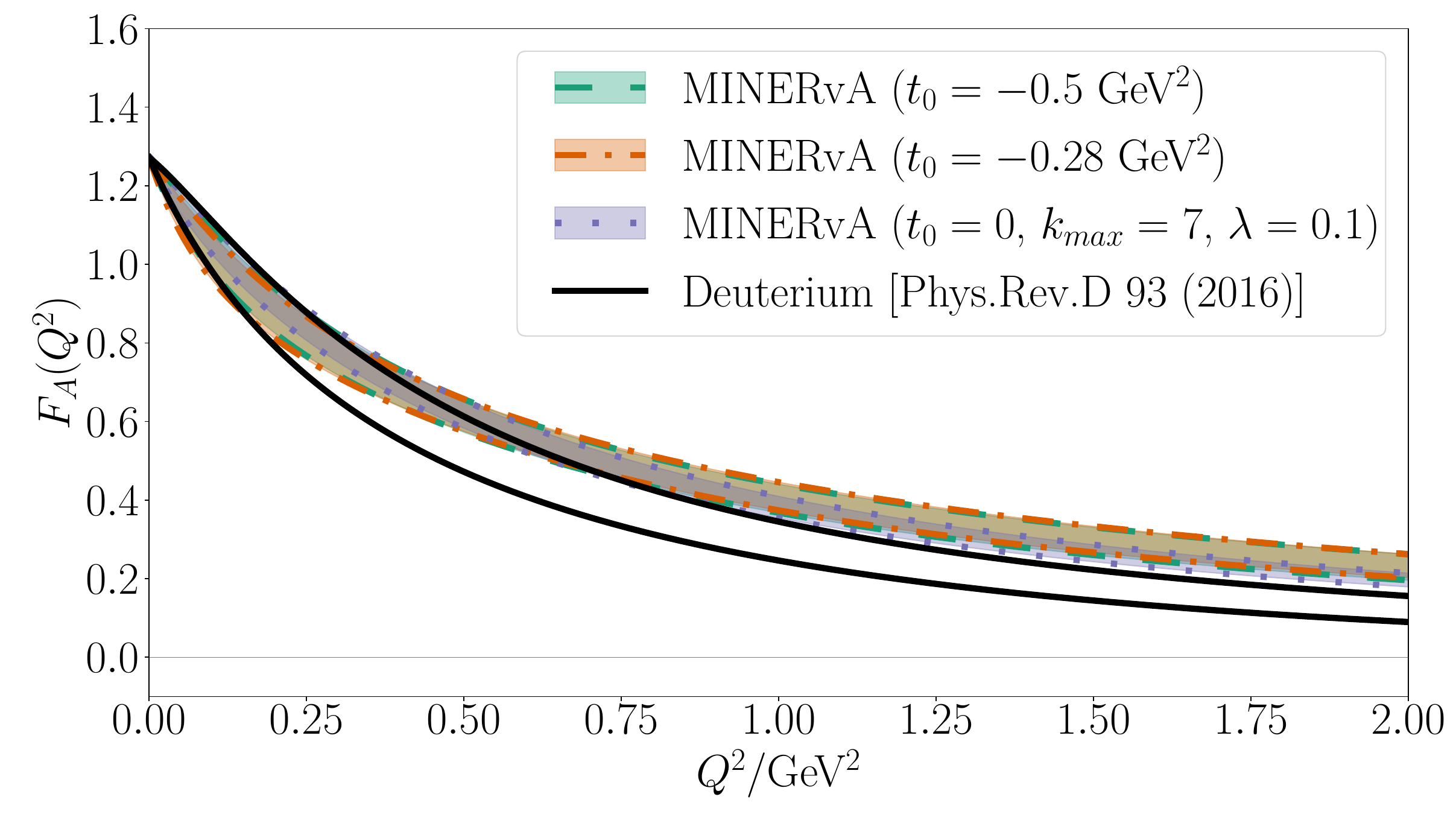}
\caption{
The same as \fgr{fig:faq2_minerva_t0}, but fit to different choices of \tz.
The fits with $\tz=-0.50~\GeVsq$ and $\tz=-0.28~\GeVsq$ have
 $\kmax=6$ and $\lambda=0$.
The fit to $\tz=0$ requires an additional change of \kmax{} and $\lambda$.
\label{fig:faq2_minerva_t0}
}
\end{figure}

\subsubsection{Vector Form Factors}
\label{sec:systematics_vector}

This subsection explores the effects of
 vector form factor parameterizations
 on the axial form factor shape and uncertainty.
In \fgr{fig:faq2_minerva_vectorff},
 the result of replacing the \bbba{} parameterization
 for the vector form factors with the \zexp{} parameterization.
As expected, no substantial difference is seen between the two choices.
The axial form factor uncertainty
 is considerably larger than the difference
 between the parameterizations.

\begin{figure}[tbp]
\includegraphics[width=0.45\textwidth]{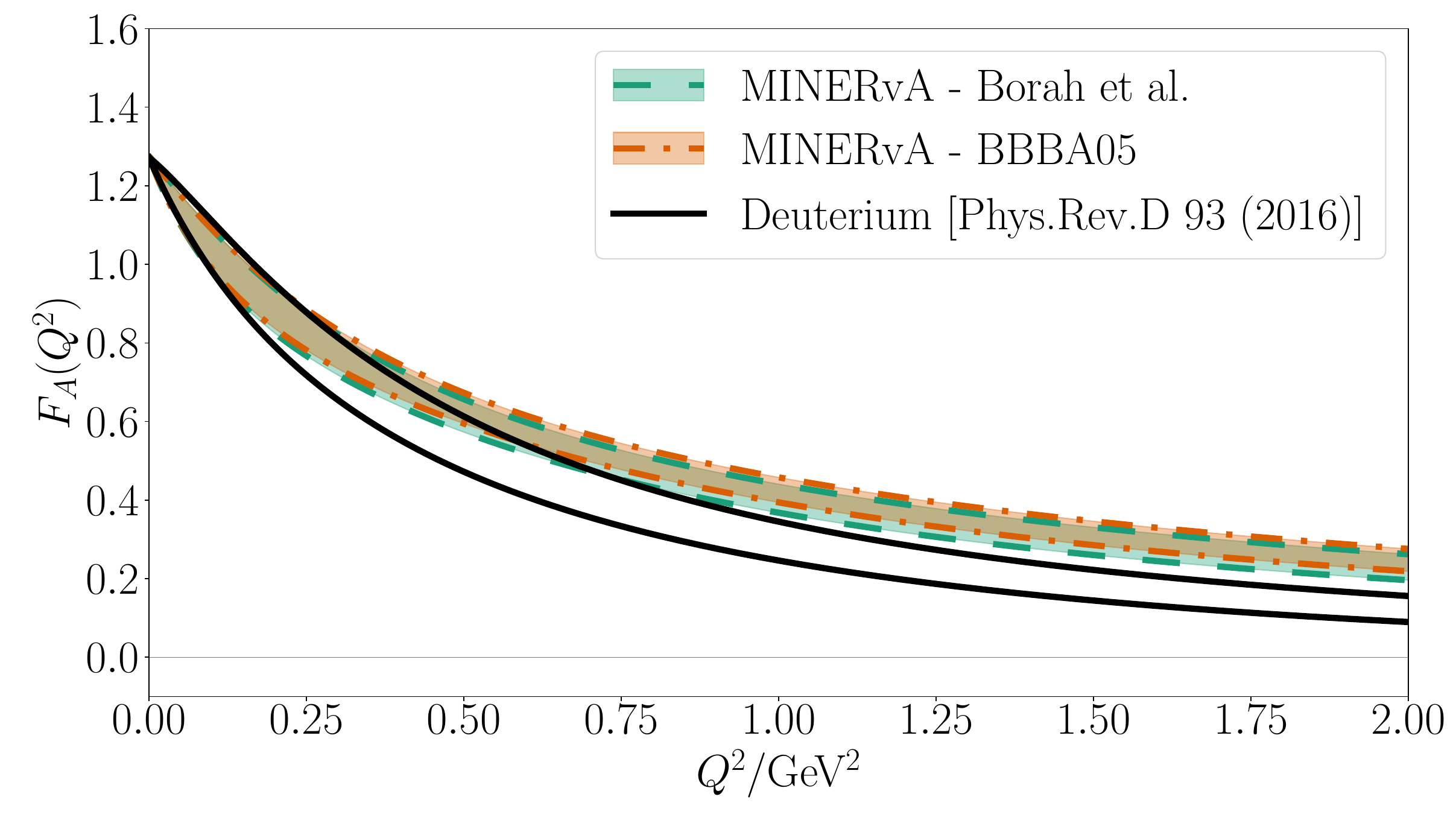}
\caption{
The \minerva{} dataset fit to both
 the \bbba{} vector form factor parameterization~\cite{Bradford:2006yz}
 (teal shading bordered by a solid line)
 and the \zexp{} vector form factor parameterization from \rfr{Borah:2020gte}
 (orange shading bordered by a dashed line).
For reference, the form factor result from \rfr{Meyer:2016oeg},
 fit to the deuterium data, is plotted
 as an unfilled region bounded by black lines.
\label{fig:faq2_minerva_vectorff}
}
\end{figure}

The uncertainty from the \zexp{} parameterization is propagated
 through the fits using principal component analysis
 with the covariance matrix reported in \rfr{Borah:2020gte}.
The mean values of the vector form factor parameters are shifted
 by the principal components extracted from the covariance matrix,
 then new fits are obtained.
After the fit, the difference between the mean values for
 the axial form factor parameters of the nominal and shifted fits
 is taken as an additional uncertainty.
The uncertainties from each of the 8 principal components are then summed
 in quadrature.

\section{Results}
\label{sec:results}

Various fit results are collected in this section.
The central values with their uncertainties are reported
 along with the means for the full set of coefficients,
 including those constrained by sum rules,
 and the covariances of the fit parameters.
The values listed here combined with the parameters
 in \tbl{tab:fitparameters} are sufficient to reproduce
 the form factor curves shown in plots in this work.
In addition, the axial radius squared is provided
 for comparison with experiments that are sensitive to low-$Q^{2}$
 behavior of the form factor.
The squared radius is related to the slope at $Q^{2}=0$
 through the relation
\begin{align}
 \rasq = -\frac{6}{g_{A}} \frac{dF_{A}}{dQ^{2}} \Biggr|_{Q^{2}=0}.
\end{align}

The fits with $\kmax=6$ and $\lambda=0$ are the recommended optimal choice
 of parameters.
This is motivated by two observations:
\begin{enumerate}
\item the L-curve study in \sct{sec:lcurve} for fits to the \minerva{} dataset
 show that moving from an unregularized $\kmax=5$ fit
 to an unregularized $\kmax=6$ fit decreases $\chi^{2}_{\rm aug}$
 by only about 0.2, which when considered in isolation is insufficient
 to justify the increase $\kmax=6$; and
\item
 although the \minerva{} fits prefer $\kmax=5$, the \pvalue{} is not
 significantly diminished at $\kmax=6$ and so comparisons with \lqcd{},
 which prefer $\kmax=6$, can be performed without adversely impacting
 the fit quality.
\end{enumerate}
For evaluating systematics due to finite truncation of the \zexp{} parameterizations,
 the values for $\kmax=7$ and $\lambda=0.1$ fits (or $\lambda=0$ for \lqcd{} alone)
 are also reported.

In all cases considered, the value of the coefficient $a_{1}$ is consistent
 when moving from a $\kmax=6$ to $\kmax=7$ fit.
The value of $a_{2}$ and \rasq{} can change by $\gtrsim2\sigma$,
 suggesting that higher-order form factor shape effects are still not
 fully captured by truncation to $\kmax=6$.
The degraded \pvalue{}s for $\kmax=7$ suggest that the data
 are not sufficiently constraining enough to justify the additional
 fit parameter and that the slight tensions might be the result
 of overfitting rather than the true data preference.

\newcommand{\phm}{\phantom{\ensuremath{-}}}
\subsection{\texorpdfstring{\minerva}{MINERvA} Hydrogen Fit Result}

The final results from fits to the \minerva{}
 data are listed in this subsection.
The $\kmax=6$, $\lambda=0$ fit yields the \zexp{} parameters
\begin{align}
 \big( a_{1}, a_{2} \big)
 &=
 \big(
 -1.65(24), 0.94(30)
 \big)
 \label{eq:finalminerva6}
\end{align}
 with the covariance matrix
\begin{align}
 \left( \begin{array}{lll}
\phm0.05554150 &    -0.03262482 \\
   -0.03262482 & \phm0.09151761 \\
 \end{array} \right) .
\end{align}
With the sum rule constraints applied,
 the central values of the full set of coefficients are
\begin{align}
 &\big( a_{0}, \dots, a_{6} \big) \nonumber\\
 &=
 \begin{array}{clll}
 \big( & \phm0.61490770, &    -1.64778080, & \phm0.94181417, \\
       & \phm0.41239729, & \phm0.36611559, &    -1.18722194, \\
       & \phm0.49976799  & \big) .\\
 \end{array}
\end{align}
The axial radius squared obtained from this parameterization is
\begin{align}
\rasq =  0.575(222)~\fmsq .
\end{align}
This choice has a well-defined $\pvalue\approx0.69$
 from $\chi^{2}\approx9.16$ and 12 degrees of freedom.

The $\kmax=7$ fit with the regularization parameter $\lambda=0.1$
 is also listed as a comparison point.
The fit central values are
\begin{align}
 \big( a_{1}, a_{2}, a_{3} \big)
 &=
 \big(
 -1.69(17), 0.81(34), 0.9(1.2)
 \big)
 \label{eq:finalminerva}
\end{align}
 with the covariance matrix
\begin{align}
 \left( \begin{array}{lll}
\phm0.02771279 &    -0.00093839 &    -0.19136214 \\
   -0.00093839 & \phm0.11854816 &    -0.12082981 \\
   -0.19136214 &    -0.12082981 & \phm1.45823056 \\
 \end{array} \right) .
\end{align}
With the sum rule constraints applied,
 the central values of the full set of coefficients are
\begin{align}
 &\big( a_{0}, \dots, a_{7} \big) \nonumber\\
 &=
 \begin{array}{clll}
 \big( & \phm0.62174048, &    -1.69373431, & \phm0.80639393, \\
       & \phm0.87442257, & \phm0.55213983, &    -2.61742963, \\
       & \phm1.85900234, &    -0.40253521  & \big) . \\
 \end{array}
\end{align}
The axial radius squared obtained from this parameterization is
\begin{align}
\rasq =  0.565(149)~\fmsq.
\end{align}
Treating the priors as additional data,
 the fit give a $\pvalue{}\approx0.96$
 from an augmented $\chi^{2}\approx9.14$ and 18 degrees of freedom.
Ignoring the priors yields $\pvalue{}\approx0.62$
 from data $\chi^{2}\approx8.97$ and 11 degrees of freedom.

\subsection{\lqcd{} Fit Result}

For comparison with the results in this work,
 the result of the sister paper~\cite{Meyer:inprep}
 is reproduced here.
The needed fit parameters match those listed in \tbl{tab:fitparameters}.
The $\kmax=6$, $\lambda=0$ fit is
\begin{align}
 \big( a_{1}, a_{2} \big)
 &=
 \big(
 -1.701(42), 0.263(90)
 \big)
 \label{eq:finallqcd6}
\end{align}
 with the covariance matrix
\begin{align}
 \left( \begin{array}{lll}
\phm0.00175929 &    -0.00294651 \\
   -0.00294651 & \phm0.00807158 \\
 \end{array} \right) .
\end{align}
With the sum rule constraints applied,
 the full set of coefficients are
\begin{align}
 &\big( a_{0}, \dots, a_{6} \big) \nonumber\\
 &=
 \begin{array}{clll}
 \big( & \phm0.72115656, &    -1.70104640, & \phm0.26324902, \\
       & \phm1.53433681, & \phm0.01061114, &    -1.49893610, \\
       & \phm0.67062898  & \big) . \\
 \end{array}
\end{align}
The axial radius squared obtained from this parameterization is
\begin{align}
\rasq =  0.369(30)~\fmsq.
\end{align}
The unknown correlations between \lqcd{} results prevent
 the assignment of a definitive \pvalue.
Instead, covariance derating~\cite{Koch:2024tit}
 is used to estimate a $99\%$ confidence interval upper bound
 on the goodness-of-fit.
This limits $\pvalue\lesssim0.65$.
Assuming the unknown correlations are identically 0,
 the \pvalue{} can be computed directly as
 $\pv\approx0.52$ from $\chi^{2}\approx7.14$
 and 8 degrees of freedom.

The $\kmax=7$, $\lambda=0$ fit is
\begin{align}
 \big( a_{1}, a_{2}, a_{3} \big)
 &=
 \big(
 -1.658(95), 0.431(347), 1.06(97)
 \big)
 \label{eq:finallqcd7}
\end{align}
 with the covariance matrix
\begin{align}
 \left( \begin{array}{lll}
\phm0.00910412 & \phm0.02592048 &    -0.08986544 \\
\phm0.02592048 & \phm0.12033231 &    -0.30490195 \\
   -0.08986544 &    -0.30490195 & \phm0.93874891 \\
 \end{array} \right) .
\end{align}
With the sum rule constraints applied,
 the full set of coefficients are
\begin{align}
 &\big( a_{0}, \dots, a_{7} \big) \nonumber\\
 &=
 \begin{array}{clll}
 \big( & \phm0.72240694, &    -1.65806627, & \phm0.43099261, \\
      & \phm1.06243035, &    -0.68256510, & \phm1.06363538, \\
      &    -1.59271086, & \phm0.65387696  & \big) . \\
 \end{array}
\end{align}
The axial radius squared obtained from this parameterization is
\begin{align}
\rasq = 0.341(62)~\fmsq.
\end{align}
In this case, derating the goodness-of-fit limits
 $\pv\lesssim0.55$
 at $99\%$ confidence.
Assuming the unknown correlations are all identically 0,
 $\pv\approx0.43$ from $\chi^{2}\approx6.99$
 and 7 degrees of freedom.

\subsection{Previous Deuterium Fit Result}

In the interest of concrete comparisons,
 the deuterium results from \rfr{Meyer:2016oeg}
 with $\tz=-0.28~\GeVsq$ are converted to a \zexp{}
 parameterization with $\tz=-0.50~\GeVsq$.
This is possible because the \zexp{} power series coefficients
 are exactly proportional to the derivatives with respect to $z$
 at $Q^{2}=-\tz$ (or $z=0$).
In other words, a \zexp{} parameterization satisfies the proportionality
\begin{equation}
 a_{k} = \frac{1}{k!} \frac{d^{k}F_{A}(z)}{dz^{k}} \Biggr|_{z=0} .
\end{equation}
Therefore, a translated \zexp{} parameterization approximating another parameterization
 can be obtained by examining the central value and derivatives
 of another form factor parameterization at $Q^{2}=-\tz$
 using the new \zexp's definition of \tz{}.
For more details, the reader is referred to~\rfr{Meyer:inprep}.

Using the above prescription for translation,
 the \zexp{} from \rfr{Meyer:2016oeg}
 with $\kmax=8$ translated to
 a parameterization with the values listed in \tbl{tab:fitparameters}
 and $\kmax=6$ gives the coefficients
\begin{align}
 \big( a_{1}, a_{2} \big)
 &=
 \big(
 -2.08(21), 1.90(37)
 \big)
 \label{eq:finalpreviousdeuterium6}
\end{align}
 with the covariance matrix
\begin{align}
 \left( \begin{array}{lll}
\phm0.04304942 & \phm0.02482393 \\
\phm0.02482393 & \phm0.13790576 \\
 \end{array} \right) .
\end{align}
With the sum rule constraints applied,
 the central values of the full set of coefficients are
\begin{align}
 &\big( a_{0}, \dots, a_{6} \big) \nonumber\\
 &=
 \begin{array}{clll}
 \big( & \phm0.54264533, &    -2.08493637, & \phm1.89831616, \\
       & \phm2.40319245, &    -5.88979056, & \phm4.14554900, \\
       &    -1.01497601  & \big) . \\
 \end{array}
\end{align}
This reproduces the form factor central value shape and magnitude with
 no more than 2\% fractional deviation
 over the entire range 0--2~\GeVsq.
The axial radius squared obtained from this parameterization is
\begin{align}
\rasq =  0.334(254)~\fmsq,
\end{align}
 which is in agreement with the published value $0.46(22)~\fmsq$.

Using the above prescription for translation,
 the \zexp{} from \rfr{Meyer:2016oeg}
 with $\kmax=8$ translated to
 a parameterization with the values listed in \tbl{tab:fitparameters}
 and $\kmax=7$ gives the coefficients
\begin{align}
 \big( a_{1}, a_{2}, a_{3} \big)
 &=
 \big(
 -2.08(21), 1.53(48), 2.8(1.8)
 \big)
 \label{eq:finalpreviousdeuterium7}
\end{align}
 with the covariance matrix
\begin{align}
 \left( \begin{array}{lll}
\phm0.04304942 & \phm0.06034134 &    -0.36481368 \\
\phm0.06034134 & \phm0.23520236 &    -0.67627605 \\
   -0.36481368 &    -0.67627605 & \phm3.27220947 \\
 \end{array} \right) .
\end{align}
With the sum rule constraints applied,
 the central values of the full set of coefficients are
\begin{align}
 &\big( a_{0}, \dots, a_{7} \big) \nonumber\\
 &=
 \begin{array}{clll}
 \big( & \phm0.54264533, &    -2.08493637, & \phm1.53196776, \\
       & \phm2.78626545, &    -3.75859858, &    -0.88298095, \\
       & \phm2.94795795, &    -1.08232059  & \big) . \\
 \end{array}
\end{align}
This reproduces the form factor shape and magnitude with
 no more than 3\% fractional deviation
 over the range 0--1~\GeVsq{}
 and no more than 10\% over the range 0--2~\GeVsq.
The axial radius squared obtained from this parameterization is
\begin{align}
\rasq =  0.434(215)~\fmsq,
\end{align}
 which is in good agreement with the published value $0.46(22)~\fmsq$.

\subsection{Combined Hydrogen--LQCD Fit Result}

In this subsection,
 the \minerva{} hydrogen data are fit simultaneous
 with the \lqcd{} results from~\rfr{Meyer:inprep}.
These results are used to create a $\Delta\chi^{2}$ comparison
 with which to assess the compatibility between the hydrogen and \lqcd.

The result for a simultaneous fit to the \lqcd{} results and the \minerva{} dataset with $\kmax=6$ and $\lambda=0$ yields
\begin{align}
 \big( a_{1}, a_{2} \big)
 &=
 \big(
 -1.717(41), 0.317(86)
 \big)
 \label{eq:finaljoint6}
\end{align}
 with the covariance matrix
\begin{align}
 \left( \begin{array}{lll}
\phm0.00165493 &    -0.00268883 \\
   -0.00268883 & \phm0.00737483 \\
 \end{array} \right) .
\end{align}
With the sum rule constraints applied,
 the full set of coefficients are
\begin{align}
 &\big( a_{0}, \dots, a_{6} \big) \nonumber\\
 &=
 \begin{array}{clll}
 \big( & \phm0.71592650, &    -1.71743430, & \phm0.31652883, \\
       & \phm1.58969769, &    -0.23282053, &    -1.27795482, \\
       & \phm0.60605663  & \big) . \\ 
 \end{array}
\end{align}
The axial radius squared obtained from this parameterization is
\begin{align}
\rasq =  0.364(30)~\fmsq.
\end{align}
The derated goodness-of-fit yields $\pv\lesssim0.68$.
With the unknown correlations set to 0,
 the goodness-of-fit is $\pv\approx0.62$
 from data $\chi^{2}\approx19.45$
 and 22 degrees of freedom.

Assuming $\kmax=7$ and $\lambda=0$,
 the result is
\begin{align}
 \big( a_{1}, a_{2}, a_{3} \big)
 &=
 \big(
 -1.776(76), 0.05(29), 2.27(77)
 \big)
 \label{eq:finaljoint7}
\end{align}
 with the covariance matrix
\begin{align}
 \left( \begin{array}{lll}
\phm0.00576457 & \phm0.01550765 &    -0.05596858 \\
\phm0.01550765 & \phm0.08687614 &    -0.19816291 \\
   -0.05596858 &    -0.19816291 & \phm0.59358115 \\
 \end{array} \right) .
\end{align}
With the sum rule constraints applied,
 the full set of coefficients are
\begin{align}
 &\big( a_{0}, \dots, a_{7} \big) \nonumber\\
 &=
 \begin{array}{clll}
 \big( & \phm0.71531031, &    -1.77604402, & \phm0.04912094, \\
      & \phm2.27409531, & \phm0.89742871, &    -5.20892371, \\
      & \phm4.03266727, &    -0.98365481  & \big) . \\
 \end{array}
\end{align}
The axial radius squared obtained from this parameterization is
\begin{align}
\rasq =  0.408(55)~\fmsq.
\end{align}
With this fit, the derated goodness-of-fit is
$\pvalue\lesssim0.65$ at 99\% confidence.
Without correlations, the goodness-of-fit is
 $p\approx0.59$ from augmented $\chi^{2}\approx18.91$
 and 21 degrees of freedom.

\begin{table}[btp]
\begin{tabular}{l|rrrr}
 & \multicolumn{2}{c}{ignore unknown}
 & \multicolumn{2}{c}{derate} \\[.25em]
Fit
 & $\chi^{2}/{\rm DoF}$ & $p_{\Delta \chi^{2}}$ 
 & $\chi^{2}/{\rm DoF}$ & $p_{\Delta \chi^{2}}$
 \\[.25em]
\hline\\[-1em]
LQCD              &      7.1/\phantom{1}8
 &&     6.0/\phantom{1}8
 \\[.25em]
\minerva           &      9.2/16
 &&     9.2/16
 \\[.25em]
\minerva+\lqcd &     19.5/22
 &&    18.5/22
 \\[.25em]
$\Delta \chi^{2}$ &      3.2/\phantom{1}1 & $0.08$
 &  3.4/\phantom{1}1 & 0.07
 \\[.25em]
\end{tabular}
\caption{
The $\Delta\chi^{2}$ fit compatibility test between the \minerva{} and \lqcd{} fits.
Due to the presence of unknown correlations between the \lqcd{} results,
 the compatibility test is split into two computations:
 the first (``ignore unknown'') assumes that all of the unknown correlations are exactly 0,
 and the second (``derate'') uses the covariance derating~\cite{Koch:2024tit}
 technique to extract a 99\% confidence interval upper bound on
 the \pvalue{} from allowed covariance variations.
For each fit, the $\chi^{2}$ and \dof{} are reported.
The $\Delta\chi^{2}$ from \eqn{eq:deltachi2} is reported in the last line
 along with a \pvalue{} for the 1~\dof{} $\Delta\chi^{2}$ check.
\label{tab:deltachi2_minervalqcd}
}
\end{table}

The $\Delta\chi^{2}$ compatibility comparison between the \minerva{} and \lqcd{}
 datasets for the $\kmax=6$ fits are shown in \tbl{tab:deltachi2_minervalqcd}.
Both the tests when ignoring the unknown correlations of the \lqcd{}
 (labeled as ``ignore unknown'') and using the covariance derating (``derating'')
 are listed.
Although there is some apparent tension in the $Q^{2}$ shape of the two fits,
 the tension only has a mild effect on the combined fit $\chi^{2}$.
The resulting combined fit is largely dominated by the \lqcd{} fit
 such that the increase in $\chi^{2}$ for the combined fit is almost
 entirely due to the \minerva{} residuals increasing.
The \pvalue{}s for both choices of correlation treatment
 are above the 5\% threshold to consider them compatible.

\subsection{Deuterium Fit Results}

This subsection lists the results for combined deuterium fits
 including the results of \anl, \bnl, \fnal, and \bebc{} datasets together.
All fits in this section assume that the normalization on the \bebc{} flux is 10\%.
The results will be used to produce error envelopes
 describing the full range of potential form factor shapes allowed
 by the deuterium data.
These results are listed for completeness
 but caution is advised when using these results
 because of the issues discussed throughout this work.

The fit results for the $\kmax=6$ fit
 with a data cut at $\Qtmin=0.06~\GeVsq$ are
\begin{align}
 \big( a_{1}, a_{2} \big)
 &=
 \big(
 -2.10(16), 2.46(20)
 \big)
 \label{eq:results_deuterium_k6_qtmin006_cen}
\end{align}
 with a covariance matrix
\begin{align}
 \left( \begin{array}{lll}
\phm0.02429039 & \phm0.02129680 \\
\phm0.02129680 & \phm0.03924659 \\
 \end{array} \right) .
 \label{eq:results_deuterium_k6_qtmin006_cov}
\end{align}
The full precision coefficients after computing sum rules are
\begin{align}
 &\big( a_{0}, \dots, a_{6} \big) \nonumber\\
 &=
 \begin{array}{clll}
 \big( & \phm0.46232604, &    -2.09518344, & \phm2.46409998, \\
       & \phm1.84891368, &    -6.31439714, & \phm4.92761426, \\
       &    -1.29337338  & \big) . \\
 \end{array}
\end{align}
The computed axial radius squared is
\begin{align}
\rasq = 0.449(194)~\fmsq.
\end{align}
This fit yields
 $\chi^{2}_{\rm aug}\approx121.0$
 and
 $\chi^{2}_{\rm data}\approx116.8$
 with 108 degrees of freedom.

The fit results for the $\kmax=6$ fit
 with a data cut at $\Qtmin=0.20~\GeVsq$ are
\begin{align}
 \big( a_{1}, a_{2} \big)
 &=
 \big(
 -1.73(15), 3.64(20)
 \big)
 \label{eq:results_deuterium_k6_qtmin020_cen}
\end{align}
 with a covariance matrix
\begin{align}
 \left( \begin{array}{lll}
\phm0.02362940 & \phm0.02199389 \\
\phm0.02199389 & \phm0.04075856
 \label{eq:results_deuterium_k6_qtmin020_cov}
 \end{array} \right) .
\end{align}
The full precision coefficients after computing sum rules are
\begin{align}
 &\big( a_{0}, \dots, a_{6} \big) \nonumber\\
 &=
 \begin{array}{clll}
 \big( & \phm0.23685629, &    -1.72881714, & \phm3.63918001, \\
       &    -2.00567449, &    -2.08272958, & \phm2.84871053, \\
       &    -0.90752562  & \big) \\
 \end{array}
\end{align}
The computed axial radius squared is
\begin{align}
\rasq = 1.091(193)~\fmsq.
\end{align}
This fit yields
 $\chi^{2}_{\rm aug}\approx101.1$
 and
 $\chi^{2}_{\rm data}\approx99.5$
 with 102 degrees of freedom.

The fit results for the $\kmax=7$ fit
 with a data cut at $\Qtmin=0.06~\GeVsq$ are
\begin{align}
 \big( a_{1}, a_{2}, a_{3} \big)
 &=
 \big(
 -2.01(13), 2.08(31), 1.6(1.1)
 \big)
\end{align}
 with a covariance matrix
\begin{align}
 \left( \begin{array}{lll}
\phm0.01810872 & \phm0.01599862 &    -0.14396435 \\
\phm0.01599862 & \phm0.09367352 &    -0.21192733 \\
   -0.14396435 &    -0.21192733 & \phm1.23483538 \\
 \end{array} \right) .
\end{align}
The full precision coefficients after computing sum rules are
\begin{align}
 &\big( a_{0}, \dots, a_{6} \big) \nonumber\\
 &=
 \begin{array}{clll}
 \big( & \phm0.44565733, &    -2.00572162, & \phm2.08155251, \\
       & \phm1.62372606, &    -2.79400350, &    -1.44885058, \\
       & \phm3.29177330, &    -1.19413349  & \big) \\
 \end{array}
\end{align}
The computed axial radius squared is
\begin{align}
\rasq = 0.657(176)~\fmsq.
\end{align}
This fit yields
 $\chi^{2}_{\rm aug}\approx120.5$
 and
 $\chi^{2}_{\rm data}\approx116.2$
 with 114 degrees of freedom.

The fit results for the $\kmax=7$ fit
 with a data cut at $\Qtmin=0.20~\GeVsq$ are
\begin{align}
 \big( a_{1}, a_{2}, a_{3} \big)
 &=
 \big(
 -2.01(13), 2.08(31), 1.6(1.1)
 \big)
\end{align}
 with a covariance matrix
\begin{align}
 \left( \begin{array}{lll}
\phm0.01810872 & \phm0.01599862 &    -0.14396435 \\
\phm0.01599862 & \phm0.09367352 &    -0.21192733 \\
   -0.14396435 &    -0.21192733 & \phm1.23483538 \\
 \end{array} \right) .
\end{align}
The full precision coefficients after computing sum rules are
\begin{align}
 &\big( a_{0}, \dots, a_{6} \big) \nonumber\\
 &=
 \begin{array}{clll}
 \big( & \phm0.44565733, &    -2.00572162, & \phm2.08155251, \\
       & \phm1.62372606, &    -2.79400350, &    -1.44885058, \\
       & \phm3.29177330, &    -1.19413349  & \big) \\
 \end{array}
\end{align}
The computed axial radius squared is
\begin{align}
\rasq = 0.657(176)~\fmsq.
\end{align}
This fit yields
 $\chi^{2}_{\rm aug}\approx102.8$
 and
 $\chi^{2}_{\rm data}\approx100.2$
 with 108 degrees of freedom.

\subsection{Summary of fits}

The set of results described in this work is shown in \fgr{fig:summary}.
As discussed in the manuscript,
 the deuterium suffers from a strong dependence on the prior assumptions.
As an attempt to capture the full spread of this dependence,
 the extreme bounds of the two $\Qtmin$ cuts,
 with the lower bound taken from
 \eqns~(\ref{eq:results_deuterium_k6_qtmin020_cen})~%
 and~(\ref{eq:results_deuterium_k6_qtmin020_cov})
 and the upper bound from
 \eqns~(\ref{eq:results_deuterium_k6_qtmin006_cen})~%
 and~(\ref{eq:results_deuterium_k6_qtmin006_cov}),
 are used to produce an uncertainty envelope for the datasets.
This spread would be even worse without the inclusion of the \bebc~dataset.
Even with the inflated uncertainty, the extreme bounds
 fall significantly below the other results at larger $Q^{2}$
 and more than $1\sigma$ for even the results of \rfr{Meyer:2016oeg}.
Above $Q^{2}\approx1.25~\GeVsq$,
 the sign of the form factor is not constrained at more than $1\sigma$.

Although the $\Delta\chi^{2}$ test described in \tbl{tab:deltachi2_minervalqcd} passes a $\pvalue\geq0.05$ test, the \minerva{} and \lqcd{} fits appear to have slightly different values across $Q^{2}$ in \fgr{fig:summary}, outside of the uncertainty of either fit.  As seen in the comparison of \eqn{eq:finalminerva6} to \eqn{eq:finallqcd6}, this is primarily due to the difference in the second-order parameter $a_{2}$, which differs by approximately two standard deviations in the two fits.  However, this tension is only at the modest $2\sigma$ level in $a_{2}$.  With this interpretation, it is easy to understand the correlation in the difference across all values of $Q^{2}$.

Another test of compatibility is the \pvalue{} obtained from the \minerva{} residuals for both the \minerva{}-only and joint fits.  For the joint fit, we find $\chi^{2}_{\minerva}\approx12.10$ for 12 degrees of freedom, yielding $\pv\approx0.44$.  This is to be compared with \minerva{} values in \tbl{tab:deltachi2_minervalqcd}, with $\chi^{2}_{\minerva}\approx12.10$ for 12 degrees of freedom and $\pv\approx0.69$.  Because of the greater precision of the \lqcd{} constraints relative to those of \minerva, the joint fit does not significantly change the \minerva{} residuals, even if the uncertainty band appears to be far from the \minerva{} result."

\begin{figure}[tbp]
\includegraphics[width=0.45\textwidth]{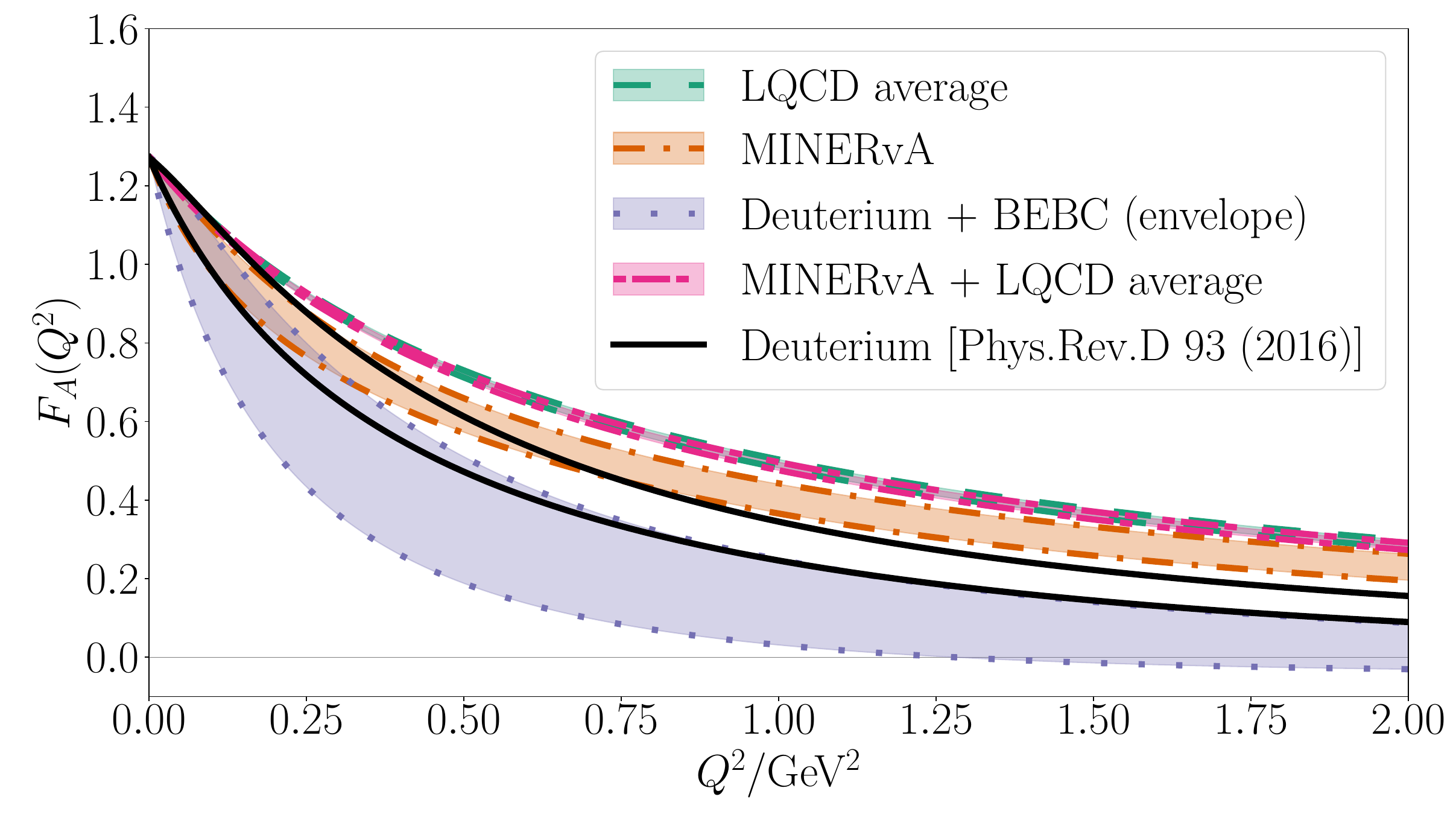} \\
\includegraphics[width=0.45\textwidth]{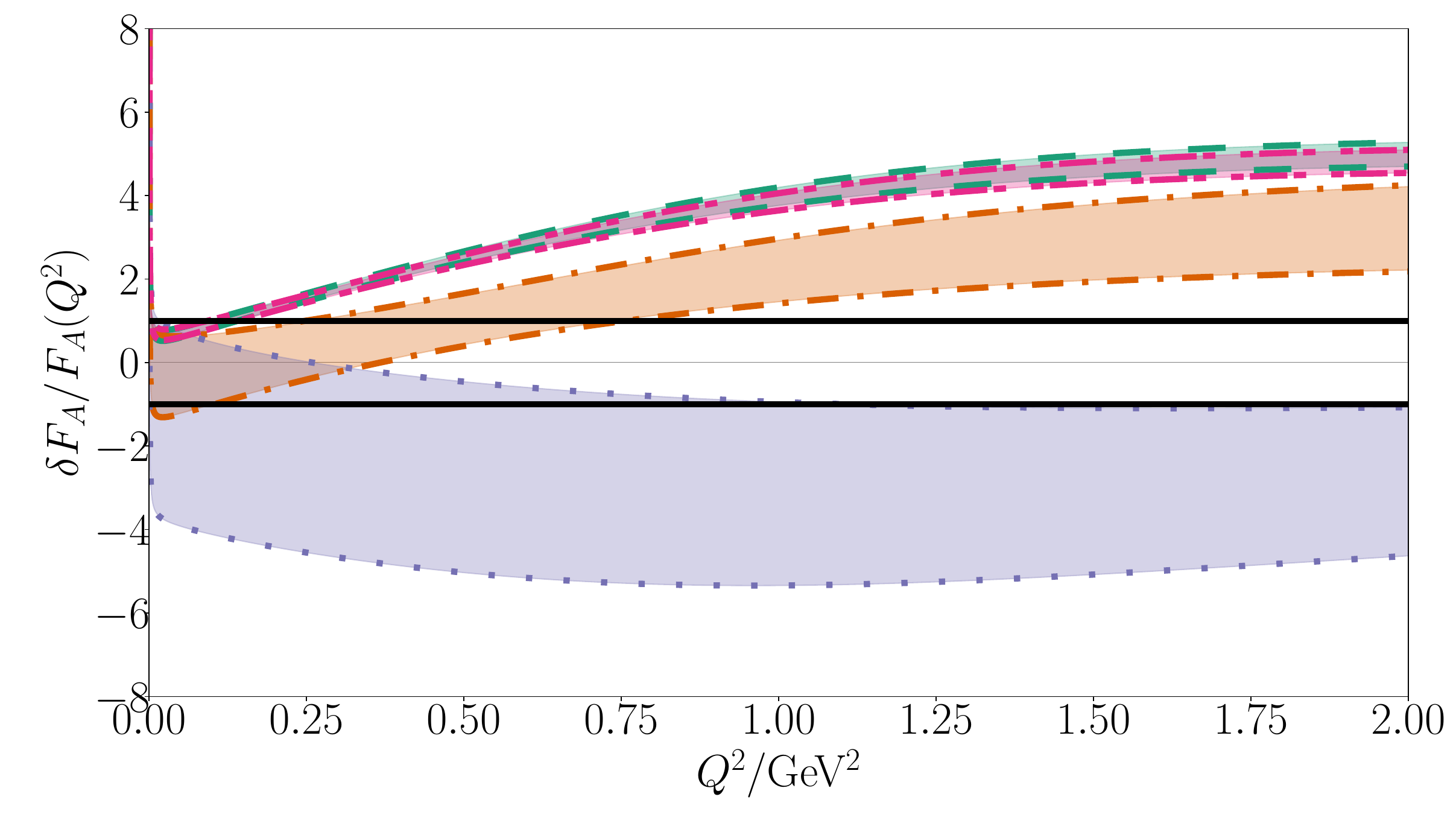}
\caption{
 The top panel shows the final choices for the axial form factor parameterizations
 after the fits discussed in this work.
 The bottom panel shows those same parameterizations normalized to the
 deuterium result of \rfr{Meyer:2016oeg}.
 The \minerva{} and \lqcd{} Average fits are obtained from
 \eqns~(\ref{eq:finalminerva6})~and~(\ref{eq:finallqcd6}), respectively.
 The bounds for the curve labeled
 ``Deuterium+BEBC (envelope)''
 are obtained from the lower bound of the
 ``Deuterium+BEBC ($\Qtmin=0.20~\GeVsq$)''
 and the upper bound of the
 ``Deuterium+BEBC ($\Qtmin=0.06~\GeVsq$)''
 curves in \fgr{fig:addition_bebc_summary}.
 The \minerva{}+\lqcd{} Average fit is taken from \eqn{eq:finaljoint6}.
 \label{fig:summary}
}
\end{figure}

\begin{figure}[tbp]
\includegraphics[width=0.45\textwidth]{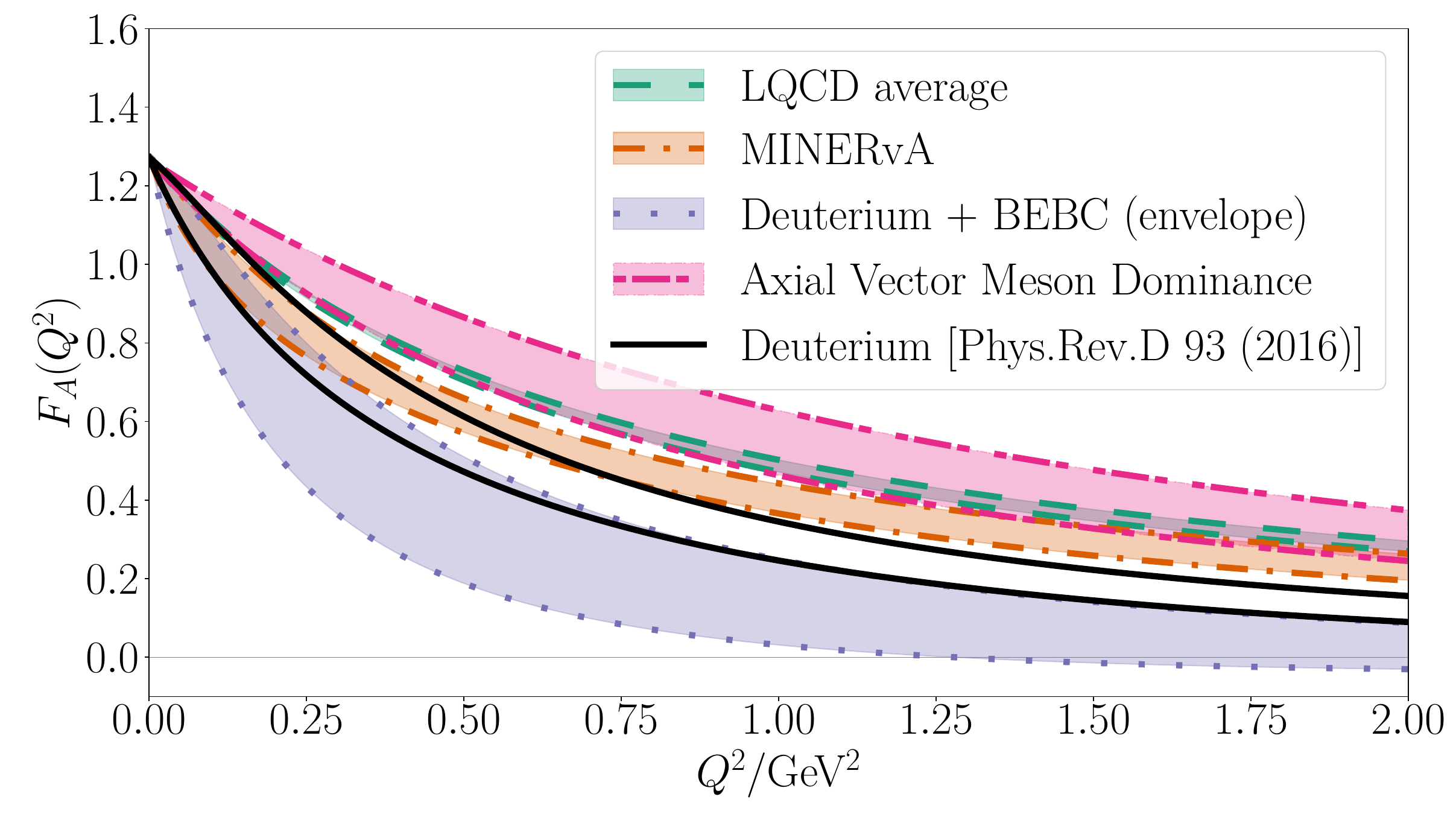} \\
\caption{
 The same as the top panel of \fgr{fig:summary},
 but with the addition of the
 Axial Vector Meson Dominance model
 (pink shaded region with a triple-dash boundary).
 The form factor curve from the combined fit, \eqn{eq:finaljoint6},
 only has minor deviations from the \lqcd{} fit and so is
 omitted for clarity.
 \label{fig:summary_avmd}
}
\end{figure}

There is another zero parameter prediction of the axial form factor in the literature based on an axial vector meson dominance model~\cite{Amaro:2015lga}.  It provides a prediction which is consistent with the \lqcd{} and \minerva{} form factors, within uncertainties of the model and these fit results, albeit with a visibly larger form factor for $Q^2>0.2~\GeVsq$ as shown in Fig.~\ref{fig:summary_avmd}.

\begin{table}[btp]
\begin{tabular}{l|lll}
& \multicolumn{3}{c}{\rasq/\fmsq} \\
Fit & $\kmax=6$ & $\kmax=7$ & Ref.~\cite{Meyer:2016oeg} \\
\hline
\minerva        & 0.575(222) & 0.565(149) \\
\lqcd           & 0.369\phantom{0}(30) & 0.341\phantom{0}(62) \\
{\rm Deuterium} & 0.334(254) & 0.434(215) & 0.46(22) \\
\minerva+\lqcd & 0.364\phantom{0}(30) & 0.408\phantom{0}(55)
\end{tabular}
\caption{
 Summary of the \rasq{} values obtained from fits in this work
 for both choices of \kmax.
 The \minerva{} fit with $\kmax=7$ also imposes the regularization $\lambda=0.1$,
 which accounts for the decrease in uncertainty in moving from $\kmax=6$ to $\kmax=7$.
 Except for the row for {\rm deuterium}, which sets $\lambda=1$,
 all other fits use $\lambda=0$.
\label{tab:ra2_summary}}
\end{table}

The values of \rasq{} obtained during this work are listed in \tbl{tab:ra2_summary}.
The smaller \rasq{} values from the $\kmax=6$ fits
 are consistent with the extraction of \rasq{} from \rfr{Hill:2017wgb},
 which claims $\rasq=0.46(16)~\fmsq$
 in agreement with this work.

Other historical assessments of $r_A^2$ appealed to the dipole parameterization for the axial form factor,
 which doesn't allow sufficient freedom to describe the form factor shape and leads to underestimated uncertainty.
Such works include deuterium scattering and pion electroproduction constraints
 from \rfr{Bodek:2007ym},
 which lists the values $\rasq=0.453(23)~\fmsq$ and $\rasq=0.454(14)~\fmsq$, respectively,
 as well as neutrino-carbon scattering from the MiniBooNE experiment~\cite{MiniBooNE:2010bsu}
 with $\rasq=0.26(7)~\fmsq$~\footnote{%
This is computed from an effective axial mass that is obtained by modifying
 the nuclear binding energy.}
 and the NOMAD
experiment~\cite{Lyubushkin:2008pe} 
with $\rasq=0.42(5)~\fmsq$.
For a comprehensive list of previous axial radius calculations,
 see \rfr{Goharipour:2025yxm}.

The result of the fit to the \minerva~hydrogen dataset
 is significantly higher than the deuterium fit result
 above $Q^{2}\approx0.25~\GeVsq$
 and above the fit result from \rfr{Meyer:2016oeg}
 above $Q^{2}\approx0.6~\GeVsq$.
However, the value of \rasq~is consistent between the hydrogen and deuterium datasets.
This suggests that the shape of the hydrogen data are not well captured
 by the dipole parameterization, which would need to follow closely with the
 result of \rfr{Meyer:2016oeg} to match the same slope at $Q^{2}=0$.
This nontrivial shape is possible to describe well with the \zexp~parameterization.
The agreement between the \rasq~values also suggests that measurements
 sensitive to only to low $Q^{2}$ behavior of the form factor could not be used
 to make strong inferences about the form factor shape at higher $Q^{2}$.

\begin{figure}[tbp]
\includegraphics[width=0.45\textwidth]{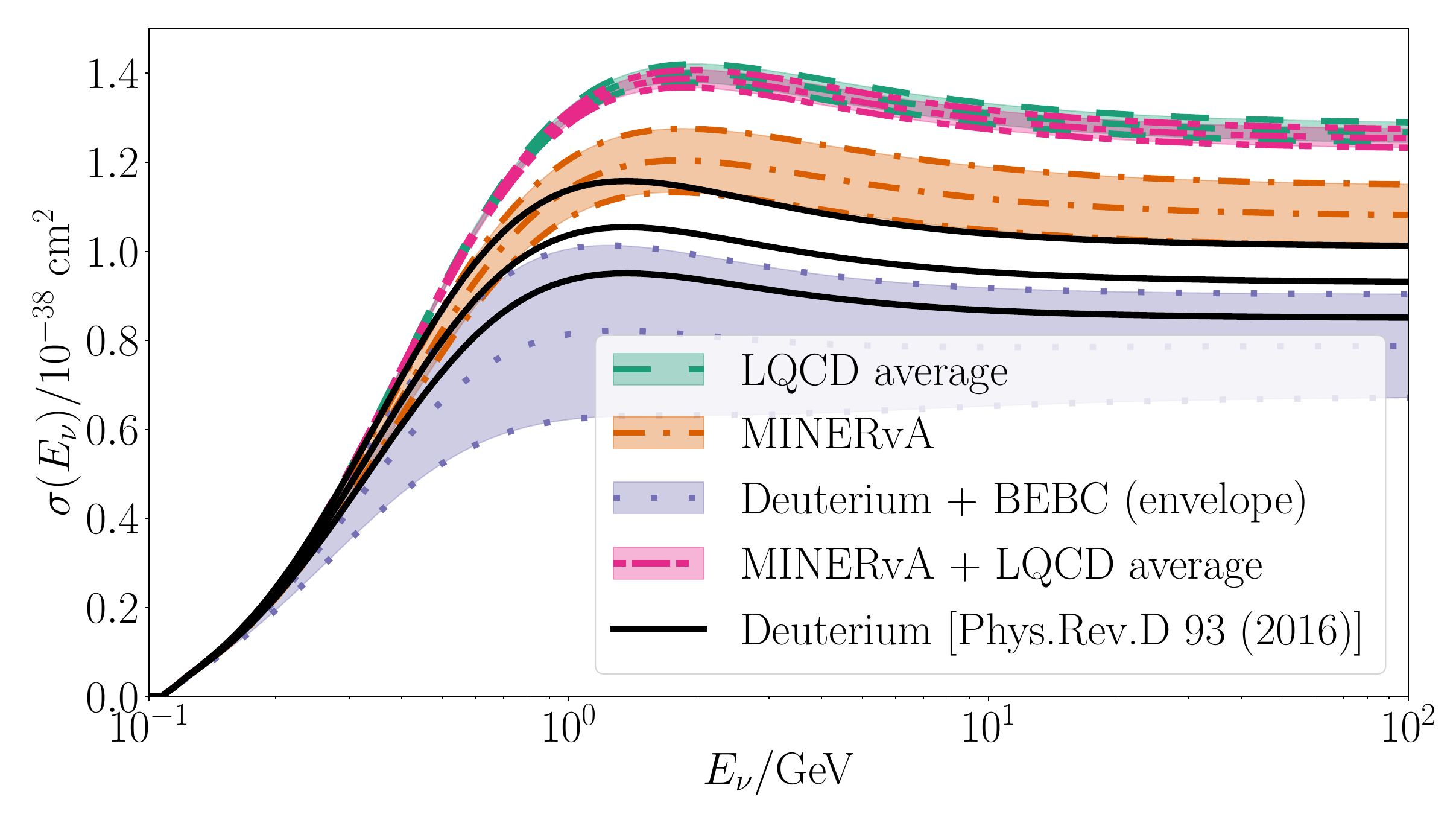} \\
\caption{
 The quasielastic cross section for the interaction $\nu_{\mu}n\to\mu^{-}p$.
 The cross section assumes the same form factor choices used to perform the fits
 in this work.
 For all of the fits, the curves for the central value and $\pm1\sigma$ uncertainties
 are shown as three separate lines, with shading between the uncertainty bounds.
 The deuterium fit (blue-violet dotted) shows the full error envelope of the two \Qtmin{} cuts.
 The previous work of \rfr{Meyer:2016oeg} (unfilled solid black) is shown for comparison.
 \label{fig:cross_section}
}
\end{figure}

The quasielastic cross section obtained from the fits in this work are shown
 in \fgr{fig:cross_section}.
The $Q^{2}$-dependence of the form factors is integrated to produce
 a total cross section.
The slower $Q^{2}$ falloff of the \lqcd{} and \minerva{} results therefore
 corresponds to a larger cross section,
 enhanced by as much as 30-40\% for the \lqcd{} average
 compared to the deuterium result.

\section{Discussion}
\label{sec:discussion}

In this work, neutrino scattering data from multiple elementary
 target sources are examined to determine the most up-to-date knowledge
 for the nucleon quasielastic axial form factor.
This work appeals to historic deuterium bubble chamber datasets,
 a recent work by the \minerva{} collaboration to extract
 the antineutrino-hydrogen scattering cross section,
 and first principles computations with \lqcd.
This builds upon the previous work of \rfr{Meyer:2016oeg}
 and was prepared in conjunction with the sister paper,
 \rfr{Meyer:inprep}.

There are a number of departures from the results of the
 previous work of \rfr{Meyer:2016oeg} that are notable.
Of these, the most significant is the heuristic used to choose
 a set of priors for the \zexp{} coefficients.
In \rfr{Meyer:2016oeg}, the priors were chosen to satisfy
 expectations from unitarity preventing higher-order coefficients
 from being unnaturally large compared to lower-order coefficients.
As a departure from that heuristic, this work uses the relationship
 between $\chi^{2}_{\rm data}$ and $\chi^{2}_{\rm penalty}$
 to choose the point of lowest curvature on an L-curve plot,
 described in \sct{sec:lcurve}.
This corresponds to a compromise between the dependence of
 the fit coefficients on prior widths versus on the data itself.

Using an L-curve heuristic for choosing prior widths in \sct{sec:minimumQ2} revealed
 a degeneracy between the data normalization and form factor shape.
The floating normalizations that were applied to overcome the unknown
 absolute normalization of the neutrino event rates also prevent
 pinning down the shape of the form factor in the absence of a strong prior.
For deuterium fits, the L-curve heuristic in \fgr{fig:lcurve_deutx} indicates
 that a milder prior constraint is preferred by the fits,
 which exposes the degeneracy that manifests
 as a strong dependence on the \Qtmin{} cutoff.
This degeneracy was not observed in the analysis of \rfr{Meyer:2016oeg}
 because of the regularization of the \zexp{} parameters with $\lambda=1$.
This work finds such a strong regularization to be too constraining.
The resulting curve is overfit and likely underestimating the uncertainty,
 much like how the dipole parameterization with only the axial mass as a free parameter
 historically led to underestimated uncertainties.

Relaxing the regularization term by decreasing $\lambda\to0$
 allows for a large range of variation of the form factor.
The effects of the aforementioned degeneracy is partially overcome in this work
 by introducing the \bebc{} dataset to the fits,
 which does have an absolute normalization up to a flux uncertainty
 and can pin the fits down to within about $2\sigma$
 of the \bebc{} central value depending on the \Qtmin{} cut.
However, the lack of absolute normalization still leads to substantial
 systematic uncertainty on the form factor shape.
This is not accounted for explicitly in the fits but is schematically represented
 by an uncertainty envelope in \fgr{fig:summary}.

If a there were a sufficiently well-motivated systematic that could account
 for the range of variation exhibited by the deuterium results,
 inclusion of this would significantly reduce the constraining power
 of the deuterium data relative to the hydrogen and \lqcd.
If this were realized, it seems plausible that the deuterium results
 would be compatible with the \minerva{} and \lqcd{} fits.

The compatibility of neutrino scattering datasets was tested
 with 1 degree of freedom $\Delta\chi^{2}$ comparisons
 outlined in \sct{sec:deltachi2_compatibility}.
Testing compatibility of the neutrino scattering datasets in \sct{sec:minimumQ2}
 shows that the datasets with deuterium targets, including the \bebc{} result,
 are consistent with each other but not with the \minerva{} hydrogen result.
The combined fits including both hydrogen and deuterium are dominated
 by the deuterium fit results.
The minimum preferred by the \minerva{} experiment disagrees with the joint-fit
 minimum (and therefore the deuterium minimum) enough to
 increase the $\Delta\chi^{2}$ and fail a compatibility test.
Although the compatibility is almost permissible for the $\Qtmin=0.06~\GeVsq$
 combined fit, that apparent compatibility is a consequence of the poorly-described
 fit data at low $Q^{2}$ that coincidentally push the combined
 fit result closer to the preferred \minerva{} hydrogen minimum.

The key difference between the single-nucleon results,
 including \minerva{} hydrogen and \lqcd{},
 versus the deuterium results is the falloff behavior
 of the form factor with respect to $Q^{2}$.
The single-nucleon results fall with $Q^{2}$ slower than the deuterium results.
This was consistently realized in the \lqcd{} results,
 which first started reporting slow $Q^{2}$ falloff as early as 2020~\cite{
 RQCD:2019jai,
 Park:2021ypf,
 Djukanovic:2022wru,
 Jang:2023zts,
 Alexandrou:2025tfv}.
Later corroborating evidence has come not only from the \minerva{}~\cite{
 MINERvA:2022vmb,
 MINERvA:2023avz}
 results explored here, but also Monte Carlo tunes~\cite{MicroBooNE:2021ccs,GENIE:2022qrc},
 Generalized Parton Distributions~\cite{Hashamipour:2019pgy,Irani:2023lol},
 and Schwinger functional methods~\cite{
 Chen:2021guo,
 Chen:2022odn}.

Given the model assumptions inherent in using a deuterium target,
 the known efficiency issues at low $Q^{2}$,
 and the lack of historical preservation of the data,
 the shape tension between the hydrogen and deuterium datasets
 calls into question the accuracy of the deuterium data.
This work has made the choice to omit the deuterium data
 as a result of these deficiencies.
Possible sources of the tension could be the lack of
 energy transfer-dependence in the deuterium correction,
 or from growing overlap between the quasielastic
 and $\Delta$ resonance responses at moderate momentum transfers~\cite{Shen:2012xz}.
Even if the deuterium data were still included in the fits,
 the systematic uncertainty that must be introduced
 to account for the lack of absolute normalization would decrease
 the pull of the deuterium data in a combined fit,
 leaving the fits to be dominated mostly by the
 \lqcd{} and \minerva{} results.

If the slower falloff with $Q^{2}$ is born out by future constraints,
 then this will have substantial impacts for neutrino oscillation analyses.
This was demonstrated in \rfr{Meyer:2022mix} by substituting a representative
 \lqcd{} result into the GENIE Monte Carlo event generator.
Neutrino oscillation experiments would see $E_{\nu}$-dependent changes to the
 event rates if the axial form factor is modified from a dipole
 to the new \lqcd{} average.
These observations warrant due caution when tuning to neutrino scattering
 data and advocate for selection of models that are flexible enough
 to account for the full range of possible variations.

The parameterizations for all of the fits obtained in this work are listed in \sct{sec:results},
 including
 the fits to the \minerva{} dataset,
 a translation of the results of \rfr{Meyer:2016oeg}
 to the parameter set used in this work,
 a summary of the \lqcd{} fit results from \rfr{Meyer:inprep},
 and a combined fit to both \minerva{} and \lqcd.
Under the L-curve heuristic, the
 \minerva{} fit tolerates unregularized fits with
 $\kmax=6$ as an optimal choice.
The \lqcd{} result similarly prefers $\kmax=6$ based on fit \pvalue{s}.
The \minerva{} dataset alone achieves a fractional uncertainty
 of 7\% at $Q^{2}=0.50~\GeVsq$, already indicating an improvement
 over the precision assumed in \rfr{Meyer:2016oeg}.
With \lqcd{} results included, the precision improves to
 a 2\% fractional uncertainty at $Q^{2}=0.50~\GeVsq$.
At the level of 1\% fractional uncertainty,
 other systematics would become relevant before further improvement
 could be achieved, such as tensions between vector form factors
 in \rfrs~\cite{Bradford:2006yz}~and~\cite{Borah:2020gte},
 or isospin corrections~\cite{Cirigliano:2022hob}.

\subsection{Recommendations for the Axial Form Factor}

Based on this work,
 we recommend replacing the deuterium results by
 the $\kmax=6$ results for either \lqcd{} or \lqcd{}+\minerva{}
 as the new best form factor parameterization.
Given the observed agreement between the \minerva{} and the \lqcd{} results,
 both of which are truly free nucleon constraints on the axial form factor,
 it appears likely that the deuterium datasets have some systematic
 effect that is not properly accounted for.
This suspicion of the deuterium bubble chamber results,
 combined with an ill-defined inflation of the uncertainty to account
 for both \Qtmin{} cuts, would likely result in the deuterium
 contributing minimally or contributing misinformation in a combined
 fit with all data sources.

The \minerva{} result achieves a reasonable constraint on the uncertainty,
 but the constraint from \lqcd{} is significantly more precise and joint fits
 between the two is dominated by the \lqcd{}.

If one wanted a ``theory-free" determination of the axial form-factor, one could simply use the \minerva{} result.  However, given the consistency between the \lqcd{} and \minerva{} result, there is no clear motivation for doing so.

Although both \lqcd{} and \minerva{} results fit well to $\kmax=7$ parameterizations,
 the improvement to $\chi^{2}$ is not large enough to overcome the decrease
 to the degrees of freedom, resulting in slightly smaller \pvalue{s}.
This suggests that the additional free parameter in the $\kmax=7$ fits
 is likely not well-informed by the data and is only working to inflate the uncertainty.

\subsection{Future Directions}

There are several possibilities of future investigation to refine the present work.
New neutrino scattering data from hydrogen and deuterium targets
 will without doubt contribute to strengthen constraints
 on the shape of the axial form factor.
Studies of modern pion electroproduction data, interpreted with the help of
 appropriately flexible model parameterizations, could provide more information
 about the low momentum transfer region of the axial form factor and the
 squared axial radius.
Simultaneous fits to the axial and vector form factors together
 could yield interesting surprises, specifically by leveraging
 the purely isovector weak interaction to pin down slight degeneracies
 in the vector form factor contributions.
\lqcd{} will also have the same benefit,
 providing linearly independent constraints on the isospin-symmetric
 vector form factors and induced pseudoscalar form factor
 in addition to the axial form factor.
In the near future, \lqcd{} could also make headway towards
 understanding deficiencies in our understanding of deuterium corrections
 at large energy and momentum transfers by producing explicit
 matrix element calculations with two-nucleon systems.
The availability of both free nucleon and deuteron responses
 would provide valuable insights about the nature of interactions
 binding nucleons together into atomic nuclei.

This work has advanced the field another step toward a precise
 axial form factor parameterization from elementary target sources.
Making use of all of available datasets is essential
 for maximizing the physics potential of upcoming
 neutrino oscillation experiments.
With the high-precision experiments currently running
 and new flagship experiments just around the corner,
 supporting theory predictions of neutrino scattering cross sections
 are of critical importance.

\section{Acknowledgments}

The authors would like to thank
 Sara Collins,
 Peter Denton,
 Lukas Koch,
 Andreas Kronfeld,
 Sasha Tomalak,
 Andr\'e Walker-Loud,
 Callum Wilkinson,
 and
 Clarence Wret
 for useful discussions.

The work of A.S.M.\ was performed under the auspices of
 the U.S. Department of Energy by Lawrence Livermore National Laboratory
 under Contract DE-AC52-07NA27344
 and the Neutrino Theory Network Program under Grant DE-AC02-07CHI11359
 and U.S. Department of Energy Award DE-SC0020250.
The work of R.J.H.\ was supported by the U.S. Department of Energy, Office of Science,
 Office of High Energy Physics, under Award DE-SC0019095 (R.J.H).

Other authors performed this work as members of the MINERvA collaboration, and their work was supported using the resources of the Fermi National Accelerator Laboratory (Fermilab), a U.S. Department of Energy, Office of Science, HEP User Facility. Fermilab is managed by Fermi Research Alliance, LLC (FRA), acting under Contract No. DE-AC02-07CH11359.
Support for participating scientists in the MINERvA collaboration was provided by NSF and DOE (USA); by CAPES and CNPq (Brazil); by CoNaCyT (Mexico); by ANID PIA / APOYO AFB180002, CONICYT PIA ACT1413, and Fondecyt 3170845 and 11130133 (Chile); 
by CONCYTEC (Consejo Nacional de Ciencia, Tecnolog\'ia e Innovaci\'on Tecnol\'ogica), DGI-PUCP (Direcci\'on de Gesti\'on de la Investigaci\'on  - Pontificia Universidad Cat\'olica del Peru), and VRI-UNI (Vice-Rectorate for Research of National University of Engineering) (Peru); NCN Opus Grant No. 2016/21/B/ST2/01092 (Poland); by Science and Technology Facilities Council (UK); by EU Horizon 2020 Marie Skłodowska-Curie Action; by a Cottrell Postdoctoral Fellowship from the Research Corporation for Scientific Advancement; by an Imperial College London President's PhD Scholarship.

\appendix

\section{Implementation of Sum Rules}
\label{sec:soYouWantToImplementTheSumRuleConstraint}

The sum rules in
 \eqns~(\ref{eq:sumrules_derivatives})~and~(\ref{eq:sumrules_intercept})
 are a linear system of equations that can be solved analytically.
The \zexp{} coefficients are first separated into two categories,
 namely the coefficients that are fit,
\begin{align}
 a = [ a_{1}, a_{2}, ..., a_{\kmax-4} ],
\end{align}
 and the coefficients that are fixed by sum rule constraints,
\begin{align}
 b = [ a_{0}, a_{\kmax-3}, a_{\kmax-2}, a_{\kmax-1}, a_{\kmax} ].
\end{align}
The intercept $g_{A}$ for the fixed coupling gets its own separate vector,
\begin{align}
 g = [ g_{A}, 0, 0, 0, 0 ].
\end{align}
Then the sum rules can be cast as a matrix equation
\begin{align}
 V b + W a - g = 0
\end{align}
 with
 square matrix $V$ of size $5\times5$
 and rectangular matrix $W$ of size $5\times(\kmax-4)$.
Implementation of the sum rules then amounts to solving for $b$,
\begin{align}
 b = -V^{-1} (W a - g).
\end{align}

The matrix $W$ has as its elements the prefactors in
 \eqns~(\ref{eq:sumrules_derivatives})~and~(\ref{eq:sumrules_intercept}),
 taking the form
\begin{align}
 W =
 \left(\begin{matrix}
 z_0 & z_0^{2} & z_0^{3} & ... & z_0^{K-4} \\
 1 & 1 & 1 & ... & 1 \\
 1 & 2 & 3 & ... & K-4 \\
 0 & 2 & 6 & ... & (K-4)(K-5) \\
 0 & 0 & 6 & ... & (K-4)(K-5)(K-6) \\
 \end{matrix}\right)
\end{align}
 where $K=\kmax$ and $z_0 = z(Q^{2}=0)$
 are used to shorten the expressions.
The matrix $V$ takes a similar form
\begin{widetext}
\begin{align}
 V =
 \left(\begin{matrix}
 1 & z_0^{K-3} & z_0^{K-2} & z_0^{K-1} & z_0^{K} \\
 1 & 1 & 1 & 1 & 1 \\
 0 & K-3 & K-2 & K-1 & K \\
 0 & (K-3)(K-4) & (K-2)(K-3) & (K-1)(K-2) & K(K-1) \\
 0 & (K-3)(K-4)(K-5) & (K-2)(K-3)(K-4) & (K-1)(K-2)(K-3) & K(K-1)(K-2) \\
 \end{matrix}\right) .
\end{align}
Once the parameters \kmax, \tc, and \tz{} are specified,
 the elements of the matrices $V$ and $W$ are fixed
 and may be computed once at the start of a fit.

Writing the inverse as a set of column vectors $v_{i}$,
\begin{align}
 V^{-1} = \frac{1}{\Delta} [ v_{0}, v_{1}, v_{2}, v_{3}, v_{4} ]
\end{align}
 with
\begin{align}
 \Delta =
 2 z_0^{K-3} \Big[ &
 z_0^{3} \left(K - 3\right) \left(K - 2\right) \left(K - 1\right)
 - 3 K z_0^{2} \left(K - 3\right) \left(K - 2\right)
 \nonumber\\
 &+ 3 K z_0 \left(K - 3\right) \left(K - 1\right)
 - K \left(K - 2\right) \left(K - 1\right)
 \Big]
 +12,
\end{align}
 the elements of the inverse matrix are
\begin{align}
 v_{0} =
\left(\begin{matrix}
- 2 K \left(K - 2\right) \left(K - 1\right)
\\
6 K \left(K - 3\right) \left(K - 1\right)
\\
- 6 K \left(K - 3\right) \left(K - 2\right)
\\
2 \left(K - 3\right) \left(K - 2\right) \left(K - 1\right)
\\
12
\end{matrix}\right),
\end{align}
\begin{align}
 v_{1} =
\left(\begin{matrix}
2 K \left(K - 2\right) \left(K - 1\right)
\\
- 6 K \left(K - 3\right) \left(K - 1\right)
\\
6 K \left(K - 3\right) \left(K - 2\right)
\\
- 2 \left(K - 3\right) \left(K - 2\right) \left(K - 1\right)
\\
2 z_0^{K - 3} \Big[
 z_0^{3} \left(K - 3\right) \left(K - 2\right) \left(K - 1\right)
 - 3 z_0^{2} K \left(K - 3\right) \left(K - 2\right)
 + 3 z_0 K \left(K - 3\right) \left(K - 1\right)
 - K \left(K - 2\right) \left(K - 1\right)
 \Big]
\end{matrix}\right),
\end{align}
\begin{align}
 v_{2} =
\left(\begin{matrix}
 - 6 \left(K - 2\right) \left(K - 1\right)
 + z_0^{K-2} \left(K - 2\right) \left(K - 1\right) \Big[
 z_0^{2} \left(K - 3\right) \left(K - 2\right)
 - 2 z_0 K \left(K - 3\right)
 + K \left(K - 1\right)
 \Big]
\\
 6 (K-1) \left( 3K -8 \right)
 -(K-1) z_0^{K-3} \Big[
 2 z_0^{3} \left(K - 4\right) \left(K - 3\right) \left(K - 2\right)
 - 3 z_0^{2} K \left(K - 4\right) \left(K - 3\right)
 + K \left(K - 2\right) \left(K - 1\right)
 \Big]
\\
 -6 (K-3) (3 K - 4)
 + (K-3) z_0^{K-3}
 \Big[
 z_0^{3} \left(K - 4\right) \left(K - 3\right) \left(K - 2\right)
 - 3 z_0 K \left(K - 4\right) \left(K - 1\right)
 + 2 K \left(K - 2\right) \left(K - 1\right)
 \Big]
\\
 6 (K-3) (K-2)
 - (K-3) (K-2) z_0^{K-3}
 \Big[
 z_0^{2} \left(K - 4\right) \left(K - 3\right)
 - 2 z_0 \left(K - 4\right) \left(K - 1\right)
 + \left(K - 2\right) \left(K - 1\right)
 \Big]
\\
 - 6 z_0^{K - 3} \left(z_0 - 1\right) \Big[
 z_0^{2} \left(K - 3\right) \left(K - 2\right)
 - 2 z_0 \left(K - 3\right) \left(K - 1\right)
 + \left(K - 2\right) \left(K - 1\right)
 \Big]
\end{matrix}\right),
\end{align}
\begin{align}
 v_{3} =
\left(\begin{matrix}
 6 \left(K - 2\right)
 -2 \left(K - 2\right) z_0^{K-2} \Big[
 z_0^{2} \left(K - 3\right) \left(K - 1\right)
 - z_0 K \left(2 K - 5\right)
 + K \left(K - 1\right)
 \Big]
\\
 -6 (3K -7)
 +z_0^{K-3} \Big[
 2 z_0^{3} \left(K - 3\right) \left(K - 1\right) \left(2 K - 7\right)
 - 6 z_0^{2} K \left(K - 3\right)^{2}
 + 2 K \left(K - 2\right) \left(K - 1\right)
 \Big]
\\
 6 (3K -8)
 +z_0^{K-3} \Big[
 - 2 z_0^{3} \left(K - 4\right) \left(K - 3\right) \left(K - 2\right)
 + 6 z_0 K \left(K - 3\right)^{2}
 - 2 K \left(K - 2\right) \left(2 K - 5\right)
 \Big]
\\
 -6 (K-3)
 +2 (K-3) z_0^{K-3} \Big[
 z_0^{2} \left(K - 4\right) \left(K - 2\right)
 - z_0 \left(K - 1\right) \left(2 K - 7\right)
 + \left(K - 2\right) \left(K - 1\right)
 \Big]
\\
6 z_0^{K - 3} \left(z_0 - 1\right)^{2} \left(z_0 (K -3) - (K - 2)\right)
\end{matrix}\right),
\end{align}
and
\begin{align}
 v_{4} =
\left(\begin{matrix}
 -2
 +z_0^{K-2} \Big[
 z_0^{2} \left(K - 2\right) \left(K - 1\right)
 - 2 z_0 K \left(K - 2\right)
 + K \left(K - 1\right)
 \Big]
\\
 6
 -z_0^{K-3} \Big[
 2 z_0^{3} \left(K - 3\right) \left(K - 1\right)
 - 3 z_0^{2} K \left(K - 3\right)
 + K \left(K - 1\right)
 \Big]
\\
 -6
 +z_0^{K-3} \Big[
 z_0^{3} \left(K - 3\right) \left(K - 2\right)
 - 3 z_0 K \left(K - 3\right)
 + 2 K \left(K - 2\right)
 \Big]
\\
 2
 -z_0^{K-3} \Big[
 z_0^{2} \left(K - 3\right) \left(K - 2\right)
 - 2 z_0 \left(K - 3\right) \left(K - 1\right)
 + \left(K - 2\right) \left(K - 1\right)
 \Big]
\\
- 2 z_0^{K - 3} \left(z_0 - 1\right)^{3}
\end{matrix}\right).
\end{align}

\end{widetext}

\bibliographystyle{apsrev4-1}
\bibliography{main}

\end{document}